\documentclass[a4paper,11pt]{article}
\pdfoutput=1
\usepackage{jheppub}

\usepackage{tikz,xcolor,hyperref}
\usepackage{soul}
\usepackage{tcolorbox}

\usepackage{amsmath, amssymb, amsthm, graphicx, epsfig, fancyhdr,epsfig, slashed}
\usepackage[normalem]{ulem}
\usepackage{tikzsymbols}
\usepackage{natbib}
\usepackage{float}
\usepackage{tcolorbox}
\usepackage[T1]{fontenc}

\newcommand{\mpl}{M_{\rm Pl}}

\newcommand{\nc}{\newcommand}

\nc{\non}{\nonumber}
\nc{\hsp}{\hspace{0.5cm}}
\nc{\lsp}{\hspace{1cm}}
\nc{\Lsp}{\hspace{2cm}}
\nc{\LLsp}{\lsp\lsp}
\nc{\lra}{\longrightarrow}
\newcommand{\beq}{\begin{equation}}  \newcommand{\eeq}{\end{equation}}
\newcommand{\bea}{\begin{eqnarray}}  \newcommand{\eea}{\end{eqnarray}}
\newcommand{\baa}{\begin{array}}     \newcommand{\eaa}{\end{array}}
\newcommand{\bit}{\begin{itemize}}   \newcommand{\eit}{\end{itemize}}
\newcommand{\ben}{\begin{enumerate}} \newcommand{\een}{\end{enumerate}}
\newcommand{\bce}{\begin{center}}    \newcommand{\ece}{\end{center}}
\newcommand{\bpm}{\begin{pmatrix}}   \newcommand{\epm}{\end{pmatrix}}
\newcommand{\bvt}{\begin{verbatim}}  \newcommand{\evt}{\end{verbatim}}
\nc{\meft}{m_{\text{eff,t}}^2(a)}
\nc{\mefx}{m_{\text{eff,x}}^2(a)}
\nc{\sign}{\text{sign}}
\title{Gravitational production of massive vectors non-minimally coupled to gravity
}

\author[a]{Bohdan Grzadkowski}
\author[b]{and Anna Socha}
\affiliation[a]{Faculty of Physics, University of Warsaw, Pasteura 5, 02-093 Warsaw, Poland}
\affiliation[b]{Laboratoire de Physique de l'\'{E}cole normale sup\'{e}rieure, ENS, Université PSL, CNRS, Sorbonne Universit\'{e}, Universit\'{e} Paris Cit\'{e}, F-75005 Paris, France}
\emailAdd{bohdan.grzadkowski@fuw.edu.pl}
\emailAdd{anna.socha@phys.ens.fr}
\date{\today}

\abstract{A quantum theory of massive Abelian vector bosons with non-minimal couplings to gravity has been studied within an evolving, isotropic, and homogeneous gravitational background. The vectors may play a role of dark matter if stabilizing $Z_2$ symmetry is imposed. In order to construct a gauge invariant theory of massive vectors that couple to the Ricci scalar and Ricci tensor, a generalization of the Stuckelberg mechanism has been invoked. Constraints that ensure consistency of the model had been formulated and corresponding restrictions upon the space of non-minimal couplings have been found. Canonical quantization of the theory in evolving gravitational background was adopted. Mode equations for longitudinally and transversally-polarized vector bosons were derived and solved numerically. Regions of exponential growth in the solutions of the mode equations have been determined and discussed in detail. The spectral energy density for the three polarizations has been calculated, and the UV divergence of the integrated total energy density has been addressed. Finally, assuming their stability, the present abundance of the vector bosons has also been calculated.
}


\begin{document}
\maketitle

\section{Introduction}
\label{intro}
The mystery of dark matter (DM) still remains unexplained. Although there exists an overwhelming experimental evidence confirming a presence of a dark sector through its gravitational effects on visible matter, a deeper understanding of its fundamental nature has yet to be achieved. In pursuit of this, a plethora of DM models have been proposed, the vast majority of which introduce not only new dark degrees of freedom but also beyond the Standard Model (BSM) couplings, for a recent review see Ref.\cite{Cirelli:2024ssz, Bozorgnia:2024pwk}. Usually, theories of DM postulate some interactions between the SM degrees of freedom, DM particles and/or mediators connecting dark and visible sectors. The proliferation of DM scenarios accompanied by BSM interactions has inspired astroparticle physicists to explore more minimal DM theories, which do not require BSM beyond the mere existence of DM particles~\cite{Chung:1998zb, Chung:1998ua, Chung:1998rq, Kolb:1998ki, 
Chung:2001cb}. \\ \\
A plausible dark matter model should not only offer a compelling explanation for its cosmological origin but also predict the existence of stable, non-relativistic DM relics with an abundance consistent with the experimental data measured by the Planck collaboration \cite{Planck:2018jri}. The absence of evidence for non-gravitational interactions of DM renders the purely gravitational dark matter scenario exceptionally intriguing and conceptually elegant. In recent years, this model has garnered significant attention, particularly due to the fact that it has been demonstrated that gravity itself can be the sole and sufficient source of dark particles \cite{Garny:2015sjg, Ema:2016hlw, 
Garny:2017kha,Kolb:2017jvz, Tang:2017hvq, Ema:2018ucl, Chung:2018ayg, Ema:2019yrd, Garcia:2020wiy,Mambrini:2021zpp, Clery:2021bwz, Lebedev:2022cic, Lebedev:2022ljz, Lebedev:2022vwf, Basso:2022tpd, Clery:2022wib}. The purely gravitational production of dark matter is a broad topic, encompassing a variety of phenomena, whose origins are rooted in gravitational interactions. For an excellent recent review, see Ref.\cite{Kolb:2023ydq}. In this work, we focus on the production of dark particles from the vacuum in the early Universe, driven solely by the non-adiabatic expansion of the background metric. This process has been explored in the context of spin-0 \cite{Chung:1998zb, Chung:1998ua, Chung:1998rq, Kolb:1998ki, 
Chung:2001cb, Kolb:2017jvz, Garcia:2023qab, Verner:2024agh, Racco:2024aac}, spin-1/2 \cite{Chung:2011ck, Kolb:2017jvz}, spin-1 \cite{Graham:2015rva, Ahmed:2020fhc, Kolb:2020fwh, Cembranos:2023qph, Ozsoy:2023gnl, Capanelli:2024pzd, Capanelli:2024rlk}, and even spin-3/2 \cite{Casagrande:2023fjk}, or spin-2 \cite{Kolb:2023dzp} particles. In most studies on dark matter, the discussion is often limited to the minimal scenario, where dark particles are only minimally coupled to gravity. However, it has been demonstrated that relaxing this assumption is well justified and and can have significant consequences for scalar particles \cite{Garcia:2023qab, Verner:2024agh}. The theory of massive spin-1 field does not preclude the existence of additional non-minimal couplings to gravity. Such possibility in the context of dark matter has been first highlighted in Ref.\cite{Kolb:2020fwh} and subsequently discussed to some extent in Ref.\cite{Cembranos:2023qph, Ozsoy:2023gnl}. Recent works \cite{Hell:2024xbv, Capanelli:2024pzd, Capanelli:2024rlk} provide a more comprehensive discussion of the phenomenology of non-minimally coupled vectors, identifying instabilities of the model and exploring its viability as a dark matter candidate. The aim of this work is to extend our previous study, where we explored the gravitational production of minimally-coupled spin-1 dark matter \cite{Ahmed:2020fhc}, by offering a more detailed analysis of the dynamics of non-minimally coupled vectors. Specifically, we allow for non-minimal couplings between the vector field and both the Ricci scalar and Ricci tensor. Additionally, we examine the dynamics of the vector field in the post-inflationary universe, considering two reheating scenarios: one with a matter-like equation of state and the other with a radiation-like equation of state. While our analysis remains closely related to the works of \cite{Capanelli:2024pzd, Capanelli:2024rlk}, it provides a more thorough study of the phenomenology and cosmic evolution of non-minimally coupled vectors, extending the work in Ref. \cite{Capanelli:2024rlk} by discussing the dynamics of the transverse modes, incorporating two distinct reheating models, and discussing the consequences of the applied regularization scheme. 
\\ \\
This article is organized as follows.
In sec.~\ref{back}, we describes dynamics of the early universe, considering the evolution of the inflaton field during and after inflation. Non-minimal couplings of vector fields to gravity are introduced in sec.~\ref{vec}. Sec.~\ref{sec:Viability} discusses constraints imposed upon non-minimal couplings. In 
this section, we derive robust limits on the parameters of the model, enabling us to identify the stable region of the theory that is free from instabilities. In sec.~\ref{evol}, we show and discuss properties of solutions of the mode equations for three physical polarizations of the vector field during and after inflation. Then, in sec.~\ref{rho}, we derive the full expression for the energy density of the non-minimally coupled spin-1 field, investigate its cosmic evolution, discuss diffrent methods of regularization, and calculate the relic abundance. Sec.~\ref{sum} contains summary of our findings and outlines potential directions for future work.
\section{Background dynamics} \label{back}
\subsection{Cosmic inflation}
\begin{figure}[h!]
    \centering \includegraphics[scale=0.5]{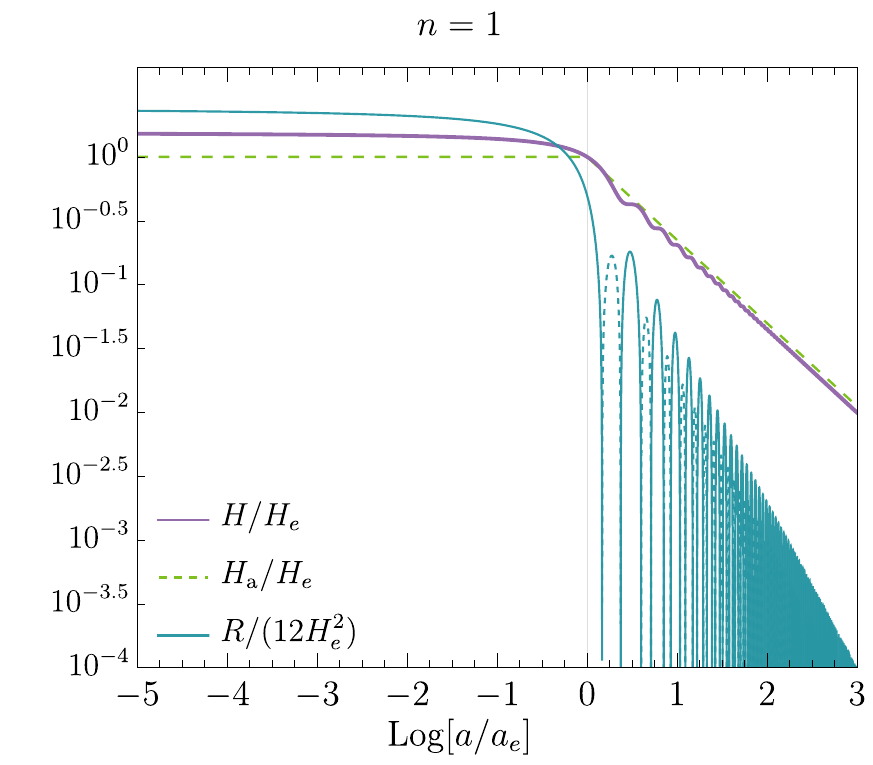}
       \includegraphics[scale=0.5]{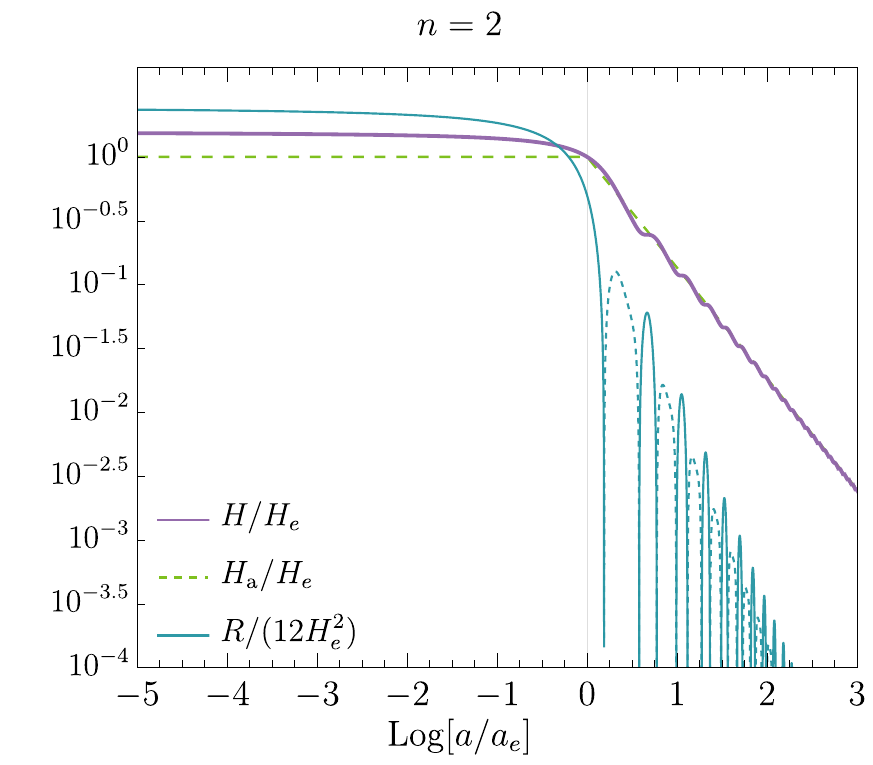}
 \caption{The Hubble rate $H$ (purple solid line) and its analytical approximation, \eqref{eq:Hubble_ap}, (green dashed line), $H_a$, as a function of the number of e-folds in two reheating scenarios: with quadratic (left) and quartic (right) inflaton potential. For completeness we also show the evolution of the Ricci scalar (blue line). The solid (dashed) parts of these curves corresponds to $R<0$ ($R>0$).}
    \label{fig:HRplot}
\end{figure}

Let us assume that the dynamics of the primordial universe is governed by a single scalar field $\phi$ with a minimal coupling to gravity. Its action is given by
\begin{align}
 & \mathcal{S}_\phi \equiv \int d^4 x \sqrt{-g} \left[ \frac{1}{2} g_{\mu \nu} \partial^\mu \phi \, \partial^\nu \phi - V(\phi) \right], \label{eq:phia}
\end{align}
where $g$ denotes the determinant of a metric tensor $g_{\mu \nu}$. Hereafter, we will consider the above action in the spatially flat FLRW metric with the following line element:
\begin{align}
 ds^2 &= dt^2 -a^2(t)d\Vec{x}^2 = a^2(\tau)\left[ d\tau^2 - d\Vec{x}^2\right], \label{FLRW}
\end{align}
with $a$ being the scale factor, and  $\tau$ denoting the conformal time coordinate. \\Above, $V(\phi)$ represents the potential term for the inflaton. As concrete examples, we explore the $\alpha-$attractor T model \cite{Kallosh:2013yoa, Kallosh:2013hoa, Kallosh:2013daa}, \begin{align}
V(\phi) \equiv  \Lambda^4  \tanh^{2n}{ \left( \frac{|\phi|}{\sqrt{6 \alpha} \mpl}\right) } \simeq \begin{cases} \Lambda^4, & |\phi| \gg \sqrt{6 \alpha} \mpl,\\
\Lambda^4 \left( \frac{|\phi|}{\sqrt{6 \alpha} \mpl} \right)^{2n}, & |\phi| \ll \sqrt{6 \alpha} \mpl,
\end{cases} \label{eq:inf_pot}
\end{align}
with $\Lambda$ representing the scale of inflation, and $n, \alpha$ being parameters of the model. Note that $\Lambda$ is bounded from above such that \cite{Ahmed:2022tfm} 
\begin{align}
\Lambda &\simeq \mpl \left(\frac{3 \pi^2}{2} r \Delta_s^2  \right)^{1/4} \lesssim  5.6 \times 10^{-3} \mpl,   \label{eq:LAMBDAconstrain}
\end{align}
where $r$ is the tensor-to-scalar ratio, and $\Delta_s^2$ denotes the amplitude
of scalar perturbations measured at the pivot scale $k_\star^{\rm CMB}$ \cite{Planck:2018jri}, i.e., 
\begin{align}
\Delta_s^2 \equiv \Delta_s^2 (k_\star^{\rm CMB}) = 2.1 \times 10^{-9}. \label{eq:PSS_Planck}
\end{align}
In what follows, we will use the benchmark values of $\alpha=1/6$ and $\Lambda = 3 \cdot 10^{-3} \; \mpl$, together with two well-motivated values for the exponent $n$ corresponding to the quadratic ($n=1$) and quartic ($n=2$) potential near the minimum.\\
Varying the action \eqref{eq:phia} with respect to the metric tensor yields the energy-momentum tensor of the inflaton field
\begin{align}
T_{\mu \nu}^\phi = \partial_\mu \phi \, \partial_\nu \phi - g_{\mu \nu} \left[ \frac{1}{2} g^{\sigma \rho} \partial_\sigma \phi \, \partial_\rho \phi - V(\phi)\right].
\end{align}
The $(0,0)$ component of $T_{\mu \nu}^\phi$ for a homogeneous inflaton field, $\phi(t,\vec{x}) \equiv \phi(t)$, takes the form
\begin{align}
\rho  = T_{0}^{0 \, \phi}  = \frac{1}{2} \dot{\phi}^{2} + V(\phi), \label{eq:phiRHO}
\end{align}
where $\dot{\phi} \equiv d \phi/d t$. The isotropy of the background metric implies that all mixed components vanish, $T_{i0}^\phi=0$, while the diagonal $(i,i)$ spatial components, representing the pressure, are
\begin{align}
p  = \frac{1}{2}\dot{\phi}^2 - V(\phi). 
\end{align}
The dynamics of the inflaton field is described by the classical equation of motion,
\begin{align}
\ddot{\phi} + 3 H \dot{\phi} + V_{, \phi} (\phi)=0,\label{eq:PhiEoM}
\end{align}
where $V_{, \phi} (\phi) \equiv \partial V(\phi) /\partial \phi$, and $H=\dot{a}/a$ denotes the Hubble rate. 
\begin{figure}[H]
    \centering
   \includegraphics[scale=0.4]{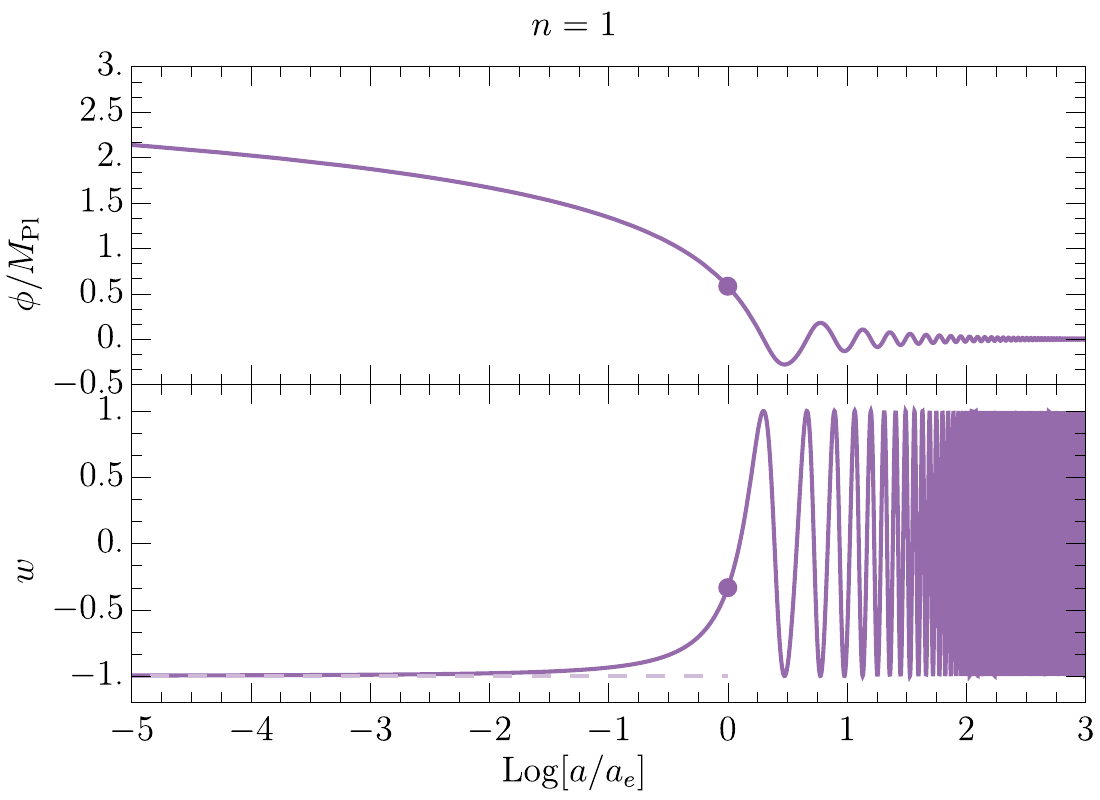}
    \includegraphics[scale=0.4]{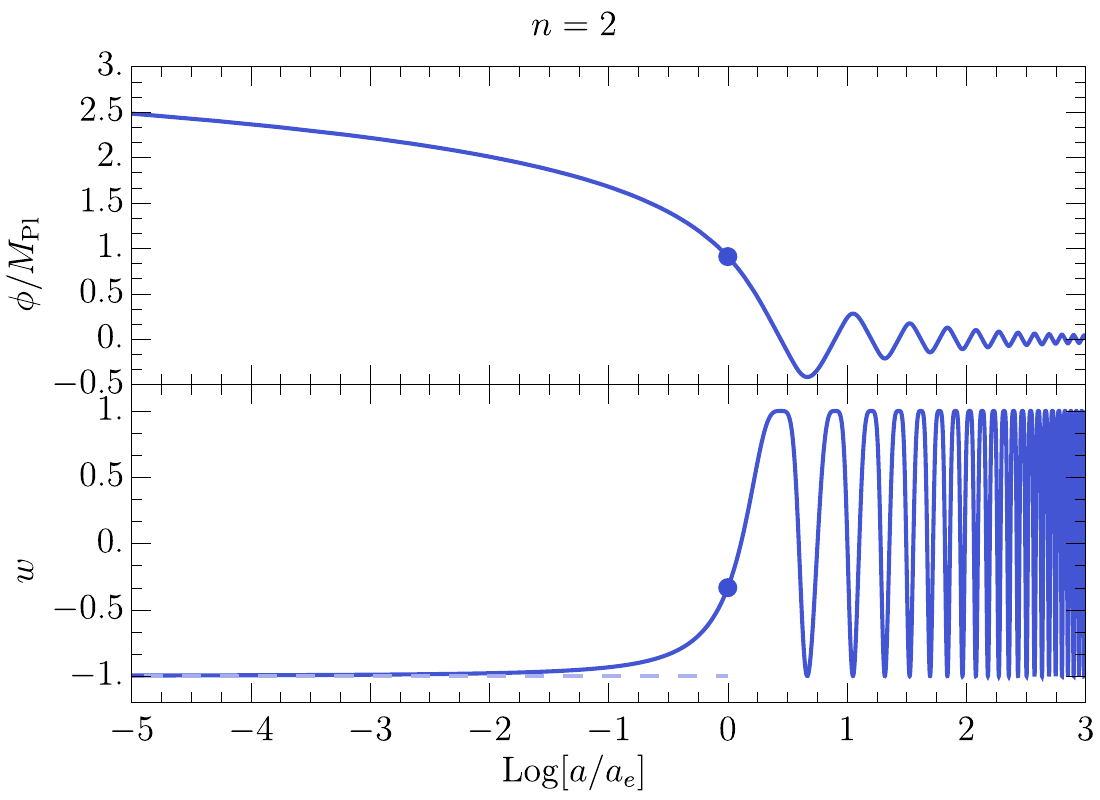}
  \caption{Upper panels: Evolution of the inflaton field during and after inflation for two choices of the inflaton potential, i.e., quadratic (left) and quartic (right). Lower panels: Evolution of the equation-of-state parameter $w$. The colored dots indicate the value of the field amplitude and the value of $w=-1/3$ at the very end of inflation. Hereafter $a_e$ denotes the scale factor at the end of inflation. The initial value of the $\phi$ field is chosen such that one gets 60 e-folds of inflation, and for the quadratic (quartic) potential it is equal $\phi_{\text{ini}}= 3.41 \mpl$ ($\phi_{\text{ini}}= 3.76 \mpl$). }
    \label{fig:phiEoM}
\end{figure}
Assuming that the total energy budget of the early universe is dominated by the inflaton energy density, the evolution of $H$ is determined by the Friedmann equation
\begin{align}
H^2 = \frac{1}{3 \mpl^2} \left( \frac{1}{2}\dot{\phi}^2 + V(\phi) \right).\label{eq:FEqn}
\end{align}
The solutions of the inflaton equation of motion for the two values of $n$ are illustrated in Fig.~\ref{fig:phiEoM}. Two phases can be identified in the dynamics of the $\phi$ field: the slow-roll period and the oscillatory stage. The termination of the former coincides with the end of inflation, which happens when $\Ddot{a}=0$. After that moment, the inflaton starts to oscillate with a decreasing amplitude. For completeness, in Fig.~\ref{fig:HRplot} we also present the corresponding evolution of the Hubble rate.
\subsection{Reheating}
In the standard cosmological scenario, the universe undergoes a reheating phase after the end of inflation, during which the energy density accumulated in the coherent oscillations of the inflaton is gradually transferred to the Standard Model sector \cite{Abbott:1982hn, Nanopoulos:1983up, Traschen:1990sw, Kofman:1994rk, Khlebnikov:1996mc, Kofman:1997yn}. This energy injection is typically achieved through direct interactions between $\phi$ and relativistic degrees of freedom. The duration of reheating is determined by the strength of the $\phi-{\rm SM}$ coupling. In various realistic models, reheating is never instantaneous; thus, it is more appropriate to distinguish the epoch of reheating preceding the early radiation-dominated era \cite{Giblin:2013kea, Garcia:2020eof, Garcia:2020wiy, Mambrini:2021zpp}. \\
The phenomenology of reheating may involve not only perturbative processes, such as decays of inflaton quanta or scattering events, but also a plethora of non-perturbative phenomena, including resonant particle production \cite{Kofman:1994rk, Kofman:1997yn, Traschen:1990sw}, tachyonic enhancement \cite{Greene:1997ge, Dufaux:2006ee, Felder:2000hj, Felder:2001kt}, and back-reactions \cite{Garcia:2023eol, Garcia:2023dyf}. This implies that the description of the post-inflationary dynamics is highly model-dependent. On the other hand, gravitational production of spin-1 fields is not very sensitive to the details of reheating, provided that the energy density of the primordial plasma remains a subdominant component of the total energy budget and that the vector field
$X_\mu$ does not have direct couplings to the visible sector. \\
For the purpose of this work, we assume that after the end of inflation, the universe undergoes the reheating phase, during which the inflaton oscillates around the minimum of its potential, and transmits its energy to the SM sector. The character of the inflaton oscillations depends on the shape of the inflaton potential near its minimum \eqref{eq:inf_pot}, whereas the efficiency of reheating, and thus its duration, is determined by the form and the strength of the inflaton-matter coupling. Remaining agnostic about the physics of reheating, we will parameterize our ignorance by introducing three quantities that characterize the duration of this epoch: $\{\bar{w}, H_{\rm rh}, T_{\rm rh} \}$,
where $\bar{w} \equiv \langle p \rangle / \langle \rho \rangle$ with $\langle \cdots \rangle$ denoting an average over one oscillation period.~\footnote{For the
power-law inflaton potential, $V\propto \phi^{2n}$, the averaged equation-of-state parameter $\bar{w}$ can be related to the exponent $n$ through the following formula~\cite{Shtanov:1994ce}:
$$
     \bar{w} \simeq \frac{n-1}{n+1}, \lsp \text{for } \lsp a \in [a_e, a_{\rm rh}], 
$$
where $a_e$ ($a_{\rm rh}$) is the scale factor at the end of inflation (reheating) while the end of reheating is defined by the condition $\rho(a_{\rm rh}) = \rho_{\rm{R}}(a_{\rm rh})$ where $\rho_{\rm R}$ denotes the energy density of radiation.} \\
Adopting \eqref{eq:phiRHO} and solving one-period averaged inflaton equation of motion \eqref{eq:PhiEoM} together with the Friedmann equation \eqref{eq:FEqn} one can find the following approximate solution for the Hubble rate in terms of the scale factor
\begin{align}
    H(a) \approx H_e \left(\frac{a_e}{a} \right)^{\frac{3}{2}(1+\bar{w})}, \label{eq:Hubble_ap}
\end{align}
where $H_e \equiv H(a_e)$. Consequently, its value at the very end of reheating is
\begin{align}
    H_{\rm rh} \equiv H(a_{\rm rh}) = H_e \left(\frac{a_e}{a_{\rm rh}} \right)^{\frac{3}{2}(1+\bar{w})}.
\end{align}
Note that the temperature of the thermal bath at $a=a_{\rm rh}$, which we refer to as \textit{reheating temperature}, is then 
\begin{align}
    T_{\rm rh} \equiv \left( \frac{90}{\pi^2 g_{\star, \rm rh}}  \mpl^2 H_{\rm rh}^2 \right)^{1/4},
\end{align}
with $g_{\star, \rm{rh}}$ being the number of relativistic degrees of freedom at the end reheating. The  consistency with the predictions of Big Bang Nucleosynthesis imposes the following lower bound on $T_{\rm rh}$ \cite{Sarkar:1995dd, deSalas:2015glj, Hasegawa:2019jsa}:
\begin{align}
    T_{\rm rh} \gtrsim 4 \; {\rm MeV}.
\end{align}
In Fig.~\ref{fig:HRplot}, we compare the evolution of $H(a)$ and the Ricci scalar $R(a)$ for two reheating models with $n=1$ and $n=2$. For completeness, we also plot the analytical approximation for the Hubble rate \eqref{eq:Hubble_ap}, which matches the numerical solution very well in the region $a> a_e$. In addition, in realistic models of inflation, $H(a)$ is not exactly constant between the onset ($H_{\text{ini}}$) and the end of inflation. Instead, it slowly decreases with a typical variation of the order of $H_{\text e} \sim 0.15 H_{\text{ini}}$. 
\section{Spin-1 spectator field with non-minimal couplings to gravity} \label{vec}
Our intention is to quantify the production of vector particles arising solely from the non-adiabatic expansion of the background metric in the early universe. To that end, we examine the dynamics of a massive spin-1 gauge boson, $X_\mu$, non-minimally coupled to gravity, whose action is
\begin{align}
     \mathcal{S}_X = \int d^4 x \sqrt{-g} \left\{ -\frac{1}{4} g^{\mu \alpha} g^{\nu \beta} X_{\mu \nu} X_{\alpha \beta} + \frac{m_X^2}{2} g^{\mu \nu} X_\mu X_\nu - \frac{\xi_1}{2}   g^{\mu \nu} RX_\mu X_\nu + \frac{\xi_2}{2}   R^{\mu \nu} X_\mu X_\nu \right\},
    \label{eq:X_action}
\end{align}
where, $X_{\mu\nu} \equiv \partial_\mu X_\nu - \partial_\nu X_\mu$, while $R_{\mu \nu}$ and $R = g_{\mu \nu} R^{\mu \nu}$ represent the Ricci tensor and scalar, respectively. The parameters $\xi_1$ and $\xi_2$ are dimensionless coupling constants that characterize the non-minimal interactions of $X_\mu$ with gravity. \\
Let us emphasize that the mass of the $X_\mu$ field, $m_X$, can be generated via the classical Stueckelberg mechanism \cite{Ruegg:2003ps}, or equivalently, by taking the appropriate limit of the Higgs mechanism \cite{Duch:2014xda}. However, in this context, we encounter not only a gauge-breaking mass term but also an additional source of gauge non-invariance due to the direct, non-minimal couplings to $R$ and $R_{\mu \nu}$. Therefore, the following generalization of the classical Stuckelberg mechanism is required:
\begin{align} \label{Stuck_action}
 \mathcal{S}_X =   
     \int d^4 x \sqrt{-g} & \left\{ -\frac{1}{4} g^{\mu \alpha} g^{\nu \beta} X_{\mu \nu} X_{\alpha \beta} \; + \right. \\ 
     &\left. +\frac12 \left[ g^{\mu\nu} - \xi_1 \frac{R}{m_X^2}g^{\mu\nu} + \xi_2
     \frac{R^{\mu\nu}}{m_X^2}\right]
     (\partial_\nu \Phi_X + m_X X_\nu)
     (\partial_\mu \Phi_X + m_X X_\mu)       \right\}, \non
\end{align}
where $\Phi_X$ stands for the Stuckelberg real scalar field. Note that the sign in front of the kinetic term for $\Phi_X$ is fixed to be positive, whereas the signs in front of the $R/m_X^2$ and $R^{\mu\nu}/m_X^2$ terms are arbitrary. The action above remains invariant under the following local $U_X(1)$ transformation:
\begin{align}
    X_\mu(x) & \to \tilde{X}_{\mu}(x) = X_\mu(x)+\partial_\mu\lambda(x), \\
    \Phi_X(x) & \to \tilde{\Phi}_X(x)=\Phi_X(x)- m_X \lambda(x).
\end{align}
By choosing the unitary gauge, i.e., selecting $\lambda(x)$ such that $\tilde{\Phi}_X = 0$, we recover the action \eqref{eq:X_action}. Accordingly, throughout the remainder of this work, we will continue to use the action \eqref{eq:X_action} in the spatially flat FLRW metric, with the line element given by \eqref{FLRW}.
\\
It is important to emphasize that in this work, the $X_\mu$ field is treated as a spectator field during inflation and reheating, meaning it does not influence neither the background metric nor the inflaton field configuration. In other words, this assumption implies that the gravitational production of dark vectors has a negligible impact on the dynamics of the primordial universe, with the dark sector becoming relevant only at later stages, such as during the post-radiation epochs. \\
Next, by going to the Fourier space,
\begin{align}
    X_\mu(\tau, x) = \int \frac{d^3 k}{(2 \pi)^3} X_\mu(\tau, k) e^{i \Vec{k}\cdot \Vec{x}},
    \label{Fourier_rep}
\end{align}
one can show, see e.g., \cite{Kolb:2020fwh}, that the action $\mathcal{S}_X$ for the two transverse modes $T = \pm$ and the single longitudinal mode $L$, representing the three physical degrees of freedom of $X_\mu$, is given by
\begin{align}
     \mathcal{S}_{\rm T} = \sum_{T=\pm} \int d \tau \int \frac{d^3 k}{(2 \pi)^3} \left\{ \frac{1}{2} |X_{ \rm T }^\prime|^2 - \frac{1}{2} [k^2+ a^2 m^2_{\rm eff, x}]|X_{ \rm T }|^2  \right\},  \label{eq:sT} \\
     \mathcal{S}_{\rm L} = \int d \tau \int \frac{d^3 k}{(2 \pi)^3} \left\{ \frac{1}{2} \frac{1}{A_{\rm L}^2}|X_{\rm L}^\prime|^2 - \frac{1}{2}a^2 m^2_{\rm eff, x} |X_{\rm L}|^2  \right\}, \label{eq:sL}
\end{align}
where $k^2 \equiv |\vec{k}|^2$, $\prime \equiv d/d \tau$, and 
\begin{align}
    A_{\rm L} &\equiv A_{\rm L}(a, k) =  \frac{\sqrt{k^2 + a^2 \meft}}{a m_{\text{eff,t}}(a)}, \label{eq:ALdef} \\
       m^2_{\rm eff, x} &\equiv \mefx = m_X^2 - \xi_1 R(a) + \frac{1}{6} \xi_2 R(a) - \xi_2 H^2(a), \label{eq:meffx} \\
            m^2_{\rm eff, t} &\equiv \meft = m_X^2 - \xi_1 R(a) + \frac{1}{2} \xi_2 R(a)+ 3 \xi_2 H^2(a). \label{eq:meff0} 
\end{align}
A few remarks are here in order. Firstly, the time component of the $X_\mu$ field does not have a kinetic term \cite{Ahmed:2020fhc}, 
making it an auxiliary degree of freedom that can be integrated out. Furthermore, the action $\mathcal{S}_T$ takes the form of the action for a scalar field with a time-dependent mass, $m^2 = \mefx$ \cite{Kolb:2023ydq}. On the other hand, from Eq.\eqref{eq:sL}, one can see that the kinetic term for the longitudinal polarization is not canonically normalized. Moreover, by examining Eq.\eqref{eq:meff0}, one observes that $\meft$ does not have a definite sign. Therefore, there may be regions in the $(\xi_1, \xi_2)$ parameter space where the kinetic term of $X_{\rm L}$ becomes negative, leading to the so-called ghost instability \cite{Kolb:2020fwh, Capanelli:2024nkf, Capanelli:2024pzd}. 
\\ \\ 
To discuss the evolution of $X_\mu$ it is instructive to perform the following transformation of its longitudinal component:
\begin{align}
    &X_L= A_{\rm L} \mathcal{X}_L, \label{eq:Ltrans}
\end{align}
such that the action $\mathcal{S}_L$ expressed in terms of the redefined field $\mathcal{X}_L$ has a proper kinetic term, 
\begin{align}
     \mathcal{S}_{\rm L} = \frac{1}{2} \int d \tau \int \frac{d^3 k}{(2 \pi)^3} &\left\{ |\mathcal{X}_{\rm L}^\prime|^2 + \frac{A_{\rm L}^\prime }{A_{\rm L}}\left(\mathcal{X}_{\rm L}^\prime \mathcal{X}_{\rm L}^* + \mathcal{X}_{\rm L}^{*\prime} \mathcal{X}_{\rm L} \right)
     - \left[ a^2 m^2_{\rm eff,x} A_{\rm L}^2 - \left( \frac{A_{\rm L}^\prime}{A_{\rm L}}\right)^2\right]|\mathcal{X}_{\rm L}|^2  \right\}.
\end{align}
Integrating by parts and dropping the boundary term, one can get rid of the second term, obtaining 
\begin{align}
    \mathcal{S}_{\rm L} = \frac{1}{2} \int d \tau \int \frac{d^3 k}{(2 \pi)^3} &\left\{ |\mathcal{X}_{\rm L}^\prime|^2 
     - \left[ a^2 m^2_{\rm eff,x} A_{\rm L}^2 + \frac{A_{\rm L}^{\prime \prime}}{A}- 2 \left( \frac{A_{\rm L}^{\prime}}{A_{\rm L}} \right)^2 \right]|\mathcal{X}_{\rm L}|^2  \right\}.
\end{align}
Then, from the above action and the action for the transversely-polarized modes, one can read out the following equations of motions\footnote{Note that the evolution of the physical longitudinal component of the vector field, can be found from Eq.\eqref{eq:chiL} together with Eq.\eqref{eq:Ltrans}.} (EoM):
\begin{align}
    X_{\rm T}^{\prime \prime } + \omega_{\rm T}^2  X_{\rm T} =0, \label{eq:EoMT} \\
    \mathcal{X}_{\rm L}^{\prime \prime } + \omega_{\rm L}^2 \mathcal{X}_{\rm L}=0, \label{eq:chiL}
\end{align}
where the time-dependent frequencies are 
\begin{align}
    \omega_{\rm T}^2 &\equiv \omega_{\rm T}^2(a, k) = k^2 +  a^2\mefx, \label{eq:omegaT} \\
    \omega_{\rm L}^2 &\equiv \omega_{\rm L}^2(a, k) =  a^2 \mefx A_{\rm L}^2(a, k)+  \frac{A_{\rm L}^{\prime \prime }(a, k)}{A_{\rm L}(a,k)} - 2 \left(\frac{A_{\rm L}^{\prime}(a, k)}{A_{\rm L}(a,k)} \right)^2. \label{eq:omegaL}
\end{align}
The above compact form of the longitudinal mode frequency conceals the complicated structure of $\omega_{\rm L}^2$ in terms of  $k, m_X, a, a^\prime, a^{\prime \prime}$
\begin{align}
    2 \left( \frac{A_{\rm L}^\prime}{A_{\rm L}} \right)^2 &= 2 \frac{(a^\prime a \, m_{\text{eff,t}}^2+m_{\text{eff,t}}^\prime m_{\text{eff,t}}\, a^2 )^2}{(k^2 + a^2 \, m_{\text{eff,t}}^2)^2} - 4 \frac{(a^\prime m_{\text{eff,t}}+m_{\text{eff,t}}^\prime a  )^2}{k^2 + a^2 \, m_{\text{eff,t}}^2} + 2 \frac{(a^\prime m_{\text{eff,t}}+m_{\text{eff,t}}^\prime a  )^2}{(a \, m_{\text{eff,t}})^2},
    \\
    \frac{A_{\rm L}^{\prime \prime}}{A_{\rm L}}&=  \frac{a^{\prime \prime}a \,m_{\text{eff,t}}^2 + m_{\text{eff,t}}^{\prime \prime} m_{\text{eff,t}} \, a^2 - (a^{\prime})^2 m_{\text{eff,t}}^2 - (m_{\text{eff,t}}^\prime)^2 a^2 }{k^2 + a^2 \, m_{\text{eff,t}}^2} - \frac{(a^\prime a \, m_{\text{eff,t}}^2+m_{\text{eff,t}}^\prime m_{\text{eff,t}}\, a^2 )^2}{(k^2 + a^2 \, m_{\text{eff,t}}^2)^2} \non \\
    & + 2 \frac{(a^\prime m_{\text{eff,t}}+m_{\text{eff,t}}^\prime a  )^2}{(a \, m_{\text{eff,t}})^2} - \frac{a^{\prime \prime} m_{\text{eff,t}} + m_{\text{eff,t}}^{\prime \prime}a + 2 m_{\text{eff,t}}^\prime a^\prime a \,m_{\text{eff,t}}}{a \, m_{\text{eff,t}}}.
\end{align}
After some straightforward algebra, one finds
\begin{align}
     \omega_{\rm L}^2
     &= k^2 \frac{m_{\text{eff,x}}^2}{m_{\text{eff,t}}^2} +  a^2 \mefx \non \\
    &
     - \frac{k^2}{k^2 + a^2 \meft} \left[ \frac{a^{\prime \prime}}{a} + \frac{m_{\text{eff,t}}^{\prime \prime}}{m_{\text{eff,t}}} + 2 \frac{a^\prime}{a} \frac{m_{\text{eff,t}}^{\prime}}{m_{\text{eff,t}}} - 3  \frac{(a^\prime m_{\text{eff,t}} + m_{\text{eff,t}}^\prime a )^2}{k^2 + a^2 \meft}
     \right], \label{eq:omegaLNM} 
\end{align}
with 
\begin{align}
m_{\text{eff,t}}^{\prime}  &=  -\frac{3}{2} \frac{H}{m_{\text{eff,t}}} \left\{ 2 H^{\prime}\left[ \left( \xi_1 - \frac{1}{2} \xi_2 \right)(3w -1) - \xi_2 \right] + 3 H \left( \xi_1 - \frac{1}{2}\xi_2 \right) w^\prime \right\}, \label{eq:meffp} \\
m_{\text{eff,t}}^{\prime \prime }  &= - \frac{3}{2} \frac{H^\prime}{m_{\text{eff,t}}} \left\{ 2 H^{\prime}\left[ \left( \xi_1 - \frac{1}{2} \xi_2 \right)(3w -1) - \xi_2 \right] + 3 H \left( \xi_1 - \frac{1}{2}\xi_2 \right) w^\prime \right\} \non \\
&- \frac{3}{2} \frac{H}{m_{\text{eff,t}}} \left\{ 2 H^{\prime \prime} \left[ \left( \xi_1 - \frac{1}{2} \xi_2 \right)(3w -1) - \xi_2 \right] + 9 H^\prime w^\prime \left( \xi_1 - \frac{1}{2} \xi_2 \right) + 3 H \left( \xi_1 - \frac{1}{2} \xi_2  \right)w^{\prime \prime} \right\} \non \\
& - \frac{9}{4} \frac{H^2}{m_{\text{eff,t}}^{3}} \left\{ 2 H^{\prime}\left[ \left( \xi_1 - \frac{1}{2} \xi_2 \right)(3w -1) - \xi_2 \right] + 3 H \left( \xi_1 - \frac{1}{2}\xi_2 \right) w^\prime \right\}^2,
\label{eq:meffpp}
\end{align}
and 
\begin{align}
    H^{\prime} &= \frac{a^{\prime \prime}}{a^2} - 2 \frac{a^{\prime 2}}{a^3}, \\
    H^{\prime \prime} &= \frac{a^{\prime \prime \prime}}{a^2} - 6 \frac{a^{\prime \prime} a^\prime}{a^3} + 6 \frac{a^{\prime 3}}{a^4}, \\
    w^\prime &= \frac{2}{3} \frac{1}{H}\left[H^\prime + 2 \frac{1}{a}\frac{H^{\prime 2}}{ H^2} - \frac{1}{a} \frac{H^{\prime \prime}}{H}\right], \\
    w^{\prime \prime} &= \frac{2}{3} \frac{1}{H} \left[ 2 H^{\prime \prime}+ 6 \frac{1}{a }\frac{H^{\prime \prime} H^\prime}{H^2} - 3 \frac{H^{\prime 2}}{H^2} - 6 \frac{1}{a} \frac{H^{\prime 3}}{H^3} - \frac{1}{a} \frac{H^{\prime \prime \prime }}{H}\right].
\end{align}
Note that $w(a) := p(a) / \rho(a) \neq \bar{w}$. 
\\ \\
Similarly, the frequency of the transversely polarized modes can be written as an explicit function of $k, m_{\rm X}, a, a^\prime$ 
\begin{align}
    \omega_{\rm T}^2  = k^2 + a^2m_{X}^2 - a^2 H^2 \left[ 3(3w-1) \left(\xi_1 - \frac{1}{6} \xi_2\right) + \xi_2\right].
\end{align}
Note that for $\xi_1, \xi_2 =0$, both effective masses are equal, i.e., $\meft = m_{X}^2 = \mefx$. Thus, $m_{\text{eff,t}}^\prime = 0 = m_{\text{eff,t}}^{\prime \prime}$, and Eq.\eqref{eq:omegaLNM} recovers the standard formula \cite{Ahmed:2020fhc, Kolb:2020fwh}
\begin{align}
    \omega^2_{\rm T} \bigg \rvert_{\xi_1 = 0 = \xi_2} &= k^2 + a^2 m_X^2, \label{eq:oTS}\\
     \omega_{\rm L}^2 \bigg \rvert_{\xi_1 = 0 = \xi_2} &= k^2 + a^2 m_X^2 - \frac{k^2}{k^2 + a^2 m_X^2}\left[ \frac{a^{\prime \prime}}{a} - 3 \frac{a^{\prime 2} m_{\rm X}^2}{k^2 + a^2 m_{\rm X}^2} \right] \label{eq:oLS}.
\end{align}

\section{Viability of non-minimal couplings to gravity} \label{sec:Viability}
The goal of this section is to determine the regions in the $(\xi_1,\xi_2)$ space for which the model of non-minimally coupled massive vectors is not ill-defined, meaning it does not suffer from any instabilities. These instabilities pertain to the longitudinal component of the $X_\mu$ field and include \cite{Capanelli:2024nkf, Capanelli:2024pzd}
\begin{itemize}
    \item ghost instability, 
    \item uncontrolled production and super-luminal propagation of short-wavelength modes. 
\end{itemize}
The former arises when the effective mass $m_{\rm eff, t}^2$, appearing in the prefactor $A_{\rm L}^2$ that multiplies the kinetic term of the $\rm{L}$ mode, becomes negative. More precisely, for modes with \mbox{$k^2 \ll |a^2 m_{\rm eff, t}^2|$}, the function $A_{\rm L}^2$ is positive. However, in the opposite case, \mbox{$k^2 \gg |a^2 m_{\rm eff, t}^2|$}, the kinetic term of the longitudinal polarization acquires a wrong sign, indicating the presence of ghosts. The latter instabilities are related to the UV behavior of the dispersion relation.   
 \subsection{Ghost instability} \label{sec:Ghost}
We aim to find values of the non-minimal couplings for which the sign 
of the kinetic term
\begin{align}
s(a) \equiv \sign\left\{\frac{a^2 \meft}{k^2 + a^2 \meft}\right\},
\label{sign}
\end{align}
remains unchanged over time for any $k^2$.\footnote{One could look for positivity regions that are parameterized by $k^2$. In other words, regions of $\xi_1,\xi_2)$ would be allowed by the positivity of the kinetic term  for $X_L$ if $k^2$ satisfied certain conditions. This is not what we are going to do here, since such restrictions upon $k^2$ is unphysical and, in general, invalidate the Fourier representation adopted in \eqref{Fourier_rep}.} To that end, it is instructive to rewrite $\meft$ as an explicit function of the scale factor $a$. Therefore, we first recast the Ricci scalar as:
\begin{align}
    R(a) =  - 6 \left( \frac{\Ddot{a}}{a} + H^2(a) \right). \label{eq:ricci}
\end{align}
Using the fact that
\begin{align}
    &\frac{\Ddot{a}}{a} = - \frac{\rho(a) + 3 p(a)}{6 \mpl^2}, && H^2(a) = \frac{\rho(a)}{3 \mpl^2},
\end{align}
where $\rho(a)$ and $p(a)$ denote the total energy density and the pressure, respectively, one finds
\begin{align}
    R(a) 
    & = 3 H^2(a)[3w(a) -1].
\end{align}
Inserting the above expression into Eqs.\eqref{eq:meff0} and \eqref{eq:meffx}, we get
 \begin{align}
    \meft = m^2_X - 3\left[\left( \xi_1 - \frac{1}{2} \xi_2\right)(3 w(a) -1)  -  \xi_2 \right]H^2(a), \label{eq:meff_t}\\
    \mefx = m^2_X - \left[3\left( \xi_1 - \frac{1}{6} \xi_2\right)(3 w(a) -1)  +  \xi_2 \right]H^2(a). \label{eq:meff_x}
 \end{align}
In general, the effective masses have two sources of time dependence: $H(a)$ and $w(a)$. The Hubble rate remains roughly constant during the de Sitter stage of inflation and subsequently decreases monotonically over time. During the reheating period, it also experiences an oscillatory contribution, as shown in Fig.~\ref{fig:HRplot}. 
The equation-of-state parameter $w(a)$ is roughly constant during inflation ($w_{\rm inf}\simeq - 1$), the eras of radiation-domination ($w_{\rm rd}=1/3$) and matter-domination ($w_{\rm md}=0$), as well as during the hypothetical epoch of kination ($w_{\rm k}=1$), changing smoothly during the transition phases. However, during the early stages of reheating, $w(a)$ oscillates rapidly between $[-1, 1]$, as illustrated in Fig.~\ref{fig:phiEoM}. In single-field inflationary models, the character of these oscillations is determined by the form of the inflaton potential near its minimum. 
\\\\
In the following subsections, we will determine for which values of $\xi_1$ and $\xi_2$ the function $s(a)$ remains positive throughout the entire evolution. We first discuss a simplified scenario by setting one of the non-minimal couplings to zero, and then we consider a more general case with $\xi_1 \neq 0$ and $\xi_2 \neq 0$. 
\subsubsection{$\xi_1 =0$}  
To begin with, let us first investigate the simplified scenario with $\xi_1 = 0$. In this case, $\meft$ reduces to 
\begin{align}
    \meft \bigg \rvert_{\xi_1 =0}= m^2_X + \frac{1}{2} \xi_2\left[3 w(a) +1\right] H^2(a).
\end{align}
Hence, the positivity of $\meft$ is guaranteed for $\xi_2$ meeting the following condition: 
\begin{align}
   2 \left( \frac{m_X}{H(a)} \right)^2 > - \xi_2 [3w(a)+1].
\end{align}
Since $H(a)$ is a decreasing function of time, the left-hand side increases over time starting at its minimal value during inflation. Therefore, the most restrictive constraint on $\xi_2$ should be imposed when the $m_X/H(a)$ ratio is smallest, i.e, at the very beginning of inflation. Consequently, the ghost instability is absent if
\begin{align}
    2 > - \eta_e\xi_2 [3w(a)+1],
\end{align}
where we have introduced the $\eta_{e}^{-1}$ parameter, defined as
\begin{align}
    \eta_{e}^{-1} \equiv \left( \frac{m_X}{H_{\text{ini}}} \right)^2.
\end{align}
The possible range of variation for $\eta_e^{-1}$ is constrained by the following requirements. First, the energy density of the $X_\mu$ field, $\rho_X$, should constitute a subdominant component of the total energy of the primordial universe. On the other hand, $\rho_X$ must be sufficiently large to eventually dominate the energy budget of the universe after the radiation-dominated phase. Assuming that gravitational production is the sole mechanism responsible for the creation of $X_\mu$, the second prerequisite implies that this mechanism must efficiently generate $X_\mu$ particles, which happens for sufficiently low mass, i.e., below the inflationary Hubble scale\footnote{Recently, it has been shown that purely gravitational production can efficiently generate super-heavy scalars with mass $\lesssim \mathcal{O}(10^3) H_e$ \cite{Verner:2024agh}, and a non-minimal coupling to gravity. This opens an intriguing possibility for an ultra-heavy spin-1 field, which we left for future work.}. Altogether, this means that $\eta_e^{-1} \in (0,1)$. Consequently, for $\xi_1 =0$, the positivity of $s(a)$ imposes the following constraint on $\xi_2$:
\begin{align}
    &\text{for } \xi_1 =0 &-\eta_e\xi_2 \gtrsim \frac{2}{3w+1} &&\text{for } w \in [-1, -1/3) &&\text{and } &-\eta_e\xi_2 \lesssim \frac{2}{3w+1} \text{ for } w \in (-1/3, 1].
\end{align}

\subsubsection{$\xi_2 =0$}  
Choosing $\xi_2 =0$ allows us to simplify Eq.\eqref{eq:meff_t} as 
\begin{align}
     \meft  \bigg \rvert_{\xi_2 =0} = m^2_X - 3 \xi_1\left[3 w (a)-1\right] H^2(a).
\end{align}
Hence, the condition $s(a)>0$ implies 
\begin{align}
   \frac{1}{3} \left( \frac{m_X}{H(a)} \right)^2  > \xi_1\left[3 w(a) -1\right].
     \label{posit2}
\end{align}
Analogously to the case discussed above, the strongest constraint on $\xi_1$ should be imposed when the ratio $m_X/H(a)$ is the smallest
\begin{align}
    \frac{1}{3} \gtrsim \eta_e\xi_1 [3w(a)-1].
\end{align}
The above condition is met trivially during radiation-domination phase, whereas in generic case one finds
\begin{align}
     &\text{for } \xi_2 =0 &\eta_e\xi_1 \gtrsim \frac{1}{3(3w-1)} &&\text{for } w \in [-1, 1/3) &&\text{and } &\eta_e\xi_1 \lesssim \frac{1}{3(3w-1)} \text{ for } w \in (1/3, 1].
\end{align}

\subsubsection{$\xi_1 \neq 0$ and $\xi_2 \neq 0$}
\begin{figure}[htb!]
    \centering
    \includegraphics[scale=0.5]{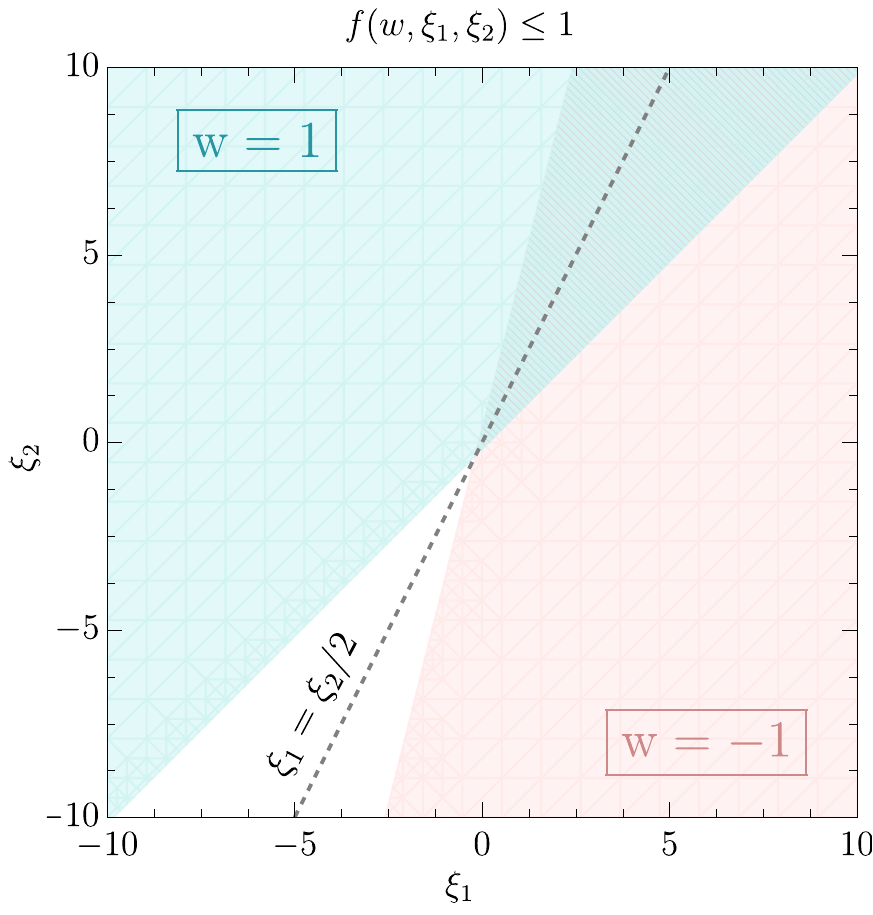}~~~~\includegraphics[scale=0.5]{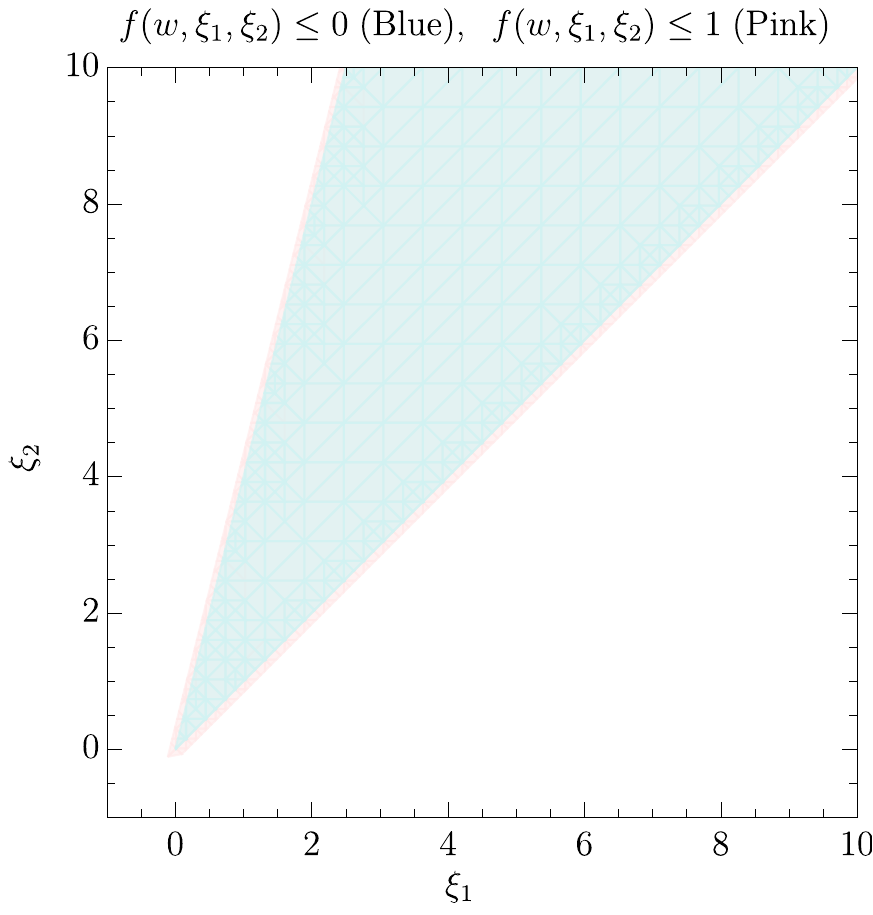}
    \caption{Left: Region in the $\xi_1-\xi_2$ parameter space satisfying $ f(w(a), \xi_1, \xi_2) \lesssim 1$, i.e. for $\eta_e^{-1}=1$, with two limiting choices of the equation-of-state parameter $w=-1$ (light pink region) and $w=1$ (light cyan region). Right: Values of $\xi_1-\xi_2$ ensuring the positivity of $\meft$ for two values of $\eta_e^{-1} \in \{0,1\}$.}
    \label{fig:xi1xi2}
\end{figure}

As already mentioned, Eq.\eqref{eq:meff_t} has two sources of time-dependency, i.e., $w(a)$ and $H(a)$. To eliminate the latter, one would need to set the non-minimal couplings to zero, i.e., $\xi_1 = 0$ and $\xi_2 = 0$. Alternatively, one can also get rid of the term $w(a)$ by adjusting  
\begin{align}
    \xi_1 =  \frac{1}{2} \xi_2. \label{eq:NME}
\end{align}
Note that for this choice, the non-minimal part of the Lagrangian density reads
\begin{align}
\sqrt{-g}{\mathcal L_{\rm X}^{\rm NM}} \equiv \sqrt{-g}\left[ - \frac{\xi_1}{2} R g_{\mu\nu} X^\mu X^\nu + \frac{\xi_2}{2} R_{\mu\nu} X^\mu X^\nu \right]=
\sqrt{-g}\xi_1 G_{\mu\nu} X^\mu X^\nu,
\end{align}
where $G^{\mu\nu}$ is the Einstein tensor, defined as $G_{\mu\nu} \equiv R_{\mu\nu} - \frac{1}{2} R g^{\mu\nu}$. This scenario has been investigated in Ref.\cite{Ozsoy:2023gnl}, where a different parametrization was applied, i.e., $\xi_1 = \alpha^2/6$. \\ 
For this choice, Eq.\eqref{eq:meff_t} simplifies as
\begin{align}
     \meft  \bigg \rvert_{\xi_1 =  \xi_2/2} = m_X^2 +6 \xi_1 H^2(a).
\end{align}
Hence, $\meft$ remains positive for $\xi_1 =\frac{1}{2} \xi_2$ provided that  
\begin{align}
  &\xi_1 \gtrsim - \frac{1}{6} \eta_e^{-1}.
\end{align}
Finally, let us discuss the most general scenario with $\xi_1 \neq \frac{\xi_2}{2}$ and $\xi_{1,2} \neq 0$. In this case, it is instructive to rewrite the condition for the positivity of $\meft$ as follows,
\begin{align}
f(w, \xi_1, \xi_2) \leq   \left( \frac{m_X}{H(a)} \right)^2, \label{eq:fCon}
\end{align}
with 
\begin{align}
    f(w(a), \xi_1, \xi_2) \equiv 3\left[\left( \xi_1 - \frac{1}{2} \xi_2\right)(3 w(a) -1)  -  \xi_2 \right]. \label{eq:mefft_con}
\end{align}
The strongest constraint on the function $f(w(a), \xi_1, \xi_2)$ comes from the ultralight vectors, for which $m_X \rightarrow 0$. In this case, $f(w(a), \xi_1, \xi_2)$ is bounded from above 
\begin{align}
    f(w(a), \xi_1, \xi_2) \lesssim 0.
\end{align}
On the other hand, for a given mass $m_X \neq 0$, $\meft$ remains positive through the whole evolution  for $\{\xi_1, \xi_2\}$ satisfying the following condition:  
\begin{align}
     &f(w(a), \xi_1,\xi_2) \lesssim \eta_e^{-1}, &\eta_e^{-1} \in (0, 1).\label{eq:fx1x2}
\end{align}
Using the fact that $w(a) \in [-1, 1]$, one can identify the region in the $\xi_1-\xi_2$ parameter space where the above condition is always satisfied for a fixed value of $\eta_e^{-1}$. This is shown in the left panel of Fig.~\ref{fig:xi1xi2} for $\eta_e^{-1}=1$. The shaded region corresponds to values of $\xi_1$ and $\xi_2$ for which condition \eqref{eq:fx1x2} holds at all times, ensuring the positivity of $s(a)$ for all modes throughout their entire evolution. In the right panel of Fig.~\ref{fig:xi1xi2}, we compare the viable regions in the $\xi_1-\xi_2$ parameter space for two limiting values of $\eta_e^{-1} \in \{0, 1\}$. One observes that an increase in the $m_X/H_{\text{ini}}$ ratio leads to a slight broadening of the area where $\meft > 0$.
\\
\subsection{UV behaviour}
Let us now examine the UV behaviour of $\omega_{\rm T}^2(a, k)$ and $\omega_{\rm L}^2(a, k)$. In the limit, $k \rightarrow \infty$, one finds
\begin{align}
    &\omega_{\rm T}^2 (a, k) \rightarrow k^2,  &\omega_{\rm L}^2 (a, k) \rightarrow k^2 \frac{\mefx}{\meft}, &&\text{as } k \rightarrow \infty. \label{eq:UVbehaviour}
\end{align}
The non-standard high-energy limit of $\omega_{\rm L}^2(a, k)$ warrants a more detailed discussion.  It is important to emphasize that, in the ghost-free region of the parameter space, the sound speed, defined by the ratio $c_s^{2} \equiv \mefx/\meft$ \cite{Capanelli:2024nkf}, can be either positive or negative, depending on the sign of $\mefx$. Additionally, in the general case, $c_s^2$ may also exceed unity \cite{Capanelli:2024nkf}. Both situations are worrying, as they may lead to: i) uncontrolled production (a phenomenon known as runaway production; see Ref.\cite{Capanelli:2024pzd, Capanelli:2024nkf}) and ii) super-luminal propagation of short-wavelength longitudinal modes \cite{Capanelli:2024nkf}.
\subsubsection{Uncontrolled production of short-wavelength modes}
As first noted in Ref.\cite{Capanelli:2024pzd}, in the region of the $\{\xi_1, \xi_2\}$ parameter space for which $\mefx$ becomes negative, the angular frequency $\omega_{\rm L}^2$ is not bounded from below for short-wavelength modes. The negativity of $\omega_{\rm L}^2$ in the UV regime results in an exponentially enhanced creation of 
large k modes. This, in particular, has dramatic consequences for the considered model, as it leads to uncontrolled tachyonic production of longitudinal modes with high momenta. Let us now examine the conditions for the occurrence of such an instability.
\\ 
\\

\begin{figure}[htb!]
    \centering
\includegraphics[scale=0.45]{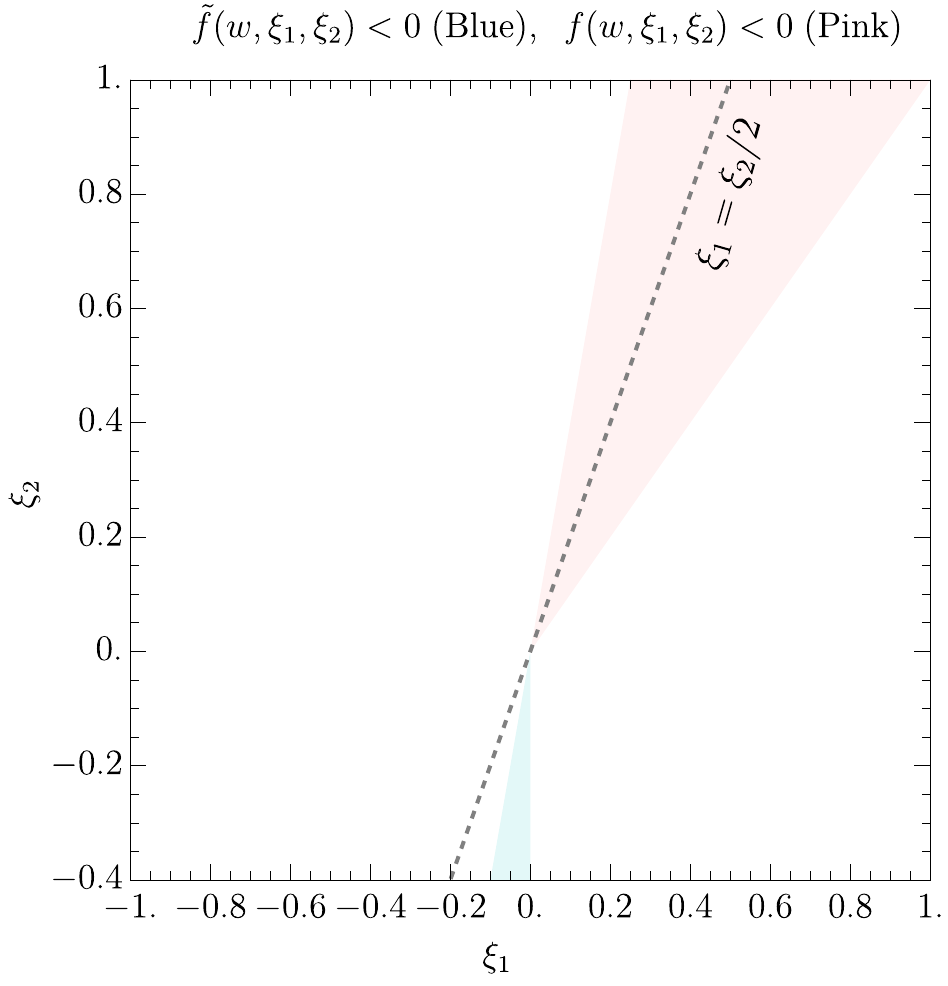}~~~~\includegraphics[scale=0.45]{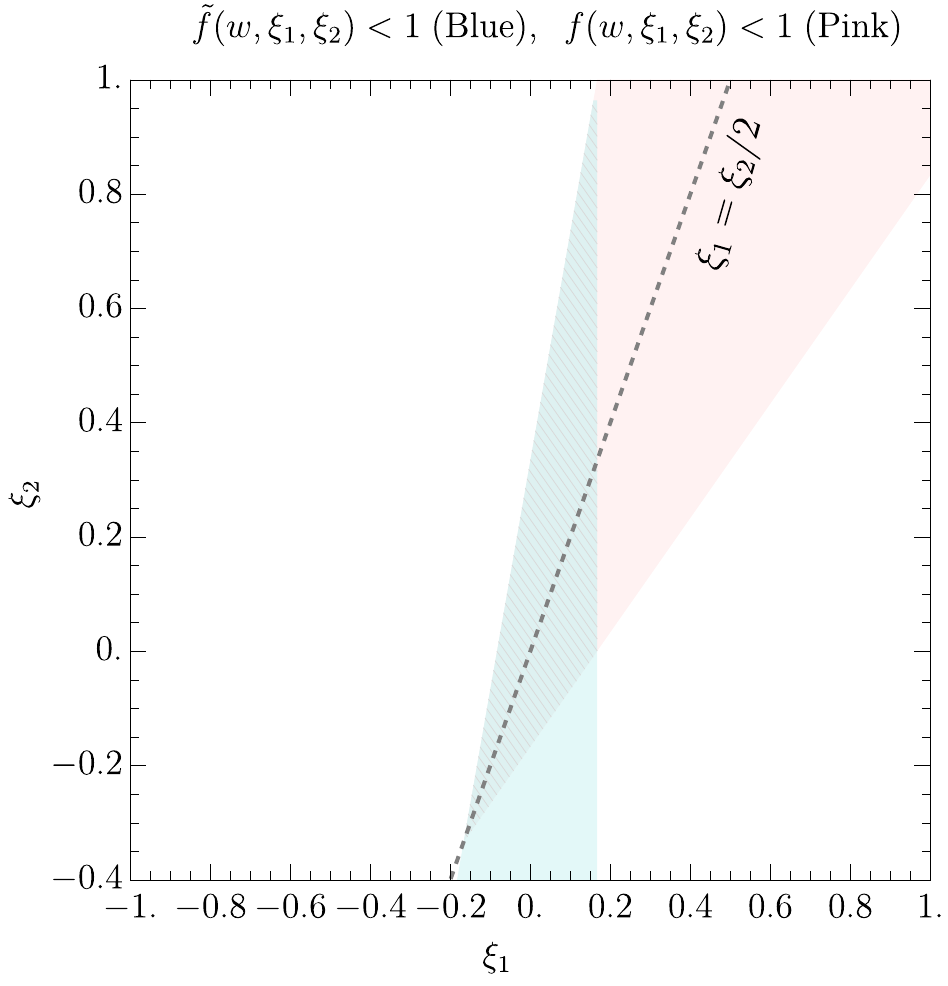}
    \caption{Regions in the $\xi_1-\xi_2$ parameter space satisfying $ f(w(a), \xi_1, \xi_2) \lesssim 
    \eta_e^{-1}$ \eqref{eq:fCon} (pink) and $ \tilde{f}(w(a), \xi_1, \xi_2) \lesssim 
    \eta_e^{-1}$ \eqref{eq:ftildeCon} (blue),  for $\eta_e^{-1}=0$ (left panel) and $\eta_e^{-1}=1$ (right panel).}
\label{fig:bothmeff}
\end{figure}

A key point to emphasize is that the occurrence of runaway production is attributable to the existence of a direct coupling between the $X_\mu$ field and the Ricci tensor. By turning off such term, i.e., setting $\xi_2 =0$, one can easily verify that 
\begin{align}
     &c_s^2 = \frac{\mefx}{\meft} = 1, &\text{for } \xi_2 =0.
\end{align}
In addition, both effective masses are also equal during the de Sitter phase of inflation, when the total energy density of the universe is dominated by the potential energy of the inflaton field and $w_{\rm inf} =-1$, i.e.,
\begin{align}
    &c_s^2 = \frac{\mefx}{\meft} = 1, &\text{for } a \lesssim a_e.
\end{align}
For non-vanishing $\xi_2$, $c_s^2$ may become negative only if $w \neq -1$ and $m_X \ll H(a)$. In this case, neglecting the mass term, one finds 
\begin{align}
    &c_s^2 \overset{m_X \ll H(a)} \simeq  \frac{3\left( \xi_1 - \frac{1}{6} \xi_2\right)(3 w(a) -1)  +  \xi_2  }{3\left[\left( \xi_1 - \frac{1}{2} \xi_2\right)(3 w(a) -1)  -  \xi_2 \right]}.
\end{align}
In the RD-phase of the universe, the above expression is always negative, regardless of the values of non-minimal couplings $\{\xi_1, \xi_2\}$, while for the MD and kination phase, $c_s^2$ becomes negative in the following regions:
\begin{align}
    &c_s^2 < 0, &\text{for } \xi_1 \in (2 \xi_2, 6 \xi_2), &&\text{and } w =0,  \\
     &c_s^2 < 0, &\text{for } \xi_1(\xi_1 - \xi_2) <0,  &&\text{and } w =1.
\end{align}
In general case, the credibility of the model can be restored in two ways. Firstly, one could impose the positivity condition on $\mefx$, analogously to the constraint that we have applied to $\meft$. Requiring $\mefx \gtrsim 0$, one gets
\begin{align}
    \tilde{f}(w, \xi_1, \xi_2) \lesssim \left( \frac{m_X}{H(a)}\right)^2, \label{eq:ftildeCon}
\end{align}
with 
\begin{align}
    \tilde{f}(w, \xi_1, \xi_2) \equiv 3[3w(a)-1] \left(\xi_1 - \frac{1}{6} \xi_2 \right) + \xi_2. \label{eq:NMCcon}
\end{align}
Adopting the similar arguments as in Sec.\eqref{sec:Ghost}, i.e., using the fact that $w(a) \in [-1, 1]$, one can find regions in the $\xi_1 - \xi_2$ parameter space for which $\mefx>0$ for a fixed value of $\eta_e^{-1}$. Hence, the conditions \eqref{eq:fCon} together with \eqref{eq:ftildeCon} determine the values of the non-minimal couplings, for which the theory is both ghost-free and does not suffer from an uncontrolled tachyonic production of short-wavelength modes. The regions in the $\xi_1-\xi_2$ parameter space, satisfying Eqs.\eqref{eq:fCon} and \eqref{eq:ftildeCon} for two limiting values of $\eta_e^{-1}$, i.e., $ \eta_e^{-1} =0$ (left panel), and $\eta_e^{-1} = 1$ (right panel), are shown in Fig.~\ref{fig:bothmeff}. Note that in the limit $m_X \rightarrow 0$, the blue and pink regions do not intersect, indicating that there is no region in the $\xi_1-\xi_2$ parameter space for which both $m_{\text{eff,t}}^2(a)$ and $m_{\text{eff,x}}^2(a)$ are positive simultaneously for an arbitrary choice of $w \in [-1, 1]$. This observation implies that longitudinal polarization of ultralight vectors non-minimally coupled to gravity would suffer either from the ghost instability, or as a result of uncontrolled tachyonic enhancement of short-wavelength modes. Hence, for $m_X \rightarrow 0$ the theory is well-defined only for $\xi_1 = 0 = \xi_2$. On the other hand, for heavier vectors there exists values of $\{ \xi_1, \xi_2 \}$ for which both effective masses remain positive throughout cosmological evolution. For a fixed $H_{\rm ini}$, the instability-free region increases with $m_X$, and its boundaries are determined by the following conditions: 
\begin{align}
    &\eta_e^{-1} > 6 \xi_1 &\land &&\eta_e^{-1} > 6 (\xi_1 -\xi_2) &&\land &&\eta_e^{-1} > 3 (\xi_2 - 4 \xi_1).  \label{eq:instabilities}
 \end{align}
\\
On the other hand, instead of imposing a positivity condition on $\mefx$, one could also introduce a UV cut-off for the model. The imposition of the cut-off scale seems well justified, as even in the minimal model, the energy-density integrals are divergent in the limit  $k\rightarrow \infty$, see Ref.\cite{Ahmed:2020fhc}. However, it is not entirely clear what the appropriate energy scale for the cut-off is. One of the more natural choices appears to be setting the cut-off scale at $\Lambda_{\rm UV} = k_e \equiv a_e H_e$. Modes with shorter wavelength, i.e, $k \gg k_e$ never exist the Hubble horizon, and can be considered as quantum fluctuations. Note that in this case, the positivity condition for $\mefx$ can be relaxed. Namely, there might be some region in the $\xi_1 - \xi_2$ parameter space, for which $\meft>0$ and $\mefx<0$ simultaneously. For such values of the non-minimal couplings, one might expect that tachyonic instability could significantly enhance the production of both longitudinal and transverse modes. Consequently, in this scenario, the predictions of the model would likely be very sensitive to the choice of the UV cut-off, which requires careful study and is thus left for future work
\subsubsection{Super-luminal propagation}
\begin{figure}[htb!]
    \centering
\includegraphics[scale=0.49]{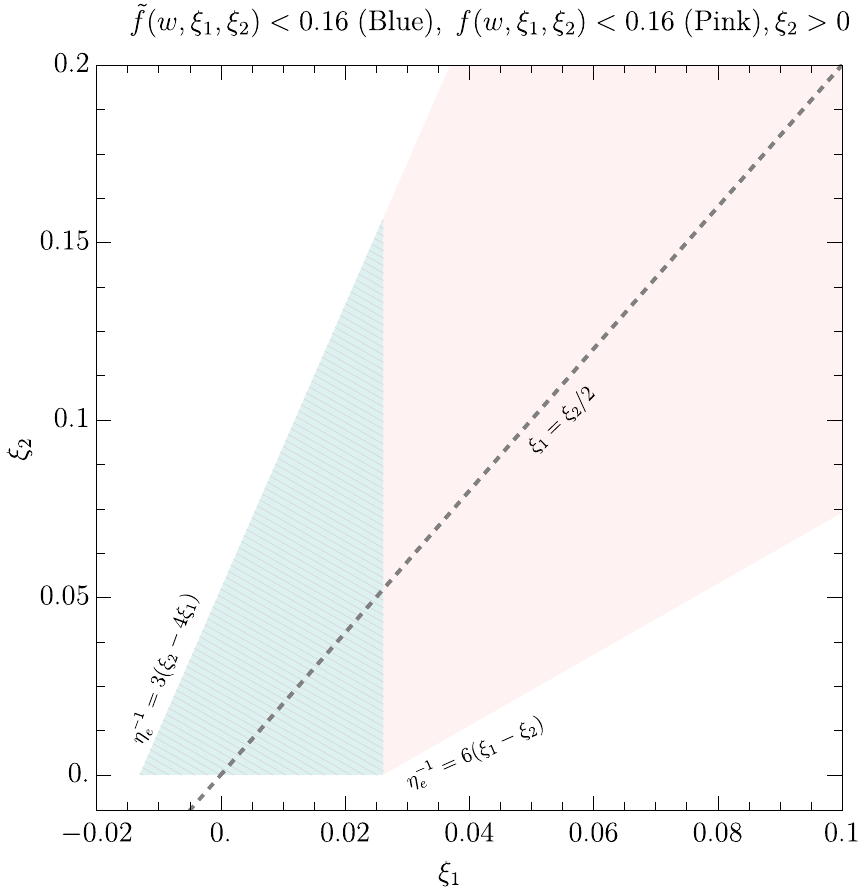}~~~~\includegraphics[scale=0.5]{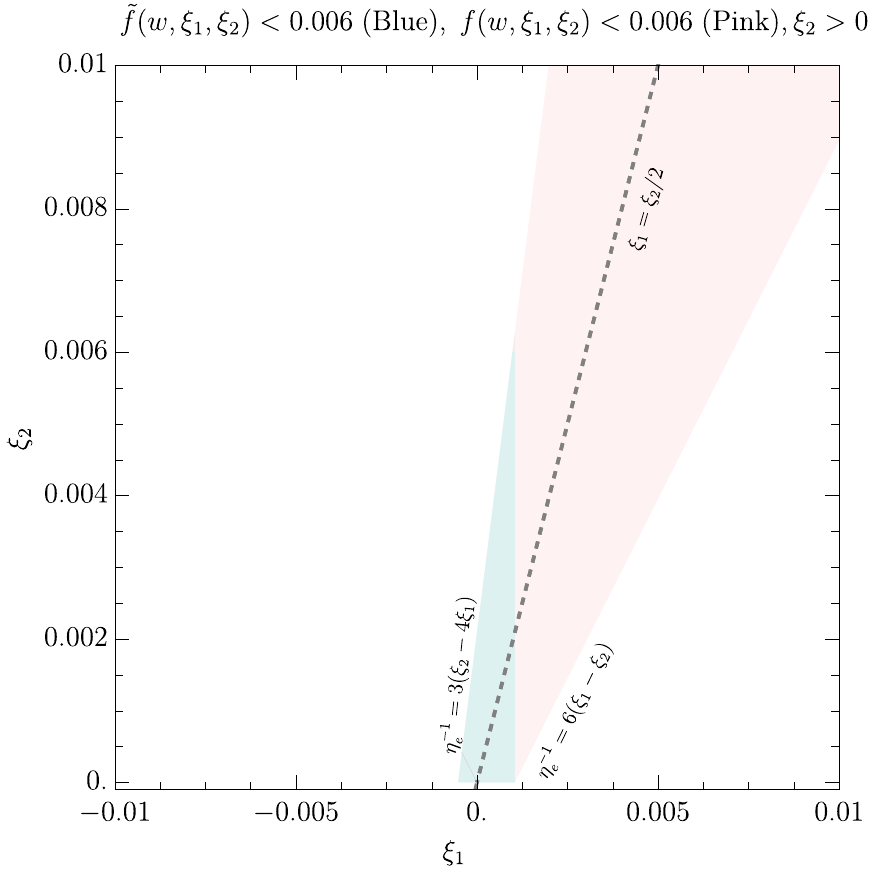}
    \caption{Instability-free region in the $\xi_1-\xi_2$ parameter described by Eq.\eqref{eq:FinalCon} for $\eta_e^{-1}=0.16$ (left panel) and $\eta_e^{-1}=0.006$ (right panel).}
\label{fig:cr}
\end{figure}

As noted in Ref.\cite{Capanelli:2024pzd}, even in the instability-free region of the model, one may encounter another problem: the super-luminal propagation of short-wavelength modes. This occurs for values of the non-minimal couplings $\{ \xi_1, \xi_2 \}$ for which $\mefx > \meft$, i.e.,
\begin{align}
 - \frac{1}{2} \xi_2 (3 w(a) -7)  \geq   0.
\end{align}
Since $3 w(a)-7 < 0$ for $w(a) \in [-1,1]$, the sound speed $c_s^2$ exceeds unity only if $\xi_2 <0$. This observation, along with the conditions \eqref{eq:instabilities}, implies that for a given value of $\eta_e^{-1}$, the model is well-defined for choices of non-minimal couplings that satisfy the following conditions: 
\begin{align}
    \xi_1 \in \left(- \frac{\eta_e^{-1}}{12},\frac{\eta_e^{-1}}{6} \right), \quad   \xi_2 \in \left[0, \frac{\eta_e^{-1}}{3} + 4 \xi_1 \right). \label{eq:FinalCon}
\end{align}
Fig.~\ref{fig:cr} illustrates the consistency region in the $\{ \xi_1, \xi_2 \}$ parameter space for two benchmark values of $\eta_e^{-1}$, i.e., $\eta_e^{-1} = 0.16$ (left panel) and $\eta_e^{-1}=0.006$ (right panel).   
\\ \\
Finally, let us emphasize that within the aforementioned region, $\omega_L^2(a,k)$ is not necessarily positive for all values of $k$. Contrarily, some modes can still be exponentially enhanced, but tachyonic production no longer leads to the uncontrolled population of short-wavelength modes. On the other hand, for $\{\xi_1, \xi_2\}$ in the region defined by Eq.\eqref{eq:FinalCon}, the purely gravitational production of transverse modes is inefficient, as the positivity condition imposed on $\mefx$ implies that 
$\omega_{\rm T}^2 (k, a)$ remains also positive for all modes. 
\begin{figure}[htb!]
    \centering
\includegraphics[scale=0.54]{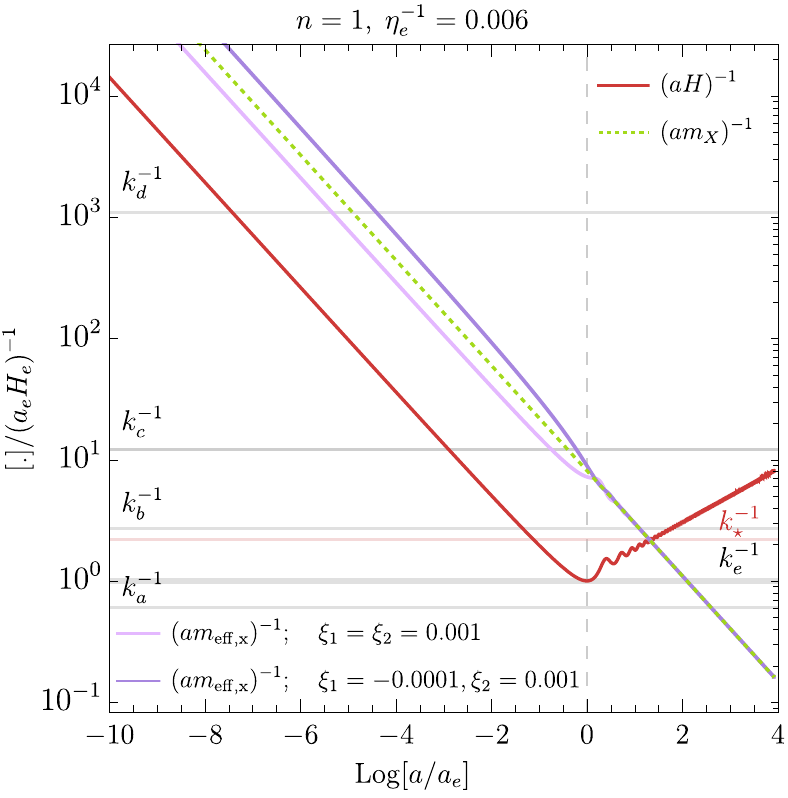}~~~~\includegraphics[scale=0.54]{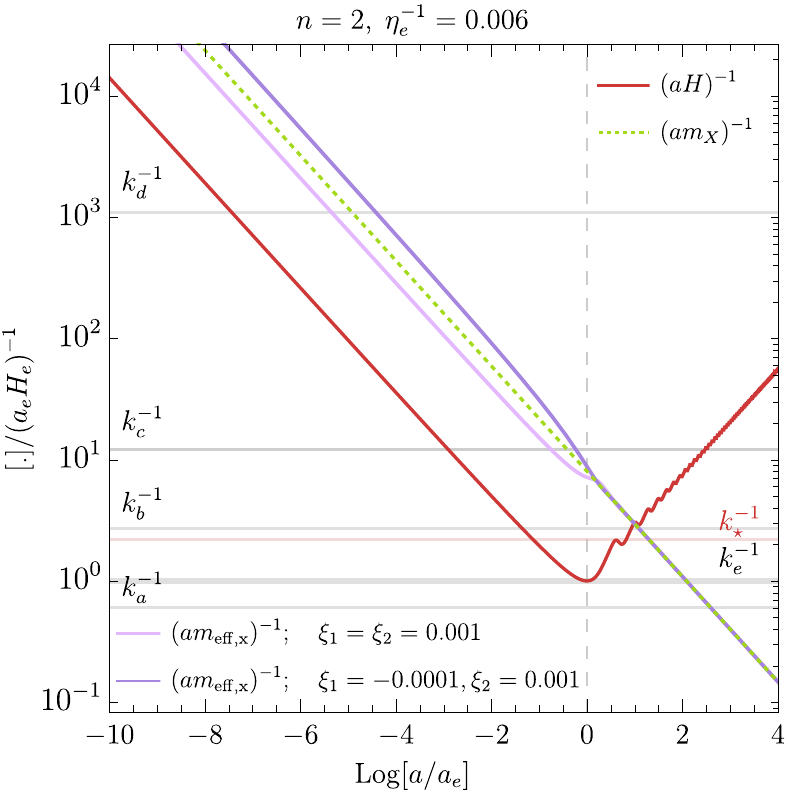}
    \caption{Evolution of different energy scales: the Hubble horizon (red line), the Compton wavelength (dashed, green), and two effective Compton wavelengths (purple lines) for the quadratic (left) and quartic (right) reheating models. Modes with wavelengths $k^{-1} \geq k_e^{-1}$ exit the horizon during inflation, cross the (effective) Compton wavelength curves, and re-enter the horizon in the post-inflationary epoch. In contrast, modes with shorter wavelengths, $k^{-1} \leq k_e^{-1}$, always remain below horizon.}
\label{fig:comovingdiag}
\end{figure}
\\
\section{Cosmic evolution of non-minimally coupled vectors} \label{evol}
Let us discuss dynamics of the non-minimally coupled $X_\mu$ field, focusing exclusively on the instability-free region of the model, described by Eq.\eqref{eq:FinalCon}. 
\subsection{Adiabaticity condition}
The amplitude of matter fields with a non-conformal couplings to gravity can be enhanced by a non-adiabatic expansion of the primordial universe \cite{Kolb:2023ydq}. More precisely, the quasi-exponential growth of the background metric during inflation, followed by its sudden change during the transition phase, may lead to a violation of the adiabaticity condition
\begin{align}
    A_\lambda (k, a) \equiv \frac{\omega_\lambda^\prime}{\omega_\lambda^2} \ll 1. \label{eq:AdiabaticCon}
\end{align}
The adiabatic parameter, $A_\lambda (k, a)$, is typically small in the remote past and future but can become large, or even divergent, during inflation and at the onset of reheating, thereby triggering efficient particle production \cite{Birrell:1982ix, Mukhanov:2007zz, Kolb:2023ydq}. Let us now examine this phenomenon in more details. \\
For the minimally-coupled vector field, the adiabatic parameters take the form
\begin{align}
    A_{\rm T} (k, a) \bigg \rvert_{\xi_1, \xi_2=0} = \frac{(a m_X)^2 a H}{\omega_{\rm T}^{3/2}} \bigg \rvert_{\xi_1, \xi_2=0},
\end{align}
and
\begin{align}
    A_{\rm L} (k, a) \bigg \rvert_{\xi_1, \xi_2=0} &= \frac{(aH)^3}{\omega_{\rm L}^{3/2}} \bigg\{ \left( \frac{m_X}{H} \right)^2 + \frac{k^2 (a m_X)^2}{(k^2 + a^2 m_X^2)^2} \bigg[ 2 + \frac{\dot{H}}{H^2} - \frac{3(am_X)^2}{k^2 + a^2 m_X^2}\bigg] \nonumber \\
    &- \frac{k^2}{k^2 +a^2 m_X^2} \bigg[ 2 + 3 \frac{\dot{H}}{H^2} + \frac{\ddot{H}}{H^3}- 3 \frac{a^2 m_X^2 \left(2 + \frac{\dot{H}}{H^2} \right)}{k^2 + a^2 m_X^2} + \frac{ 3 a^4 m_X^4}{(k^2 + a^2 m_X^2)^2}]  \bigg\}\bigg \rvert_{\xi_1, \xi_2=0}.
\end{align}
One observes that in the relativistic limit, i.e., $(am_X)^{-1} \gg k^{-1}$, the dispersion relation \eqref{eq:oLS} for the longitudinal component of the vector field becomes
\begin{align}
    &\omega_{\rm L}^2 \bigg \rvert_{\xi_1, \xi_2=0} \approx k^2 - 2 a^2 H^2, &(am_X)^{-1} \gg k^{-1},
\end{align}
where we have neglected terms proportional to $\dot{H}$. Hence, for a given mode, $A_{\rm L}(k, a)$ diverges near its horizon crossing, defined by the condition $a_{\rm h.c} H(a_{\rm h.c}) = k$, see Fig.~\ref{fig:comovingdiag}, or more precisely, at the moment when $k \simeq \sqrt{2} a H $. This initiates tachyonic enhancement of the long-wavelength modes. The growth continues until $k$ crosses the Compton wavelength $(a m_X)^{-1}$, which happens at $a_{\rm C.c} m_X= k$. After that, the evolution of the super-horizon longitudinal modes becomes adiabatic, as $A_{\rm L}(k, a)$ eventually becomes small. \\ On the other hand, for short-wavelength modes one finds
\begin{align}
    &\omega_{\rm L}^2 \approx k^2 - a^2 H^2 \bigg[ 2 + \frac{\dot{H}}{H^2} \bigg] = k^2 - \frac{a^2 H^2}{2} (1+ 3 w), & k^{-1}, (a H)^{-1} \ll (a m_X)^{-1},
\end{align}
and hence
\begin{align}
    &A_{\rm L} (k, a) \bigg \rvert_{\xi_1= \xi_2 =0} \approx - \frac{(aH)^3}{[k^2 - \frac{a^2 H^2}{2} (1+ 3 w)]^{3/2}} \bigg[2 + 3 \frac{\dot{H}}{H^2} + \frac{\ddot{H}}{H^3} \bigg], &k^{-1}, (a H)^{-1} \ll (a m_X)^{-1},
\end{align}
where we have neglected the mass of the $X_\mu$ field. In this case, the frequency $\omega_{\rm L}^2$ does not pass through zero during the de Sitter stage of inflation, and since $\dot{H}, \ddot{H} \rightarrow 0$, the adiabatic parameter is strongly suppressed during this phase. However, for $w>-1/3$, i.e., after the end of inflation, $A_{\rm L}$ diverges when
$k^2 = a^2 H^2(1+ 3 w)/2$. Furthermore, rapid oscillations of the $\phi$ field lead to a sudden, non-adiabatic change in $a, \dot{a}, \ddot{a}$, amplifying $A_{\rm L}$ during the transition from the inflationary phase to the reheating epoch. \\
In contrast, $\omega_{\rm T}^2$ never passes through zero. For relativistic (non-relativistic) modes, the adiabatic parameter scales as $A_{\rm T} \simeq (a m_X)^2 a H/ k^3$ ($A_{\rm T} \simeq a H/k$), which implies that $A_{\rm T} \ll A_{\rm L}$. Therefore, for the transverse components, the largest departures from adiabaticity occur at the end of inflation, when $(a H)^{-1}$ reaches a minimum. \\
In the presence of non-minimal couplings, the adiabatic parameters acquire more complicated structure, i.e., 
\begin{align}
    A_{\rm T} (k, a) = \frac{(a H)^3 }{\omega_{\rm T}^{3/2}} \bigg\{ \frac{m_X^2}{H^2} - \bigg[3(3w-1) (\xi_1 - \frac{1}{6} \xi_2) + \xi_2 \bigg] \left( 1 + \frac{\dot{H}}{H^2} \right) \bigg\},
\end{align}
\begin{align}
    &A_{\rm L} (k, a) = \frac{(aH)^3}{2\omega_{\rm L}^{3/2}} \bigg\{ \frac{c_s^{2 \prime}}{a H} \frac{k^2}{a^2 H^2} + 2 \left( \frac{m_{\rm eff, x}^2}{H^2} + \frac{m_{\rm eff, x} m_{\rm eff, x}^\prime}{a H^3}\right) + \frac{2 k^2 \left(\frac{m_{\rm eff, t}^2}{H^2} + 2 \frac{m_{\rm eff, t} m_{\rm eff, t}^\prime}{a H^3} \right)}{(k^2 + a^2 m_{\rm eff, t}^2)^2}\nonumber \\
    &\times \bigg[ \frac{a^{\prime \prime}}{a} + \frac{m_{\rm eff, t}^{\prime \prime}}{m_{\rm eff, t}} + 2 \frac{a^\prime }{a} \frac{m_{\rm eff, t}^{\prime}}{m_{\rm eff, t}}  - 3\frac{(a^\prime m_{\rm eff, t} - m_{\rm eff, t}^\prime a)^2}{k^2 + a^2 m_{\rm eff, t}^2}  \bigg] - \frac{k^2}{k^2 + a^2 m_{\rm eff, t}^2} \nonumber \\
    &\times \frac{1}{(a H)^3}\bigg[\frac{a^{\prime \prime \prime}}{a} - \frac{a^{\prime \prime} a^\prime}{a^2} + \frac{m_{\rm eff, t}^{\prime \prime \prime}}{m_{\rm eff, t}} - \frac{m_{\rm eff, t}^{\prime \prime} m_{\rm eff, t}^\prime}{m_{\rm eff, t}^2} + 2 \frac{a^{\prime \prime} a - a^{\prime 2}}{a^2} \frac{m_{\rm eff, t}^\prime}{m_{\rm eff, t}} + 2 \frac{a^\prime}{a}  \frac{m_{\rm eff, t}^{\prime \prime} m_{\rm eff, t} - m_{\rm eff, t}^{\prime 2}}{m_{\rm eff, t}^2} \nonumber \\
    &-6 \frac{(a^\prime m_{\rm eff, t} - m_{\rm eff, t}^\prime a) (a^{\prime \prime} m_{\rm eff, t} - m_{\rm eff, t}^{\prime \prime} a )}{k^2 + a^2 m_{\rm eff, t}^2} + 6 \frac{(a^\prime m_{\rm eff, t} - m_{\rm eff, t}^\prime a)^2 (a^\prime a  m_{\rm eff, t}^2 + a^2  m_{\rm eff, t}^\prime  m_{\rm eff, t}) }{(k^2 + a^2 m_{\rm eff, t}^2)^2} \bigg] \bigg\}.
\end{align}
Neglecting terms involving derivatives of the effective masses, one finds
\begin{align}
A_{\rm T} (k, a) &= \frac{(a m_X)^2 a H}{\omega_{\rm T}^{3/2}}, \\
    A_{\rm L}(k, a) &\simeq  \frac{(aH)^3}{\omega_{\rm L}^2} \bigg\{ \left( \frac{m_{\rm eff, x}}{H} \right)^2 + \frac{k^2 (a m_{\rm eff, t})^2}{(k^2 + a^2 m_{\rm eff, t}^2)^2} \bigg[ 2 + \frac{\dot{H}}{H^2} - \frac{3(a m_{\rm eff, t})^2}{k^2 + a^2 m_{\rm eff, t}^2}\bigg] \nonumber \\
    &- \frac{k^2}{k^2 +a^2 m_{\rm eff, t}^2} \bigg[ 2 + 3 \frac{\dot{H}}{H^2} + \frac{\ddot{H}}{H^3}- 3 \frac{a^2 m_{\rm eff, t}^2 \left(2 + \frac{\dot{H}}{H^2} \right)}{k^2 + a^2 m_{\rm eff, t}^2} + \frac{ 3 a^4 m_{\rm eff, t}^4}{(k^2 + a^2 m_{\rm eff, t}^2)^2}]  \bigg\}.
\end{align}
The inclusion of non-minimal terms does not qualitatively change the behavior of super-horizon modes. For these modes, the most significant departures from adiabaticity occur at horizon exit for longitudinal polarization and at the end of inflation for transverse polarizations. However, the presence of direct couplings between $X_\mu$ and $R, R_{\mu \nu}$ has a more significant effect on sub-horizon modes, as it alters the dispersion relation of the longitudinal modes in the large-k limit 
\begin{align}
    &\omega_{\rm L}^2  \approx \frac{m_{\rm eff, x}^2}{m_{\rm eff, t}^2}k^2 - \frac{a^2 H^2}{2} (1+ 3 w).
\end{align}
Therefore, one expects that for certain values of the non-minimal couplings, the violation of adiabaticity may be more pronounced compared to the minimal case, leading to a more efficient gravitational production of large-k modes.
\subsection{Initial conditions}
During inflation the energy budget of the universe was dominated by the potential energy of the inflaton field, implying $w_{\rm inf} = -1$. Hence, during the de-Sitter phase of inflation the dispersion relations \eqref{eq:omegaT} and \eqref{eq:omegaL} become 
\begin{align}
     \omega_{\rm T, inf}^2 &\equiv \omega_{\rm T}^2 \bigg \rvert_{a<a_e}  = k^2 + a^2 m_X^2 - 3 a^2 H^2 (\xi_2 - 4 \xi_1), \\
    \omega_{\rm L, inf}^2 &\equiv \omega_{\rm L}^2 \bigg \rvert_{a<a_e}  = k^2 + a^2 m_{\text{eff},\text{inf}}^2 - \frac{k^2}{k^2 + a^2 m_{\text{eff},\text{inf}}^2} \left[ \frac{a^{\prime \prime}}{a} +\frac{m_{\text{eff},\text{inf}}^{\prime \prime}}{m_{\text{eff},\text{inf}}} \right. \non \\
    &\left. \quad \quad \quad \quad \quad \quad \quad \; \; +2 \frac{a^\prime }{a} \frac{m_{\text{eff},\text{inf}}^\prime}{m_{\text{eff},\text{inf}}} - 3 \frac{(a^\prime m_{\text{eff},\text{inf}} + a m_{\text{eff},\text{inf}}^\prime \,)}{k^2 + a^2 m_{\text{eff},\text{inf}}^2}\right], 
\end{align}
where 
\begin{align}
    m_{\text{eff},\text{inf}}^2 \equiv m_X^2 - 3(\xi_2 - 4 \xi_1)H^2.
\end{align}
In the infinite past, i.e., $\tau \rightarrow - \infty, a\rightarrow 0, a^\prime \rightarrow 0 \dots$, implying 
\begin{align}
    \lim_{\tau \rightarrow - \infty} \omega_{\rm T, inf}^2 (a) =  \lim_{\tau \rightarrow - \infty} \omega_{\rm L, inf}^2 (a) = k^2.
\end{align}
Hence, we will assume that all modes initially are in the Bunch-Davies vacuum
\begin{align}
    &X_{\rm BD} \equiv \lim_{\tau \rightarrow -\infty} X_{\rm T} = \lim_{\tau \rightarrow -\infty} \mathcal{X}_{\rm L} =  \frac{1}{\sqrt{2k}} e^{-i k \tau},  \\
     &X_{\rm BD}^\prime  \equiv \lim_{\tau \rightarrow -\infty} X^\prime_{\rm T} = \lim_{\tau \rightarrow -\infty} \mathcal{X}^\prime_{\rm L} =   \frac{-ik}{\sqrt{2k}} e^{-i k \tau}.
\end{align}
It is important to highlight that as long as $w = -1$, both frequencies are equal, $\omega_{\rm L}^2 = \omega_{\rm T}^2 = k^2$, in the limit $k \rightarrow \infty$. Therefore, except for the last few e-folds of inflation ($N \sim -2$), $\omega_{\rm T}^2$ and $\omega_{\rm L}^2$ remain positive for large-k modes, regardless of the values of the non-minimal couplings. This implies that uncontrolled tachyonic production of short-wavelength modes does not occur during the de Sitter stage of inflation and requires a deviation from the pure de Sitter equation of state, i.e., $w_{\rm inf} = -1$.

\subsection{The transverse modes}
\begin{figure}[htb!]
    \centering
\includegraphics[scale=0.4]{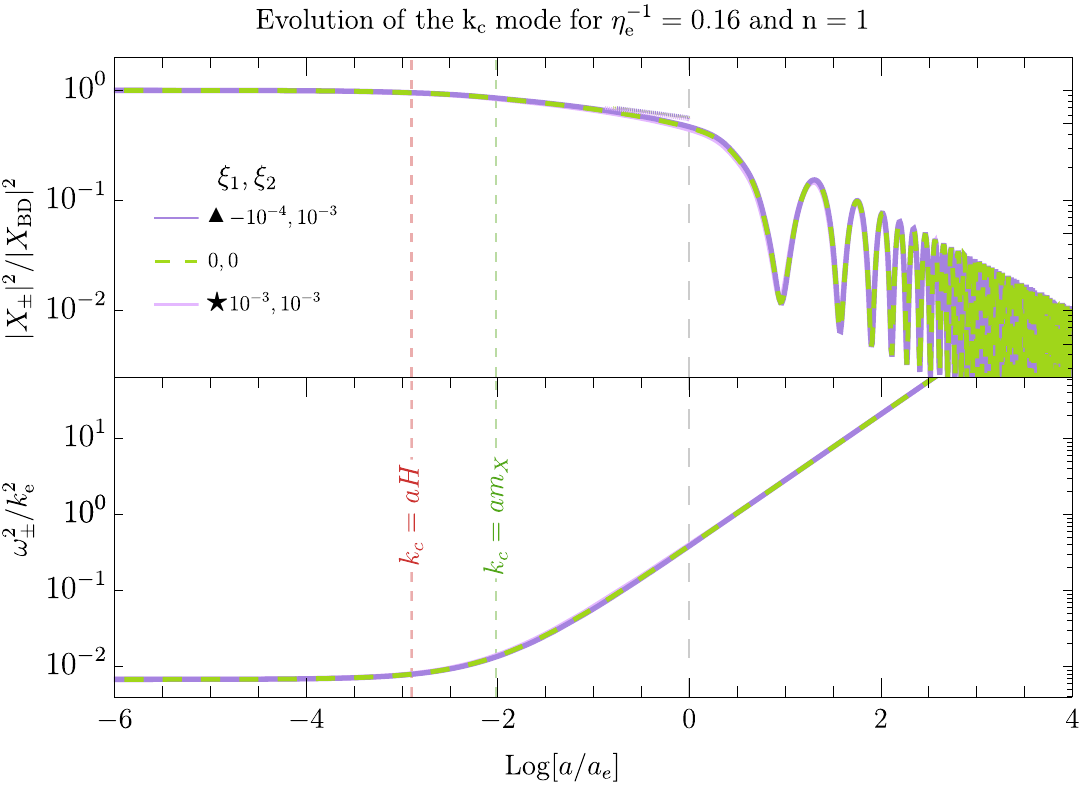}~~~~\includegraphics[scale=0.4]{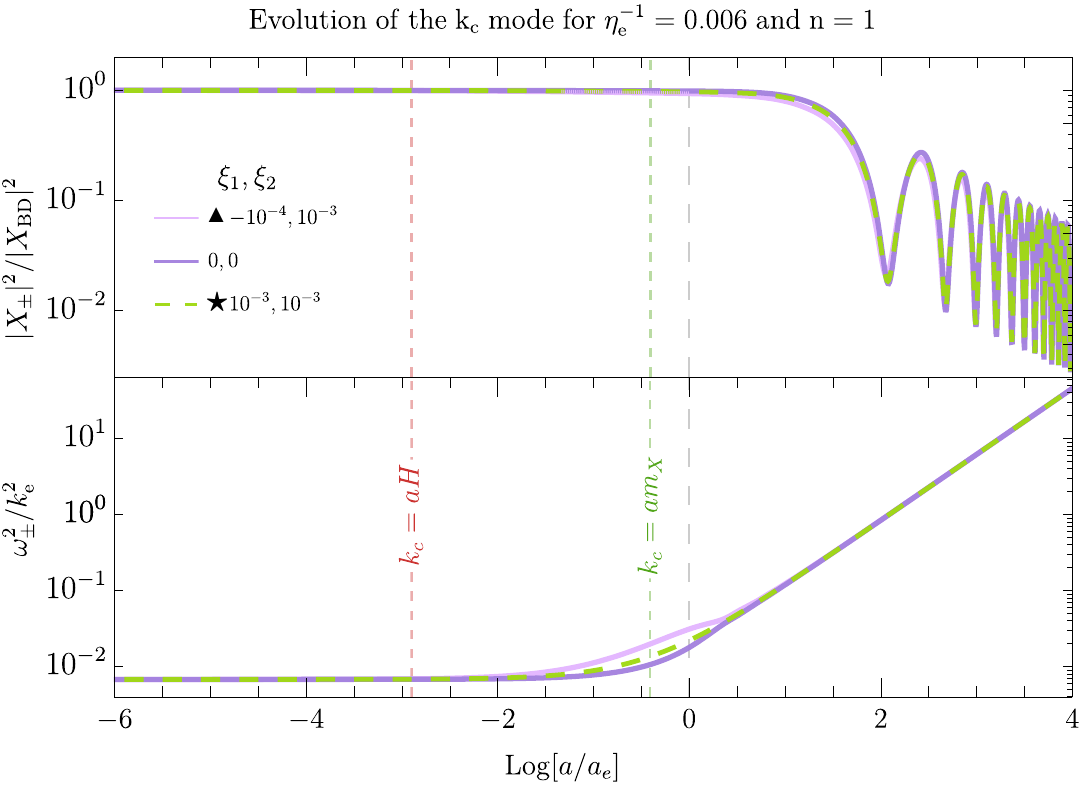}
    \caption{Upper panels: Evolution of the $k_c$ mode of the transversely-polarized vectors with mass $m_X = 5 \cdot 10^{12} \; \rm{GeV}$ (left panel) and $m_X = 10^{12}\; \rm{GeV}$ (right panel), assuming quadratic ($n=1$) inflaton potential during reheating. Densely-dashed lines correspond to the analytical solution Eq.\eqref{eq:ASolXT}. Lower panels: Evolution of the transverse frequency $\omega_{\rm \pm}^2/k_e^2$.}
\label{fig:XTPlot}
\end{figure}

Note that for $\xi_1, \xi_2$ meeting the condition \eqref{eq:instabilities}, the angular frequency $\omega_{\rm T}^2$ remains positive for all momentum modes. Indeed, one can easily check that   
\begin{align}
    k^2 + a^2 H_{\rm ini}^2 \left( \eta_e^{-1} - 3 \frac{H^2(a)}{H_{\rm ini}^2}(\xi_2 - 4 \xi_1 )\right) >0, 
 \end{align}
 as $(H(a)/H_{\rm ini})^2 \leq 1$ throughout the whole evolution. This indicates that in the instability-free region of the model, transverse modes does not experience tachyonic growth.  
\\ 
The approximate analytical solution to Eq.\eqref{eq:EoMT} can be derived in the pure de Sitter limit. Namely, during the de Sitter stage of inflation, the scale factor evolves as $a_{\rm dS} = - 1/(\tau H_{\rm dS})$ with $H_{\rm dS}=$const., which, in turn, implies that $m_X^2 - 3H_{\rm dS}^2(\xi_2 - 4 \xi_1)$ remains constant, making $a_{\rm dS}$ the only source of time variation in $\omega_{\rm T, inf}^2$. Equation \eqref{eq:EoMT} with $\omega_{\rm T}^2 = \omega^2_{\rm T, \rm inf}$ resembles the equation of motion of a conformally coupled scalar field, with the solution given by: 
\begin{align}
    &X_{\rm T}(a_{\rm dS}) = \frac{e^{i\pi(2 \nu+ 1)/4}}{2}\sqrt{\frac{\pi}{a_{\rm dS} H_{\rm dS}}} H_\nu^{(1)}\left(\frac{k}{a_{\rm dS} H_{\rm dS}} \right), &&\nu \equiv \sqrt{\frac{1}{4}- \frac{m_X^2 - 3H^2_{\rm dS}(\xi_2 - 4 \xi_1)}{H^2_{\rm dS}}}. \label{eq:ASolXT}
\end{align}
In Fig.~\ref{fig:XTPlot} we plot the amplitude squared of the $k_c$ mode (c.f. Fig.~\ref{fig:comovingdiag}) as a function of the number of e-folds for two values of $\eta_e^{-1}$ and different non-minimal couplings. The numerical solutions match the analytical (densely-dashed lines) approximation very well for $a<a_e$. The slight discrepancy between the two arises from the fact that, in a realistic inflationary scenario, the Hubble rate is not constant but slowly evolves over time. After inflation, $|X_{\rm T}|^2$ enters the oscillatory regime once the Hubble rate drops below $m_X$. In this region, the frequency, $\omega_{\rm T}^2 \approx a^2 m_X^2$, becomes a slowly-varying function, and the solution to the mode equation takes the JWKB form
\begin{align}
    &X_{\rm T}(a) \simeq \frac{c_1}{\sqrt{2 \omega_{\rm T}}} \exp{\left(- i \int d\tilde{\tau} \; \omega_{\rm T} \right)} +  \frac{c_2}{\sqrt{2 \omega_{\rm T}}} \exp{\left( i \int d\tilde{\tau} \;\omega_{\rm T} \right)}, \label{eq:JWKB}
\end{align}
From the above, one extracts the late-time behaviour of the envelope: $|X_{\rm T}|^2 \sim a^{-1}$. Note that the inclusion of the non-minimal terms does not affect the evolution of transverse modes. 

\subsection{The longitudinal mode}
The intricate structure of $\omega^2_{\rm L}$ 
indicates a non-trivial evolution of the longitudinally polarized modes. A comprehensive discussion of the solutions to the mode equation \eqref{eq:chiL} for the minimally coupled model can be found in Refs.\cite{Graham:2015rva,
Ahmed:2020fhc, Kolb:2020fwh}. The inclusion of non-minimal couplings significantly influences the dispersion relation of the longitudinal modes, resulting in a more complex time dependence. In particular, taking into account the non-minimal terms introduces additional contributions to $\omega_{\rm L}^2$, which are proportional to the scale factor and its derivatives. To analyze the solutions to the mode equation, it is convenient to distinguish five distinct scales: the distance to  the comoving Hubble horizon $(a H)^{-1}$, wavenumber of a given mode $1/k$, and the Compton wavelengths $(a m_X)^{-1}$, $(a m_{\rm eff, x})^{-1}$, and $(a m_{\rm eff, t})^{-1}$. The angular frequency $\omega_{\rm L}^2$ is controlled by different energy scales at different moments. Thus, the evolution of $\mathcal{X}_{\rm L}$ at a given time is determined either by its momentum $k$ or by one of the other energy scales. As one can see from Fig.~\ref{fig:comovingdiag}, the effective Compton wavelengths only $x$-type is shown deviate from the $(a m_X)^{-1}$ curve only at sufficiently early times, typically during inflation. This suggests that the presence of terms proportional to $\xi_1, \xi_2$ should not have a substantial impact on the evolution of the longitudinal modes. Instead, one might expect that the inclusion of non-minimal terms will only slightly modify the solutions obtained in the minimally coupled model and will not alter their qualitative behavior. 
\\ \\
\begin{figure}[htb!]
    \centering
\includegraphics[scale=0.4]{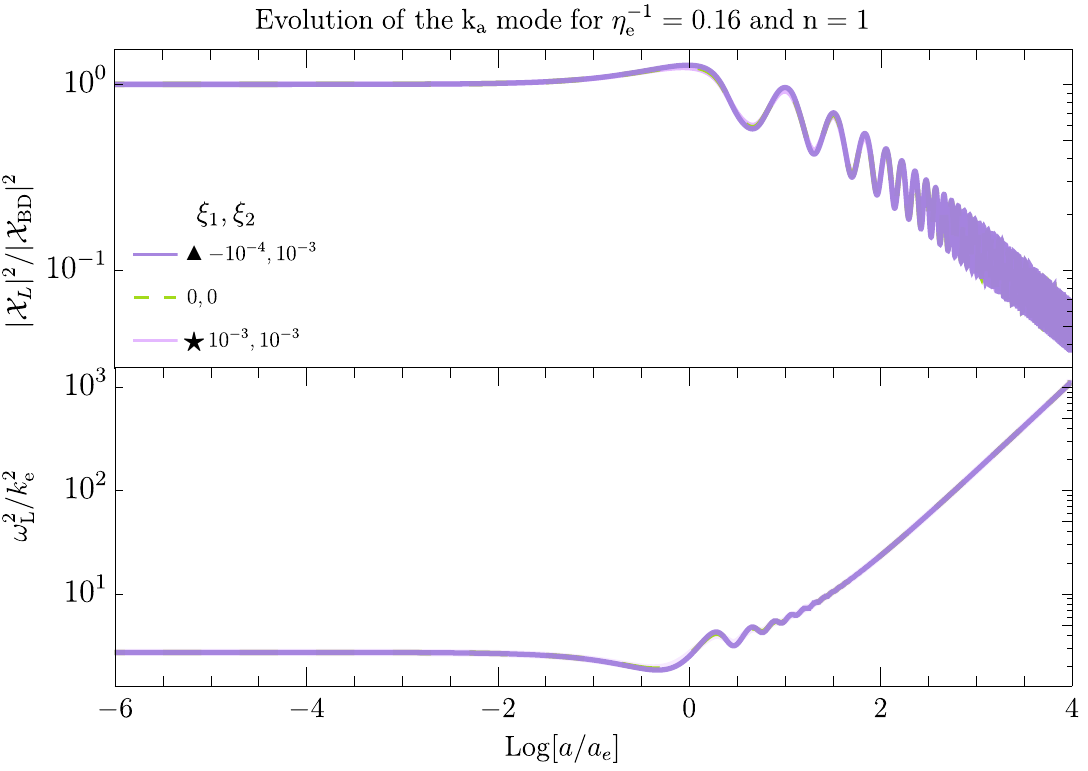}~~~~\includegraphics[scale=0.4]{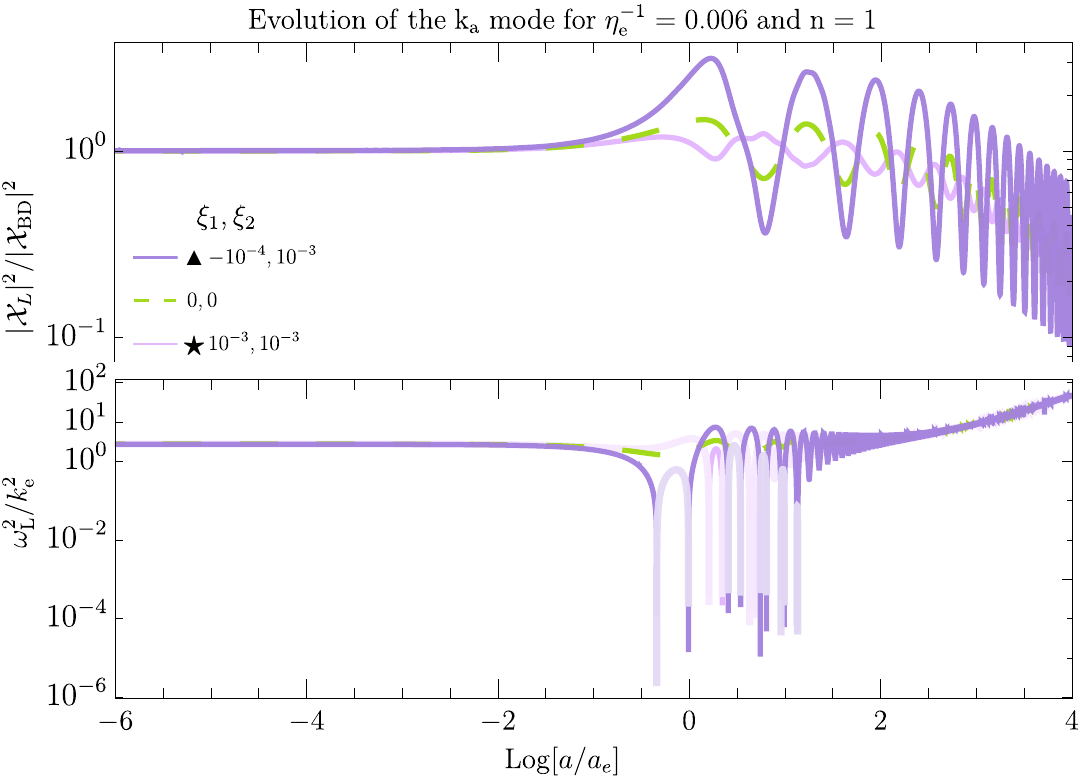} \\
\includegraphics[scale=0.4]{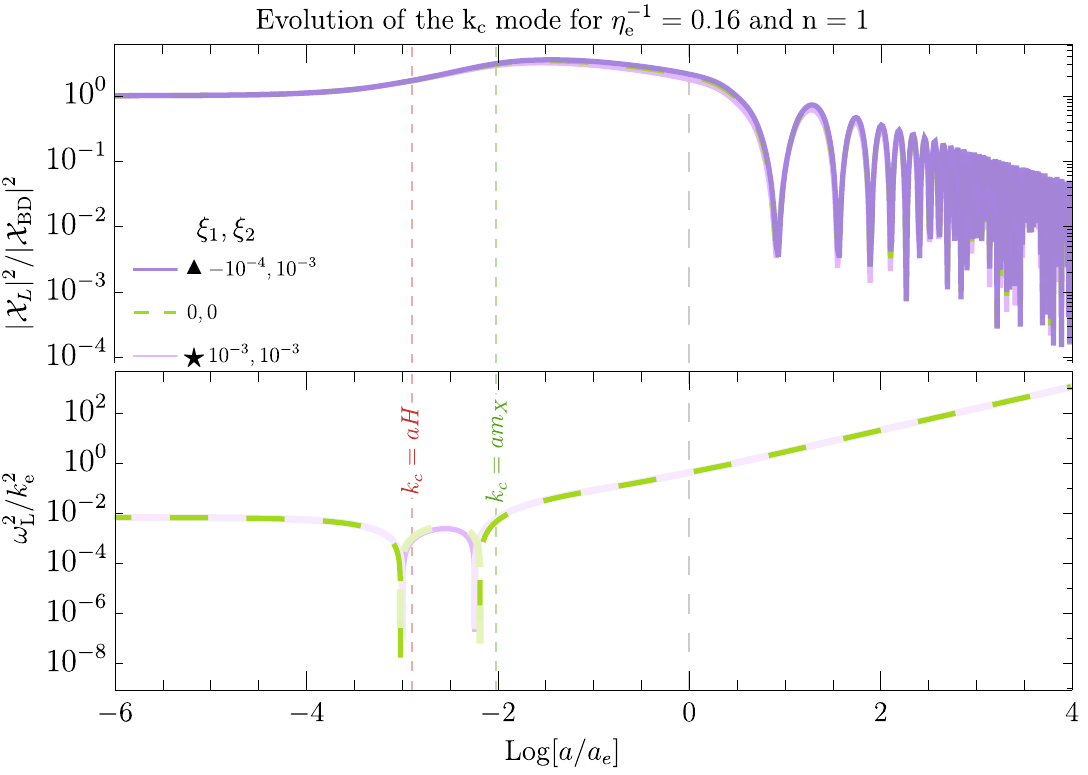}~~~~\includegraphics[scale=0.4]{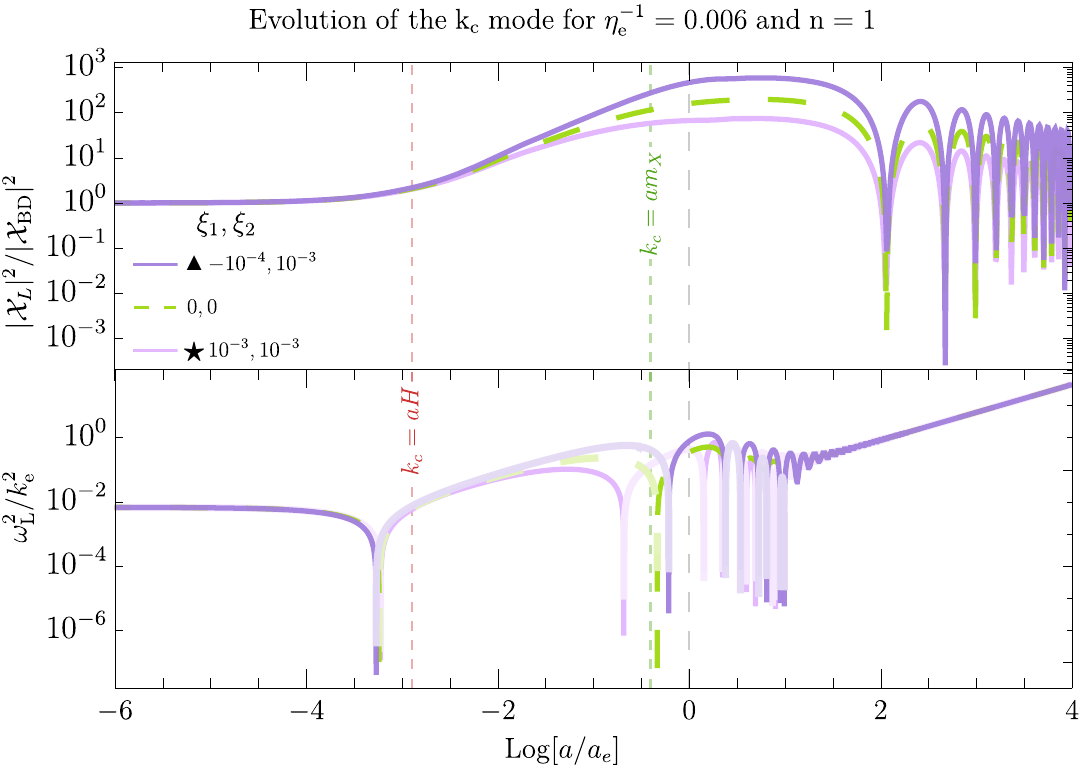} \\
\includegraphics[scale=0.4]{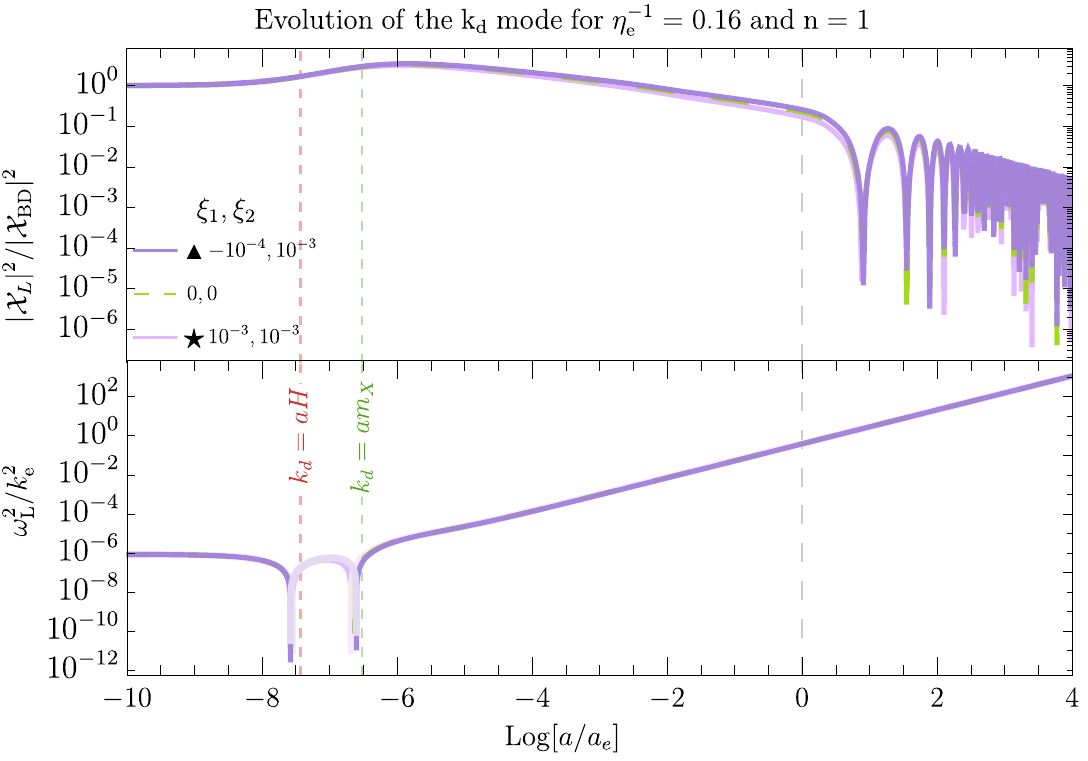}~~~~\includegraphics[scale=0.4]{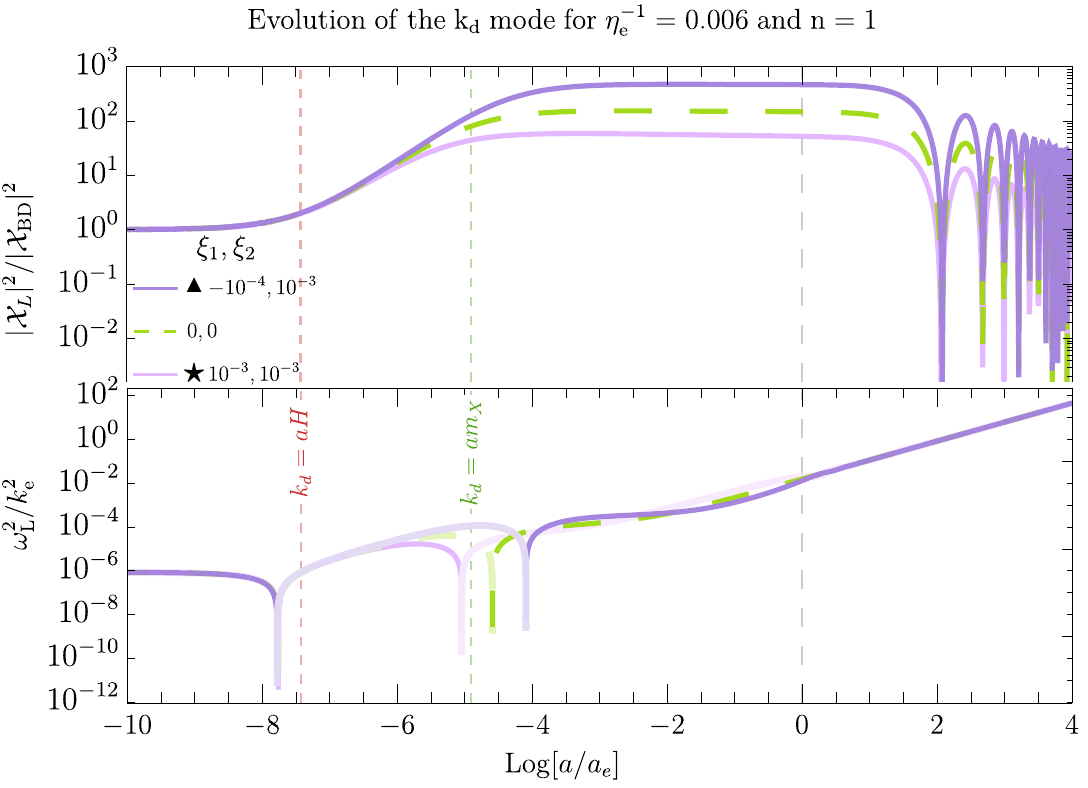}
    \caption{Upper panels: Evolution of the $k_a, k_c$ and $k_d$ modes of the longitudinal component of the $X_\mu$ field with mass $m_X = 5 \cdot 10^{12} \;\rm{GeV}$ (left) and $m_X = 10^{12} \; \rm{GeV}$ (right) for different values of $\xi_1, \xi_2$, assuming quadratic inflaton potential during reheating, i.e., $n=1$. Lower panels: The longitudinal frequency $\omega_{\rm \pm}^2/k_e^2$ as a function of the number of e-folds. The onset of tachyonic growth coincides with the frequency becoming negative (pale parts of the $\omega_{\rm L}^2$ curves), which, in turn corresponds to the horizon exit. Depending on the choice of the non-minimal couplings, the enhancement can be either more pronounced (triangle), or suppressed (star) compared to the minimal case ($\xi_1 = 0 = \xi_2$). For a given values of $\xi_1, \xi_2$  lighter vectors experience stronger amplification.}
\label{fig:XLmassPlot2}
\end{figure}
In Fig. \eqref{fig:XLmassPlot2}, we compare numerical results obtained for the minimal model with $\xi_1, \xi_2 = 0$ (dashed green curves) with its non-minimal extension for three different momentum modes: $k_d$, $k_c$, and $k_a$, cf. Fig.\eqref{fig:comovingdiag}. Notably, the inclusion of non-minimal couplings does not affect the overall behavior of $\mathcal{X}_{\rm L}$. Initially, the amplitude of all modes remains constant, and it is independent on the values of $\xi_1, \xi_2$. Modes with the wavenumber longer than $(a_e H_e)^{-1}$ ($k_c$ and $k_d$) exit the horizon at some point during inflation, as indicated by the red dashed vertical line. From that point onward, $\omega_{\rm L}^2$ becomes negative, and the solution to the mode equation takes the form \cite{Ahmed:2020fhc}
\begin{align}
    &\mathcal{X}_{\rm L} \sim \frac{H}{k^{3/2}} a, &\text{for } a \in (a_{h.c}, a_{C.c}).
\end{align}
The tachyonic growth of a given mode is terminated when the effective Compton wavelength falls below its wavenumber (indicated by the green dashed vertical line). As a result, heavier vectors undergo weaker amplification, since the time interval between $a_{h.c}$ and $a_{C.c}$ is shorter for them.
Depending on the choice of the non-minimal couplings, the enhancement can either be more significant or suppressed compared to the minimal case. For fixed values of $\xi_1$ and $\xi_2$, this effect is more pronounced for lighter species, as the contribution of the non-minimal terms to their effective masses remains relevant for a longer period. One of the most remarkable properties of longitudinal modes is that their amplitude remains frozen in the interval between $a_{C.c}$ and the point at which the Hubble rate falls below $m_X$. After that, the dynamics of the longitudinal modes begin to resemble the one of the transverse polarization. Specifically, one observe that $|\mathcal{X}_{\rm L}|^2$ starts to oscillate as described by Eq.\eqref{eq:JWKB}, with an increasing frequency $a m_X$, and a gradually decreasing amplitude that scales as $\sim a^{-1}$. In contrast, modes with wavelength shorter than $(a_e H_e)^{-1}$ 
never exit the horizon and undergo tachyonic amplification. For these modes, gravitational effects become relevant at the very onset of reheating. Near the transition point, rapid, quasi-periodic oscillations of the derivatives of the scale factor induce a sudden short-scale variation of the dispersion relation. This, in turn, violates the adiabaticity condition \eqref{eq:AdiabaticCon}, triggering strong parametric resonance for modes with momentum $k \sim \mathcal{O}(10)$, cf. the evolution of the $k_a$ mode in Fig.\eqref{fig:XLmassPlot2}. For $\xi_1 = 0 = \xi_2$ and $m_X \ll H_e$, the frequency of large-k longitudinal modes becomes 
\begin{align}
    \omega_{\rm L}^2 \approx k^2 - \frac{a^{\prime \prime}}{a} = k^2 + \frac{1}{6} a^2 R. 
\end{align}
Interestingly, the same dispersion relation is also observed for light scalars with the minimal coupling to gravity, see Ref \cite{Verner:2024agh}. To gain a better understanding of the resonance, it is convenient to determine the time structure of the Ricci scalar. To that end, we use the fact that during inflation 
\begin{align}
    R \simeq \frac{\dot{\phi}^2 - 4 V(\phi)}{ \mpl^2}, \label{eq:riccisol}
\end{align}
and that for the power-law potential $V(\phi) \sim \phi^{2n}$ \cite{Ahmed:2022tfm}
\begin{align}
    \phi \simeq \phi_e \left( \frac{a_e}{a}\right)^{\frac{3}{n+1}} \times \left[ \mathcal{I}_z^{-1} \left( \frac{1}{2n}, \frac{1}{n}\right) \right]^{\frac{1}{2n}}, \label{eq:phisol}
\end{align}
where $\phi_e$ denote the field value at the end of inflation, $\mathcal{I}_z^{-1}$ is the inverse of the regularized incomplete beta function, and $z = 1 - 4 (t-t_e) /\mathcal{T} $ with $\mathcal{T}$ being the (time-dependent) period of the oscillations. For the quadratic and quartic potential, $\mathcal{I}_z^{-1}$ becomes  \cite{Greene:1997fu, Lebedev:2022vwf}
\begin{align}
    &\cos{[m_\phi (t -t_e)]}, &n=1, \\
    &{\text cn}\left(x, \frac{1}{\sqrt{2}} \right),  &n=2,
\end{align}
with 
\begin{align}
     & x \equiv \sqrt{\lambda}  \phi_e \frac{a_e}{a} (t-t_e),
\end{align}
and $m_\phi = \Lambda^2 / (\sqrt{3 \alpha} \mpl)$ is the mass of the inflaton, $\lambda = (\Lambda/ \mpl)^4 / (9 \alpha^2)$ denotes its quartic coupling, and $\text{cn}$ is the elliptic cosine. Inserting the solution \eqref{eq:phisol} into Eq.\eqref{eq:riccisol}, one gets
\begin{align}
    R &= - \frac{m_\phi^2 \phi_e^2}{2 \mpl^2}  \left(\frac{a_e}{a} \right)^3 \left[1+ 3 \cos{[2 m_\phi(t-t_e)]} - 3 \frac{H}{m_\phi^2} \sin{[2 m_\phi(t-t_e)]}-  \frac{9}{2} \frac{H^2}{m_\phi^2} \cos^2{[2 m_\phi(t-t_e)]}   \right] , \nonumber \\
    &\simeq - \frac{m_\phi^2 \phi_e^2}{2 \mpl^2} \left(\frac{a_e}{a} \right)^3 \left[1+ 3 \cos{[2 m_\phi(t-t_e)]} \right], \quad \text{for } n=1, \\
    R &\simeq - \phi_e^4 \lambda \left(\frac{a_e}{a} \right)^4 \bigg\{ {\text cn}^4\left(x, \frac{1}{\sqrt{2}} \right) - \frac{1}{\lambda} \left( \frac{a}{a_e} \right)^2 \frac{1}{\phi_e^2}\left[ H {\text cn} \left( x, \frac{1}{\sqrt{2}} \right) + \nonumber \right.  \\
    &+ \sqrt{\lambda} \phi_e \frac{a_e}{a} [1 + H(t-t_e)] \text{dn} \left(x, \frac{1}{\sqrt{2}} \right) \text{sn} \left(x, \frac{1}{\sqrt{2}} \right)\bigg]^2 \bigg\} , \nonumber \\
    &\simeq - \frac{\lambda}{2} \phi_e^4 \left( \frac{a_e}{a} \right)^4  \left[ 1 - 3 \text{cn}^4\left( x, \frac{1}{\sqrt{2}} \right) \right], \quad \text{for } n=2,
\end{align}
where we have neglected rapidly-dying terms, suppressed by the powers of $H$. During reheating, $R$ acquires an oscillatory contribution, whose envelope decays as a power law of the scale factor, with a more rapid decrease for $n=2$, cf. Fig.\eqref{fig:HRplot}. Due to the fact that efficiency of the resonance is triggered by the evolution of the Ricci scalar, short-wavelength modes are mostly produced at the very onset of reheating, when the amplitude of $R$ is the highest.

\begin{figure}[htb!]
    \centering
\includegraphics[scale=0.3]{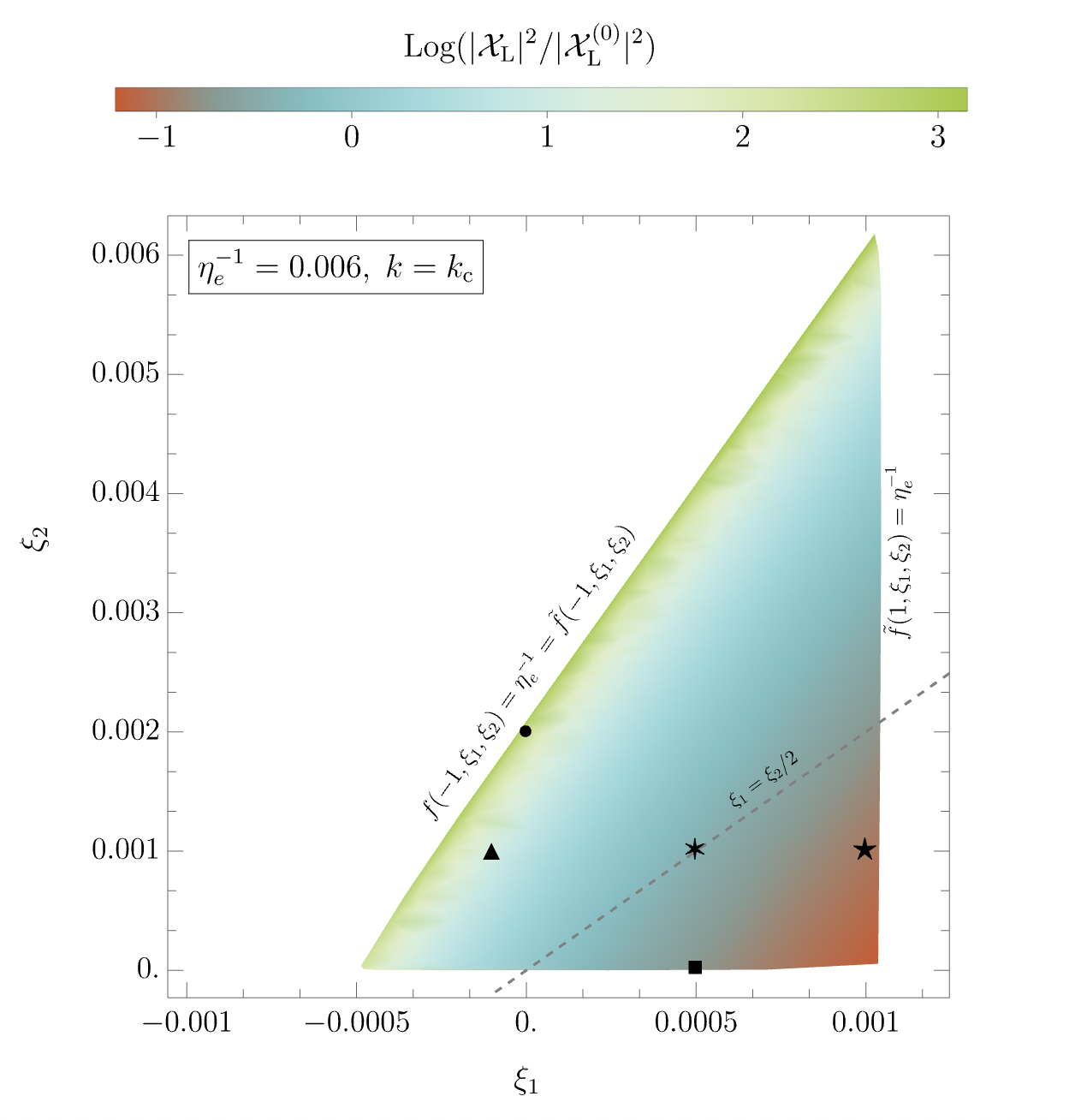}
\includegraphics[scale=0.44]{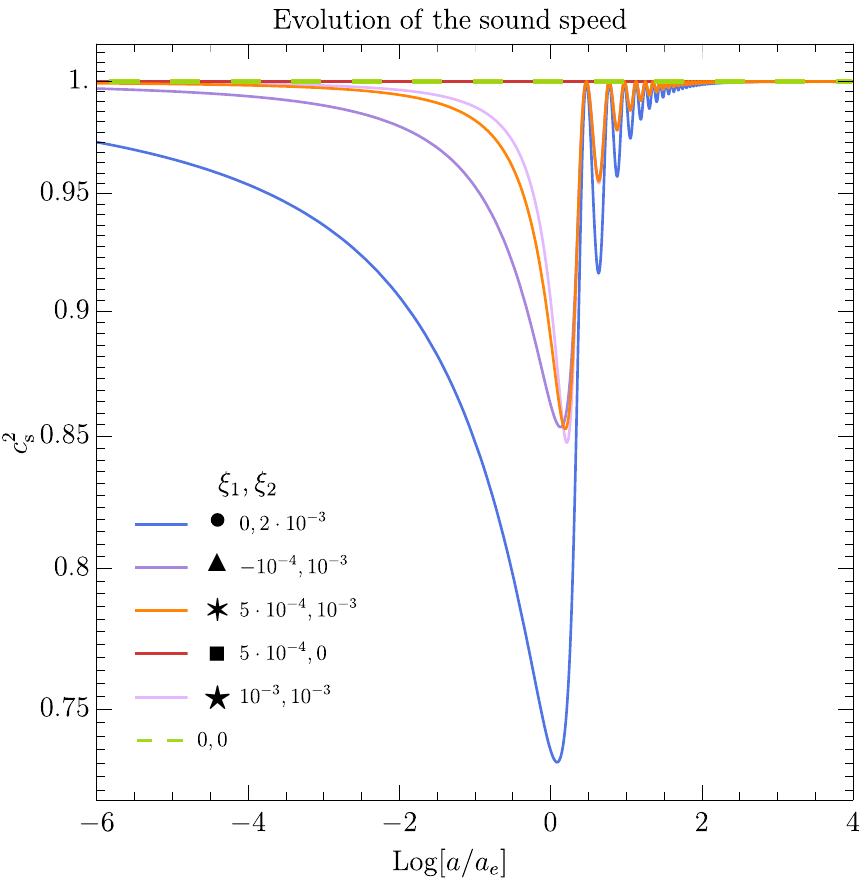}
    \caption{Left: The maximal value of the squared amplitude  $|\mathcal{X}_{\rm L}(\xi_1, \xi_2)|^2 \equiv |\mathcal{X}_{\rm L}|^2$ normalized to $|\mathcal{X}_{\rm L}(0,0)|^2 \equiv |\mathcal{X}_{\rm L}^{(0)}|^2$.  Different symbols represent several benchmark values of the non-minimal couplings ($\xi_1, \xi_2$): box ($5 \cdot 10^{-4}$,0), triangle (-$10^{-4}$, $10^{-3}$), point (0, $2 \cdot 10^{-3}$), five-point star($10^{-3}$,$10^{-3}$), six-point star ($5 \cdot 10^{-4}$, $10^{-3}$). Right: Evolution of the sound speed for selected values of $\xi_1, \xi_2$.}
\label{fig:scan}
\end{figure}

The inclusion of the non-minimal terms alters the large-k limit of $\omega_{\rm L}^2$, such that
\begin{align}
    \omega_{\rm L}^2 \approx \frac{m_{\rm eff, x}^2}{m_{\rm eff, t}^2} k^2 - \bigg[ \frac{a^{\prime \prime}}{a} + \frac{m_{\rm eff, t}^{\prime \prime}}{m_{\rm eff, t}} + 2 \frac{a^\prime}{a} \frac{m_{\rm eff, t}^\prime}{m_{\rm eff, t}} \bigg].
\end{align}
In this case, the efficiency of the resonance becomes sensitive to the values of  $\xi_1, \xi_2$, as the inclusion of non-minimal terms introduces an additional dependence of $\omega_{\rm L}^2$ on $a^\prime, a^{\prime \prime}, \dots$ In particular, for the specific values of $\xi_1, \xi_2$, we observe an amplification of the amplitude of sub-horizon modes relative to the minimal model with $\xi_1 = 0 =\xi_2$. More rigours treatment requires analysis of the stability/instability charts of the (quasi)-periodic differential equations. A detailed analysis is however beyond the scope of this work.
\\ \\
Although, Fig.\eqref{fig:XLmassPlot2} illustrates the generic features of the non-minimal model, it is instructive to explore the allowed parameter space in greater detail. In the left panel of Fig.\eqref{fig:scan}, we map the variation of the squared amplitude within the instability-free triangle. Specifically, we plot the ratio $|\mathcal{X}_{\rm L}^{\rm max} (\xi_1, \xi_2)|^2/ |\mathcal{X}_{\rm L}^{\rm max}(0,0)|^2$ for the $k_c$ mode, where $|\mathcal{X}_{\rm L}^{\rm max} (\xi_1, \xi_2)|^2$ represents the maximum squared amplitude for a given set of $\xi_1$ and $\xi_2$, as a function of the non-minimal couplings. Note that the $k_c$ mode becomes super-horizon, and for the chosen value of $\eta_e^{-1}$, it intersects the (effective) Compton wavelength curve during inflation. Consequently, its amplitude is maximized around $a_{C.c}$, prior to the end of inflation, $a_{C.c} < a_e$.  

In general, depending on the values of the non-minimal couplings two scenarios emerge: i) $|\mathcal{X}_{\rm L}^{\rm max} (\xi_1, \xi_2)|^2 > |\mathcal{X}_{\rm L}^{\rm max} (0,0)|^2$, and ii) $|\mathcal{X}_{\rm L}^{\rm max} (\xi_1, \xi_2)|^2 < |\mathcal{X}_{\rm L}^{\rm max} (0,0)|^2$. The strongest enhancement of the amplitude occurs along the green hypotenuse, described by the curve $f(-1, \xi_1, \xi_2) = \eta_e^{-1} = \tilde{f}(-1, \xi_1, \xi_2)$. For $\xi_1, \xi_2$ lying exactly on this line, one finds
\begin{align}
    m_{\rm eff, inf}^2 =  \eta_e^{-1} \left[ 1 -  \left(\frac{H}{H_{\rm ini}} \right)^2 \right]  H_{\rm ini}^2,
\end{align}
which implies that $|\mathcal{X}_{\rm L}^{\rm max} (\xi_1, \xi_2)|^2$ is maximized for the non-minimal couplings for which both effective masses vanish at the very onset of inflation. Note however that their ratio, defining the sound speed $c_s^2$, remains finite, as demonstrated in the right panel of Fig.\eqref{fig:scan}, and goes to unity in the remote past, when the evolution of the metric becomes more de Sitter-like. It is important to highlight that the inclusion of non-minimal couplings can amplify the squared amplitude by up to three orders of magnitude compared to the minimal case. The opposite effect is observed for $\xi_1, \xi_2$ lying near the intersection of the $\xi_2 = 0$ and $\tilde{f}(1, \xi_1, \xi_2) =\eta_e^{-1}$ curves. In the red region, $|\mathcal{X}_{\rm L}^{\rm max} (\xi_1, \xi_2)|^2$ might be suppressed by more than an order of magnitude relative to its minimal value.  

In the right panel of Fig.\eqref{fig:scan}, we plot the evolution of the sound speed $c_s^2$ as a function of the number of e-folds for benchmark values of the non-minimal couplings. In general, $c_s^2$ deviates from unity for $\xi_2 \neq 0$. During the de Sitter phase of inflation all curves approach a constant limit $c_s^2 \rightarrow 1$. Shortly before the end of inflation, curves corresponding to non-zero values of $\xi_2$ starts to decrease. The depth of the minimum is the largest for $\xi_1, \xi_2$ that are close to the $f(-1, \xi_1, \xi_2) = \eta_e^{-1} = \tilde{f}(-1, \xi_1, \xi_2)$ line. After inflation ends, $c_s^2$ 
starts oscillating with a damped amplitude. Eventually, the evolution of the sound speed becomes static. 

\section{The energy density of non-minimally coupled vector field} \label{rho}
\begin{figure}[htb !]
    \centering
\includegraphics[scale=0.4]{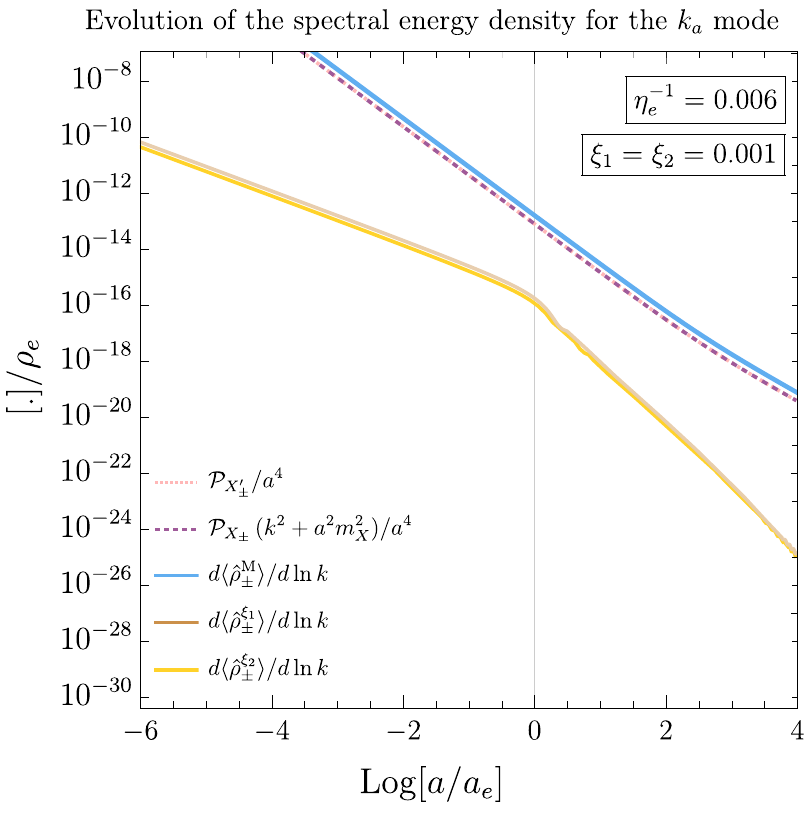}~~~~\includegraphics[scale=0.4]{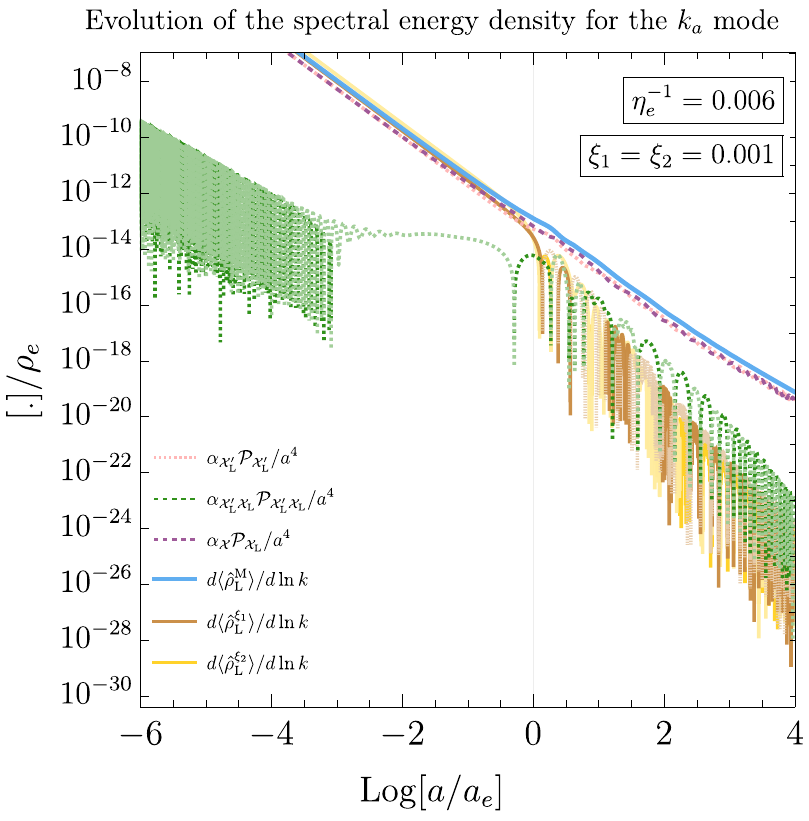} \\
\includegraphics[scale=0.4]{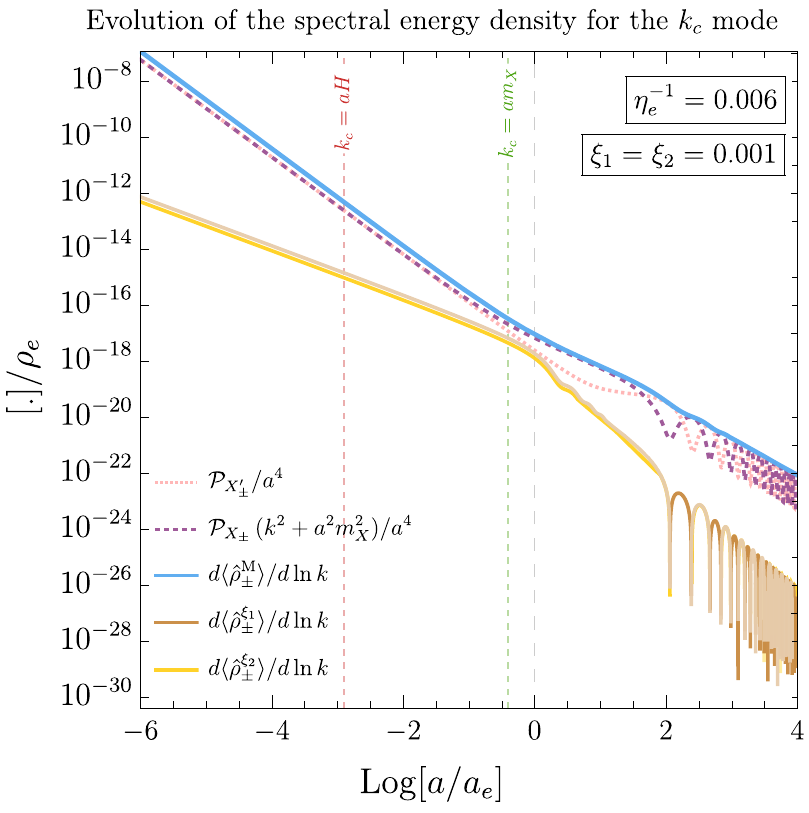}~~~~\includegraphics[scale=0.4]{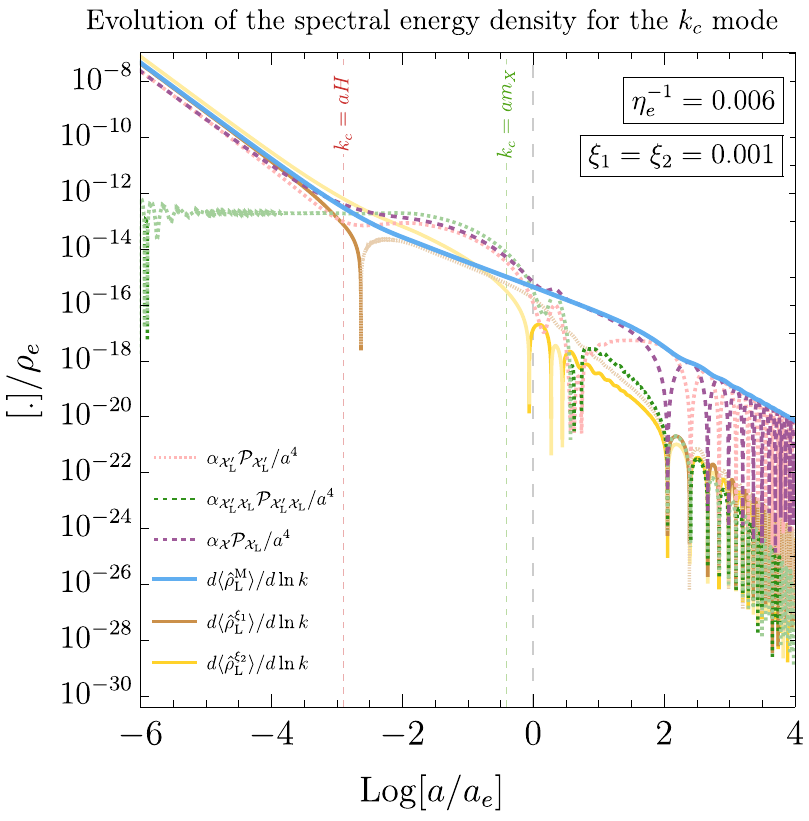} \\
\includegraphics[scale=0.4]{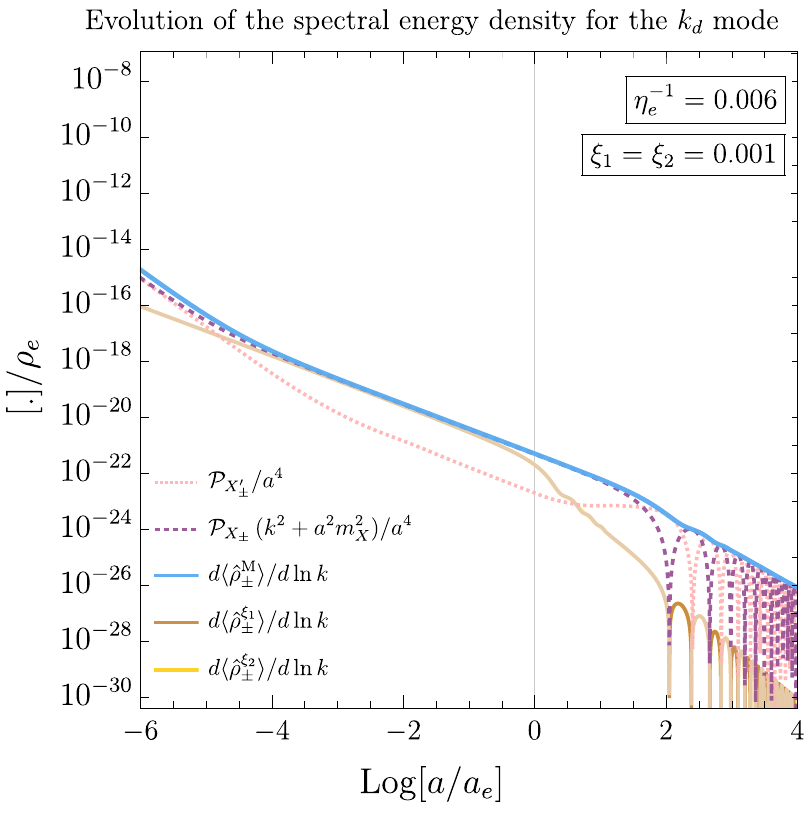}~~~~\includegraphics[scale=0.4]{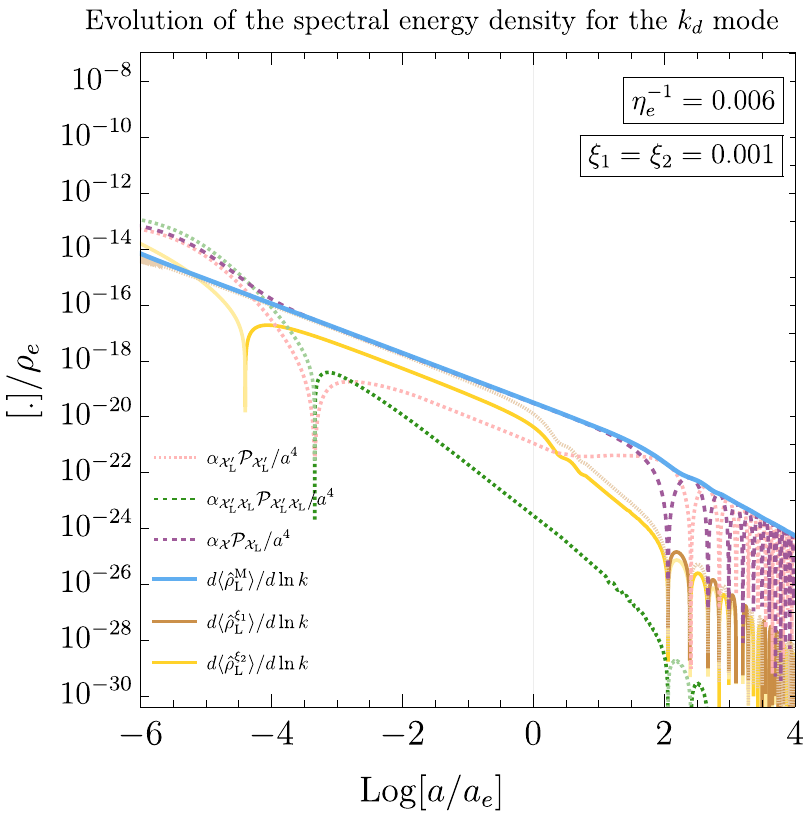}
    \caption{
    Evolution of different components of the spectral energy density for transverse (left column) and longitudinal (right column) polarizations, illustrated for three modes: the sub-horizon $k_a$ mode (first row) and two super-horizon modes $k_c$ (middle row) and $k_d$ (last row). Each component is normalized to the inflaton energy density at the end of inflation. Above, $\alpha_{\mathcal{X}^\prime_{\rm L}} \equiv a^2 m_X^2/(k^2 + a^2 m_X^2) A_{\rm L}^2$, $\alpha_{\mathcal{X}^\prime_{\rm L}\mathcal{X}_{\rm L}} \equiv a^2 m_X^2/(k^2 + a^2 m_X^2) A_{\rm L} A_{\rm L}^\prime $ and $\alpha_{\mathcal{X}_{\rm L}} \equiv a^2 m_X^2 A_{\rm L}^2 + a^2 m_X^2/(k^2 + a^2 m_X^2) A_{\rm L}^{\prime 2} $. Pale regions of the curves correspond to negative values of the respective quantities. }
\label{fig:XLmassPlot}
\end{figure}

\begin{figure}[htb!]
    \centering
\includegraphics[scale=0.5]{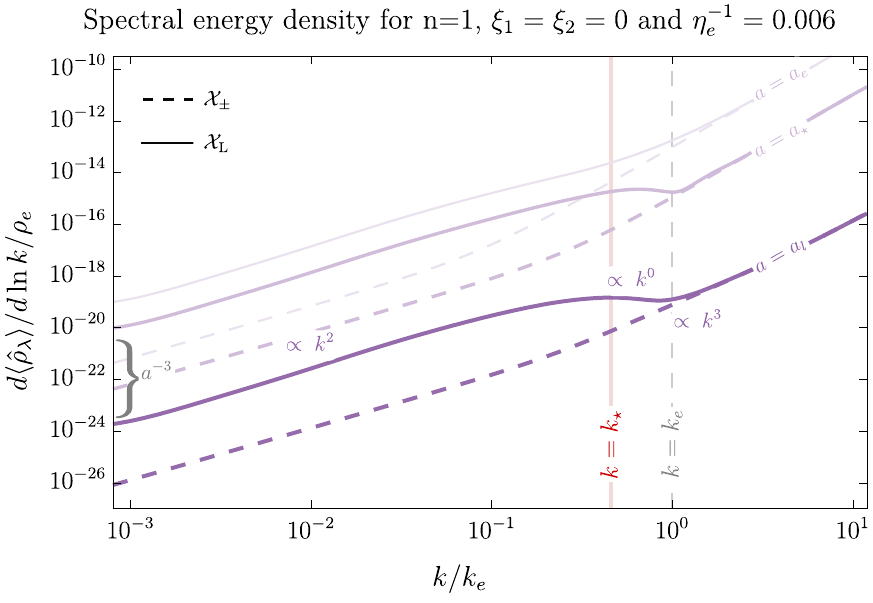}
\includegraphics[scale=0.5]{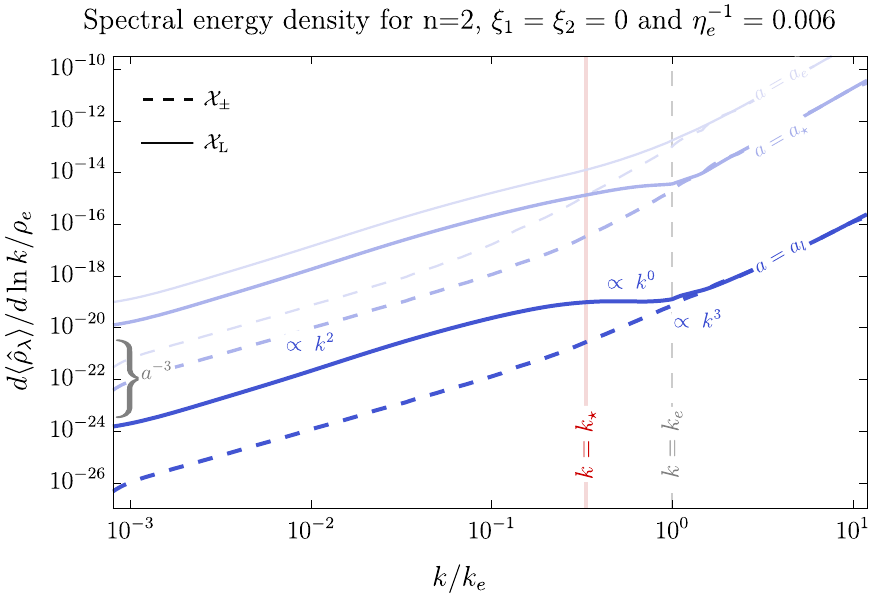}
    \caption{Spectral energy density as a function of momentum for quadratic (left) and quartic (right) reheating model, shown for the longitudinal (solid) and transverse (dashed) components of the minimally-coupled vector field ($\xi_1 =\xi_2 =0$). The spectra are normalized to the inflaton energy density at the end of inflation, $\rho_e \equiv 3 \mpl^2 H_e^2$. }
\label{fig:rhoLT0}
\end{figure}

Varying the action \eqref{eq:X_action} with respect to the metric tensor, we obtain the energy-momentum tensor for the $X$ field. The complete derivation can be found in Appendix \eqref{sec:EMT}; here, we provide only the final result. \\ The stress-energy tensor of the non-minimally coupled vector field has three constituents:
\begin{align}
    T_{\mu \nu}^{X} & \equiv T_{\mu \nu}^{\rm M} + T_{\mu \nu}^{\xi_1} + T_{\mu \nu}^{\xi_2},
\end{align}
that is, the minimal part  
\begin{align}
    T_{\mu \nu}^{\rm M}    &=g_{\mu \nu} \left( \frac{1}{4} g^{\rho \sigma} g^{\alpha \beta} X_{\rho \alpha} X_{\sigma \beta} - \frac{m_X^2}{2} g^{\alpha \beta} X_\alpha X_\beta \right) - g^{\alpha \beta}X_{\mu \alpha} X_{\nu \beta} + m_X^2 X_\mu X_\nu,
\end{align}
and two non-minimal terms,
\begin{align}
     T_{\mu \nu}^{\xi_1} = \xi_1 \bigg[ - R X_\mu X_\nu - G_{\mu \nu} g^{\rho \sigma} X_\rho X_\sigma - g_{\mu \nu}g^{\rho \sigma} g^{\alpha \beta} \nabla_\sigma \nabla_\rho (X_\alpha X_\beta) + g^{\rho \sigma}\nabla_\mu \nabla_\nu (X_\rho X_\sigma) \bigg],
\end{align}
\begin{align}
    T_{\mu \nu}^{\xi_2}  &=  \frac{\xi_2}{2} \bigg[ - g_{\mu \nu}g^{\alpha \rho} g^{\beta \sigma}R_{\rho \sigma}X_\alpha X_\beta  + 2 g^{\rho \sigma}R_{\nu \sigma} X_\mu X_\rho +2 g^{\rho \sigma}R_{\mu \sigma} X_\nu X_\rho  +  g^{\rho \sigma }\nabla_\rho \nabla_\sigma(X_\mu X_\nu) \non \\
    & +  g_{\mu \nu} g^{\lambda \rho}  g^{\kappa \sigma} \nabla_\lambda \nabla_\kappa(X_\rho X_\sigma)- g^{\lambda \sigma} \nabla_\mu \nabla_\sigma(X_\lambda X_\nu) - g^{\lambda \sigma}\nabla_\nu \nabla_\sigma(X_\lambda X_\mu) \bigg],
\end{align}
proportional to $\xi_1$ and $\xi_2$, respectively.
After quantizing the vector field one can use the above expressions to calculate the expectation value of the $X_\mu$ field energy density $\langle{\hat{\rho}_X} \rangle$, see Appendix \eqref{sec:EnergyDensity}. It includes contributions from longitudinal and transverse modes, each of which is composed of three parts, as shown in Eqs.~\eqref{eq:edL}-\eqref{eq:edT} in terms of the power spectra defined in Eqs.~\eqref{pow_spec_a}-\eqref{pow_spec_c}. 
Hereafter, we will discuss the properties of differential energy densities
\beq
\frac{d}{d\ln k}\langle \hat{\rho}_{L,\pm}^{M,\xi_1,\xi_2} \rangle,
\label{spect_dens}
\eeq
which we refer to as \textit{spectral energy density}, as functions of $a$ for a fixed momentum, e.g., for $k_a$, $k_c$ and $k_d$ in Fig.~\ref{fig:XLmassPlot}, and/or
as functions of momentum $k$ at definite times, e.g., Figs.~\ref{fig:rhoLTMNM}  and \ref{fig:rhoLT0}.
\\ \\ 
Depending on the momentum $k$ of the modes, different components may dominate the total energy density. However, at \textit{late times}, the largest contribution to $\langle{\hat{\rho}_X} \rangle$ comes from the standard minimal term $\langle{\hat{\rho}_X^{\rm M}} \rangle$ (blue curves in Fig.~\ref{fig:XLmassPlot}). Specifically, it can be observed that, for all modes, the non-minimal components (yellow and brown curves) are strongly redshifted after the end of inflation. Such suppression originates from the additional $H^2$ factor multiplying the non-minimal terms, see also Ref.\cite{Ozsoy:2023gnl}. In particular, after the end of inflation the Hubble rate decreases as a power-law with the scale factor, according to Eq.\eqref{eq:Hubble_ap}. Hence, at \textit{late times} we apply the following approximation 
\begin{align}
    \langle \hat{\rho}_X \rangle \simeq \langle \hat{\rho}_X^{\rm M} \rangle. 
\end{align}
\subsection{Super-horizon modes}
\begin{figure}[htb!]
    \centering
\includegraphics[scale=0.5]{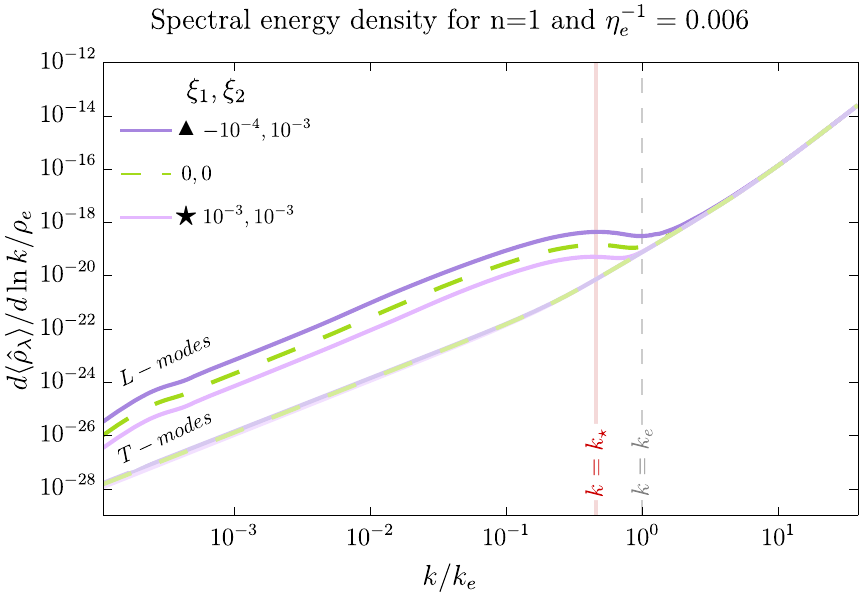}
\includegraphics[scale=0.5]{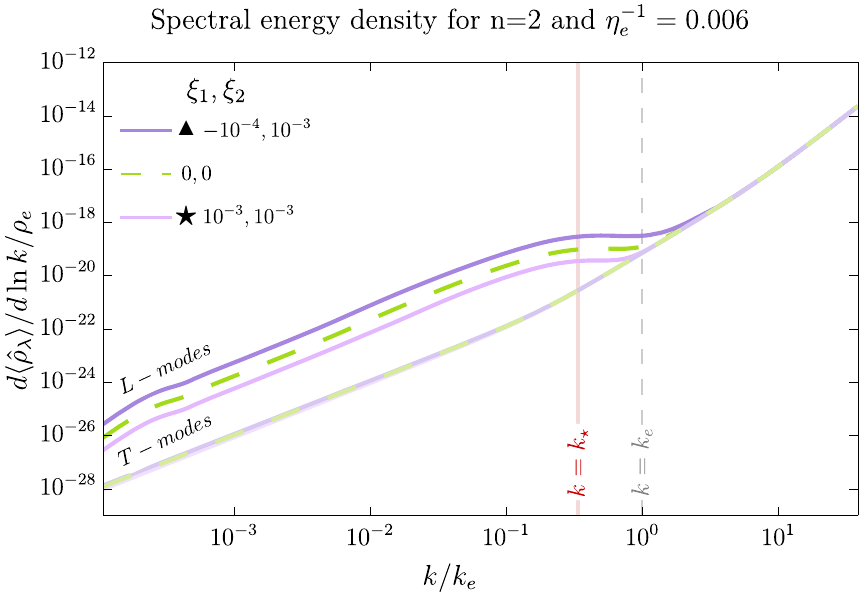}\\
\includegraphics[scale=0.5]{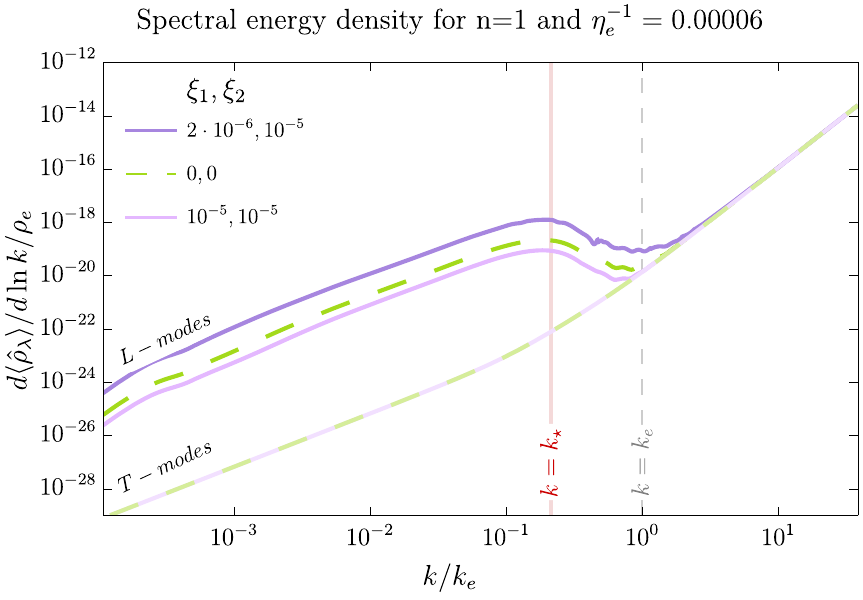}
\includegraphics[scale=0.5]{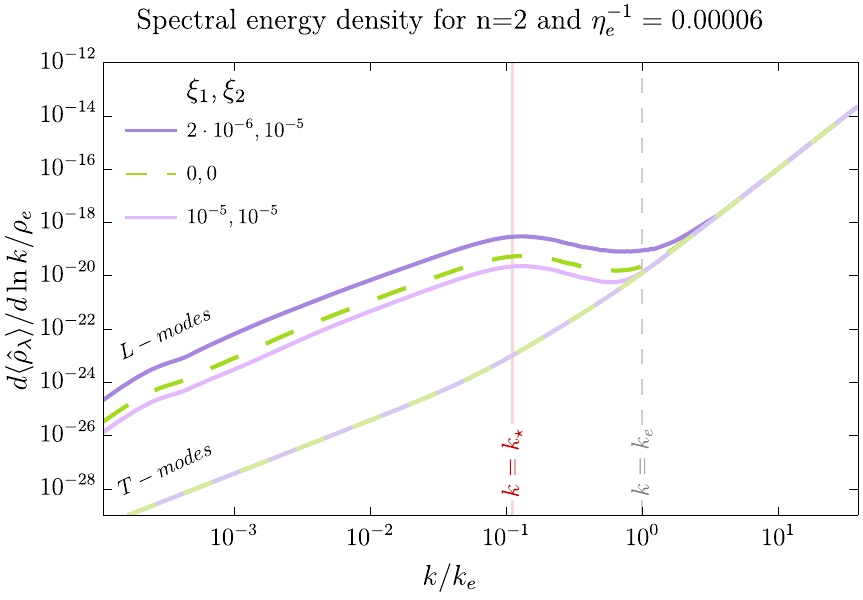}\\
    \caption{Spectral energy density of the longitudinal (L) and transverse (T) components of the minimally (dashed green) and non-minimally  (violet curves) coupled vector field for the quadratic (left panel) and quartic (right panel) reheating model at \textit{late times}.}
\label{fig:rhoLTMNM} 
\end{figure}

For super-horizon modes, e.g., $k_c$, $k_d$ in Fig.~\ref{fig:XLmassPlot}, the spectral energy density of the longitudinal polarization is several orders of magnitude larger than $d \langle \hat{\rho}_\pm \rangle/ d \ln{k}$. Longitudinal modes with long-wavelength, i.e., $k^{-1} \gtrsim k_e^{-1}$, are amplified by tachyonic instability. Consequently, the energy density of the low-k modes is dominated by the energy of the longitudinal component of the $X_\mu$ field, i.e., 
 \begin{align}
     &\langle \hat{\rho}_X \rangle \simeq \langle \hat{\rho}_{\rm L}^{\rm M} \rangle, &\text{for } k \lesssim k_e.   
\end{align}
In Fig.~\ref{fig:rhoLT0}, we present the spectral energy density of the longitudinal and transverse polarizations as a function of $k$ in three different moments of time at three different moments in time: at the end of inflation $a = a_e$, at $a=a_\star$, defined by the following condition:
\begin{align}
    m_X = H(a_\star), 
\end{align}
and at \textit{late times}, i.e, $a=a_{l}$, when the evolution of modes becomes adiabatic\footnote{Note that  $a_\star$ and $a_l$ depend on $m_X$}. 
Initially, $d \langle \hat{\rho}_{\rm L} \rangle / d \ln{k}$ scales as $k^2$ for all super-horizon modes, and as $k^4$ for short-wavelength modes with $k^{-1} \lesssim k_e^{-1}$. Then, around $a=a_\star$,
one observes the formation of a peak in the spectrum, centered around $k \simeq k_\star$. The $k_\star$ mode reenters the horizon at the moment when the comoving Compton wavelength $(am_X)^{-1}$ crosses the Hubble sphere, such that
\begin{align}
    k_\star \equiv a_\star m_X.
\end{align}
The position of the peak is determined by the evolution of the Hubble rate during reheating. Comparing the two models shown in Fig.~\ref{fig:rhoLT0}, we see that for a given mass, the wavelength of the $k_\star$ mode is longer for the quartic reheating scenario, c.f. Fig.~\ref{fig:comovingdiag}. The long-wavelength part of the spectrum, i.e., $k^{-1} > k_\star^{-1}$, grows as $k^{2}$. The scaling with $k$ in the central part, i.e., for $k^{-1} \in (k_e^{-1}, k^{-1}_{\star})$ is model dependent. Namely, for $m_X > H_{\rm rh}$ one finds that $d \langle \hat{\rho}_{\rm L} \rangle / d \ln{k} \propto k^{\frac{2(3 \bar{w} -1)}{3\bar{w}+1}}$ \cite{Ahmed:2020fhc}, which gives $d \langle \hat{\rho}_{\rm L} \rangle / d \ln{k} \propto k^{-2}$ and $d \langle \hat{\rho}_{\rm L} \rangle / d \ln{k} \propto k^0$ in the quadratic and quartic reheating scenarios, respectively. At \textit{late times}, i.e., when the evolution of the $X_\mu$ field becomes static, its energy density drops as $a^{-3}$, indicating that $X_\mu$ behaves like non-relativistic matter.
\\
A qualitatively similar structure is also observed for vectors with non-minimal gravitational coupling. Fig.~\ref{fig:rhoLTMNM} presents the spectral energy density of the $X_\mu$ field as a function of $k$ for different choices of $\xi_1, \xi_2$ at \textit{late times}. The long-wavelength part of $d \langle \hat{\rho}_{\rm L} \rangle / d \ln{k}$ has a peak around $k = k_\star$, whose position shifts towards the left as $\eta_e^{-1}$ decreases. Depending on the values of $\xi_1, \xi_2$, the energy density of the non-minimally coupled spin-1 
field may either exceed or be lower than the energy density of the minimally coupled vectors. For instance, for $\xi_1 = \xi_2 = 10^{-3}$ ($\xi_1 = - 10^{-4}, \xi_2 = 10^{-3}$), the inclusion of non-minimal couplings contributed to the decrease (increase) of the energy density stored in the $X_\mu$ field compared to the minimal case. Such behavior is consistent with the relative amplification observed in Fig.~\ref{fig:scan}. Finally, it is worth highlighting that the observed effects depends only on the relative ratio $\xi_{1,2}/\eta_e^{-1}$, see also \cite{Capanelli:2024pzd}. 

\subsection{Sub-horizon modes}
Initially, the spectral energy density of the sub-horizon modes increases as $k^4$ for both the longitudinal and transverse components of the $X_\mu$ field, see Ref.\cite{Ahmed:2020fhc}. Later,  the scaling changes to $k^{3}$ as the short-wavelength modes become non-relativistic. Figure ~\ref{fig:rhoLTMNM} shows that all curves converge in the region $k^{-1} \lesssim k_e^{-1}$, regardless of the values of the non-minimal couplings. This behavior results in the well-known UV divergence of the expectation value of the energy density, which requires regulation.
To render the energy density UV finite, various regularization schemes can be employed. Below, we briefly discuss three of them. 
\begin{itemize}
    \item UV cut-off \\
The simplest strategy relies on introducing a UV cut-off scale $\Lambda$. This approach, however, involves a fairly arbitrary choice of the cut-off scale, which can introduce ambiguity into the model and its predictions.
Physical observables, such as the present abundance of the 
$X_\mu$ field, are expected to be independent of the specific value of $\Lambda$. One of the more intuitive choices for the cut-off scale seems to be $\Lambda = \mathcal{O}(1) k_e$, since only super-horizon modes are efficiently amplified by tachyonic instability. Hence, such a choice is well-justified if the dominant contribution to the total energy density comes from the super-horizon modes. Contrarily, if other production mechanisms efficiently create short-wavelength modes, a larger cut-off should be considered. This, however, increases the complexity of the computational analysis.
\item Normal ordering \\
Another method to regularize the energy density is based on normal ordering \cite{Chung:2004nh, Kolb:2023ydq}. In this case, the physical expectation value of the energy density is calculated with respect to the initial vacuum state, e.g., the Bunch-Davies vacuum, whereas the normal ordering is performed with respect to the late-time ladder operators. At \textit{late time}, when the evolution of the modes becomes adiabatic, the total energy density can be approximated by the following formula \cite{Kolb:2023ydq}:
\begin{align}
    \langle : \! \hat{\rho}_{X} \!:\rangle =   \langle : \! \hat{\rho}_{\rm L} \!:\rangle + \langle :\! \hat{\rho}_{\rm T} \! : \rangle, 
\end{align}
with
\begin{align}
     &\langle : \! \hat{\rho}_{\rm L} \!:\rangle \simeq  \lim_{\tau \rightarrow \infty}\langle \hat{\rho}_{\rm L} \rangle \approx \frac{1}{a^4} \int \frac{d^3 k}{(2 \pi)^3} (k^2 + a^2 m_X^2)^{1/2} \lim_{\tau \rightarrow \infty}|\beta_{k}^{\rm L}|^2, \\
     & \lim_{\tau \rightarrow \infty}|\beta_{k}^{\rm L}|^2 = \frac{1}{2 \omega_{\rm L}} |\mathcal{X}_{\rm L}^\prime|^2 + \frac{\omega_{\rm L}}{2} |\mathcal{X}_{\rm L}|^2 - \frac{1}{2}, 
\end{align}
and 
\begin{align}
     &\langle : \! \hat{\rho}_{\rm T } \! : \rangle \simeq \lim_{\tau \rightarrow \infty} \langle \hat{\rho}_{\rm T}  \rangle \approx \frac{1}{a^4} \sum_{\rm T = \pm} \int \frac{d^3 k}{(2 \pi)^3} (k^2 + a^2 m_X^2)^{1/2} \lim_{\tau \rightarrow \infty} |\beta_{k}^{\rm T}|^2, \\&\lim_{\tau \rightarrow \infty} |\beta_{k}^{\rm T}|^2 = \frac{1}{2 \omega_{\rm T}} |\mathcal{X}_{\rm T}^\prime|^2 + \frac{\omega_{\rm T}}{2} |\mathcal{X}_{\rm T}|^2 - \frac{1}{2}. 
\end{align}
Note that contributions from modes with very short wavelengths, for which $\omega_{\rm L} = \omega_{\rm T} \approx k$, vanish thanks to the $1/2$ factor in $|\beta_k|^2$.
 This factor thus plays a crucial role in eliminating UV divergences of the energy density spectrum. \\ \\ 
For future convenience, it is instructive to introduce the comoving number density spectra
\begin{align}
    &\mathcal{N}_k^{\rm L} \equiv \frac{k^3}{2 \pi^2} |\beta_k^{\rm L}|^2,  && \mathcal{N}_k^{\rm T} \equiv \sum_{\rm T = \pm}\frac{k^3}{2 \pi^2} |\beta_k^{\rm T}|^2.
\end{align}
\begin{figure}[htb!]
    \centering
\includegraphics[scale=0.5]{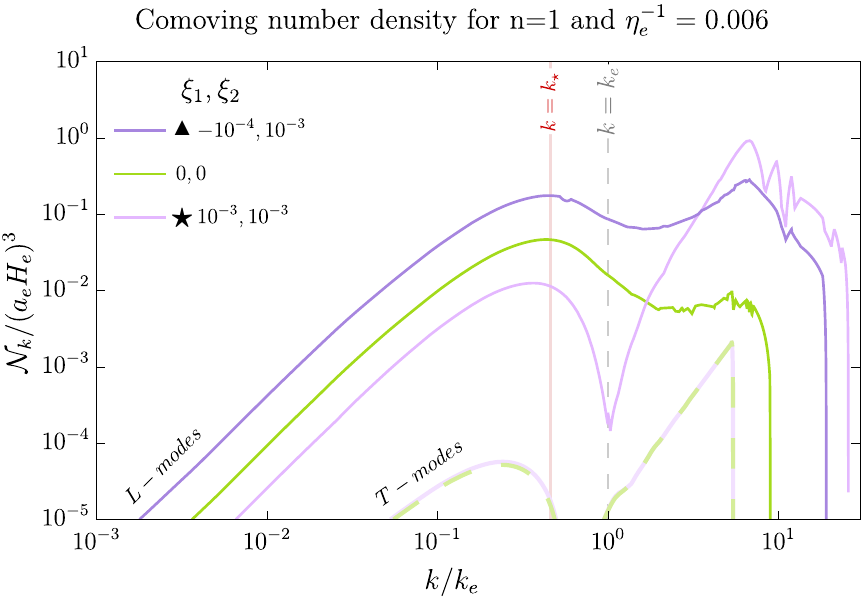}
\includegraphics[scale=0.5]{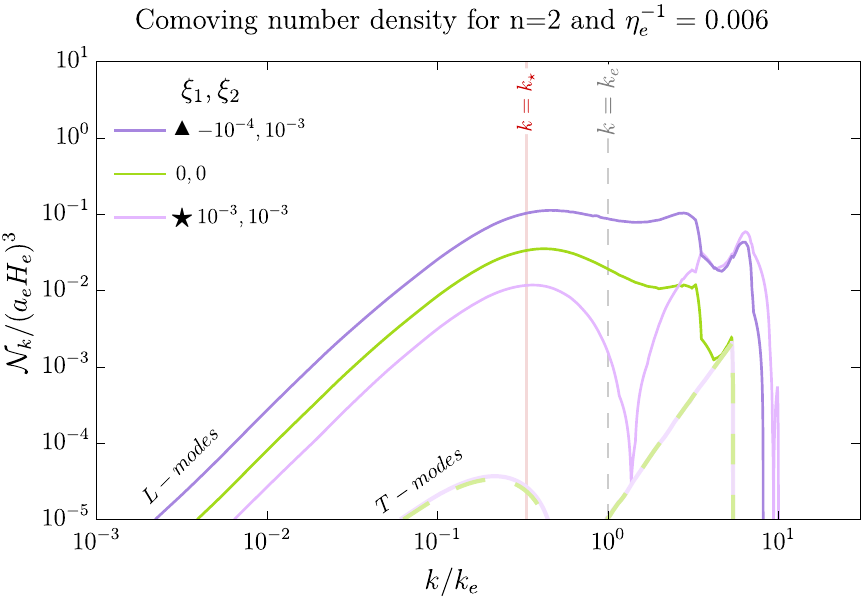}\\
\includegraphics[scale=0.5]{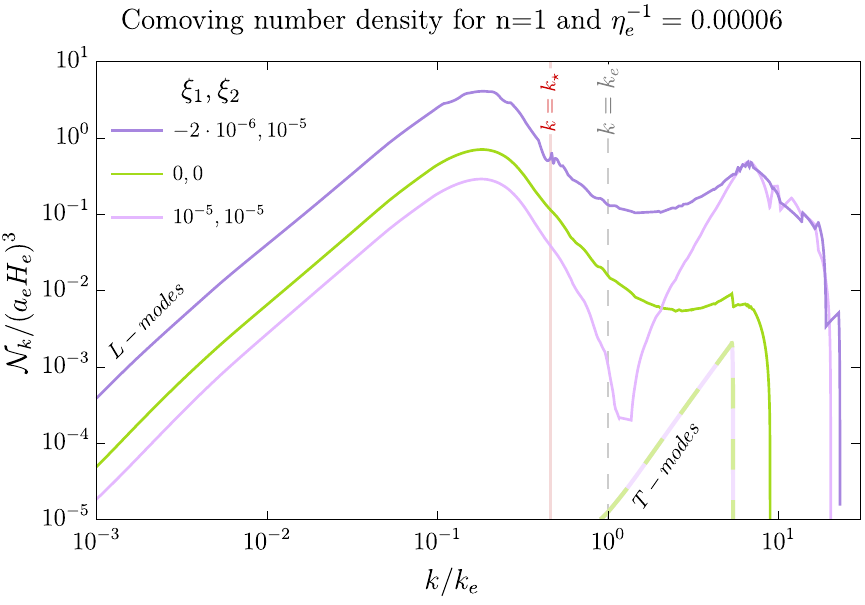}
\includegraphics[scale=0.5]{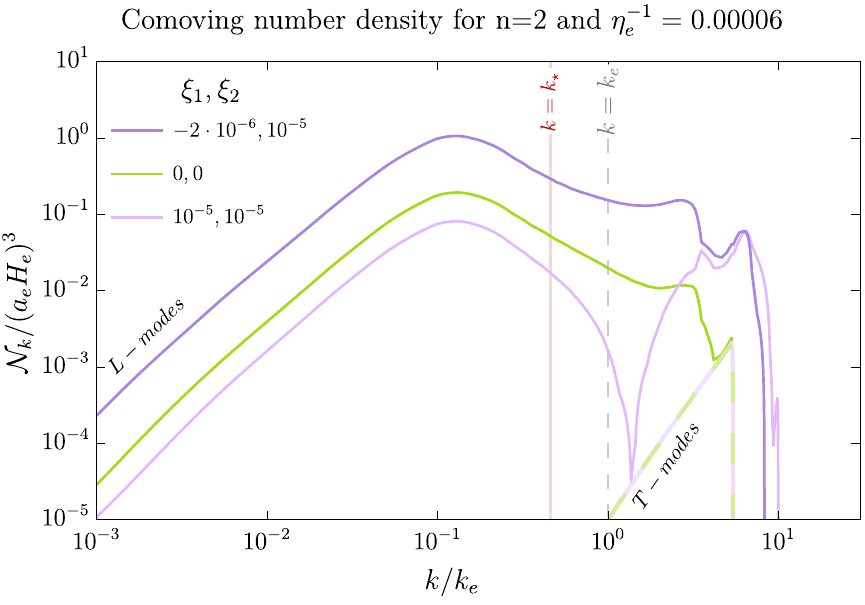}\\
    \caption{Normal-ordered comoving number density normalized to $(a_e H_e)^3$ for the longitudinal (L) and transverse (T) components of the minimally (green) and non-minimally (violet curves) coupled vector field in the quadratic (left panel) and quartic (right panel) reheating model.}
\label{fig:Nk}
\end{figure}
At \textit{sufficiently late times}, when $a^\prime, a^{\prime \prime}$ become negligible and the Hubble rate falls below $m_X$, the dispersion relations for the longitudinal and transverse polarizations converge to the same asymptotic limit, $\omega_{\rm L} \approx a \sqrt{(k/a)^2 + m_X^2} \approx \omega_{\rm T}$. In the non-relativistic regime, this simplifies to $\omega_{\rm L} \approx a m_X \approx \omega_{\rm T}$, and the solutions to both mode equations oscillate with decreasing amplitude, according to Eq.\eqref{eq:ASolXT}. Consequently, at \textit{late times,}\footnote{Typically for $a \gtrsim e^3 a_\star$.} the Bogolubov coefficients $|\beta_k^{\rm L}|^2, |\beta_k^{\rm T}|^2$ become constant, so does the comoving number density.\\ \\ In Fig.~\ref{fig:Nk}, we plot the comoving number density as a function of $k/k_e$ for two reheating scenarios. The spectra are evaluated at \textit{late times}, when $|\beta_k|^2$ and thus $\mathcal{N}_k$ remain static for the adopted values of $\eta_e^{-1}$ and the non-minimal couplings. Notably, regularization via normal ordering implies a non-trivial structure of the UV tails. In the reheating model with a quadratic inflaton potential, a central UV peak emerges around $k/k_e \simeq 6$. Interestingly enough, a similar finding has been  reported recently in Ref.\cite{Verner:2024agh} in the context of superheavy scalars non-minimally coupled to gravity. On the other hand, for $n=2$ there are two main peaks in the large-k part of the spectrum. In both considered models, the comoving number density of the transverse components is subdominant regardless of the values of non-minimal couplings. 

At \textit{late times}, one can express the spectral energy density in terms of the comoving number density as follows
\begin{align}
      &\langle : \! \hat{\rho}_{\rm L} \!:\rangle \simeq m_X \langle : \! \hat{n}_{\rm L} \!:\rangle, & a^3 \langle : \! \hat{n}_{\rm L} \!:\rangle= \int d \ln{k} \; \mathcal{N}_k^{\rm L} \equiv \mathcal{N}_{\rm L}, \label{eq:NL}
\end{align}
and 
\begin{align}
      &\langle : \! \hat{\rho}_{\rm T} \!:\rangle \simeq m_X \langle : \! \hat{n}_{\rm T} \!:\rangle, & a^3 \langle : \! \hat{n}_{\rm T} \!:\rangle= \sum_{\rm T =\pm} \int d \ln{k} \; \mathcal{N}_k^{\rm T} \equiv \mathcal{N}_{\rm T},
      \label{eq:NT}
\end{align}
where $\mathcal{N}_{\rm L}, \mathcal{N}_{\rm T}$ do not depend on time. Note that this result confirms a matter-like scaling of the $X_\mu$ field's energy density at \textit{late times}. \\ 
\begin{figure}[htb!]
    \centering
\includegraphics[scale=0.5]{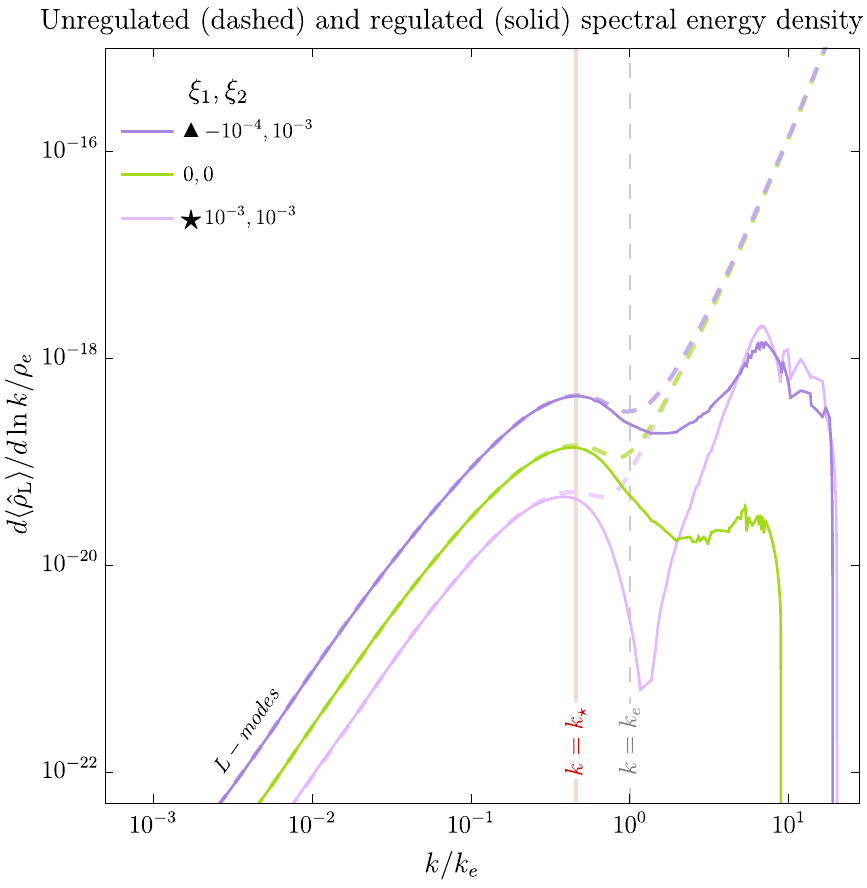}
\includegraphics[scale=0.5]{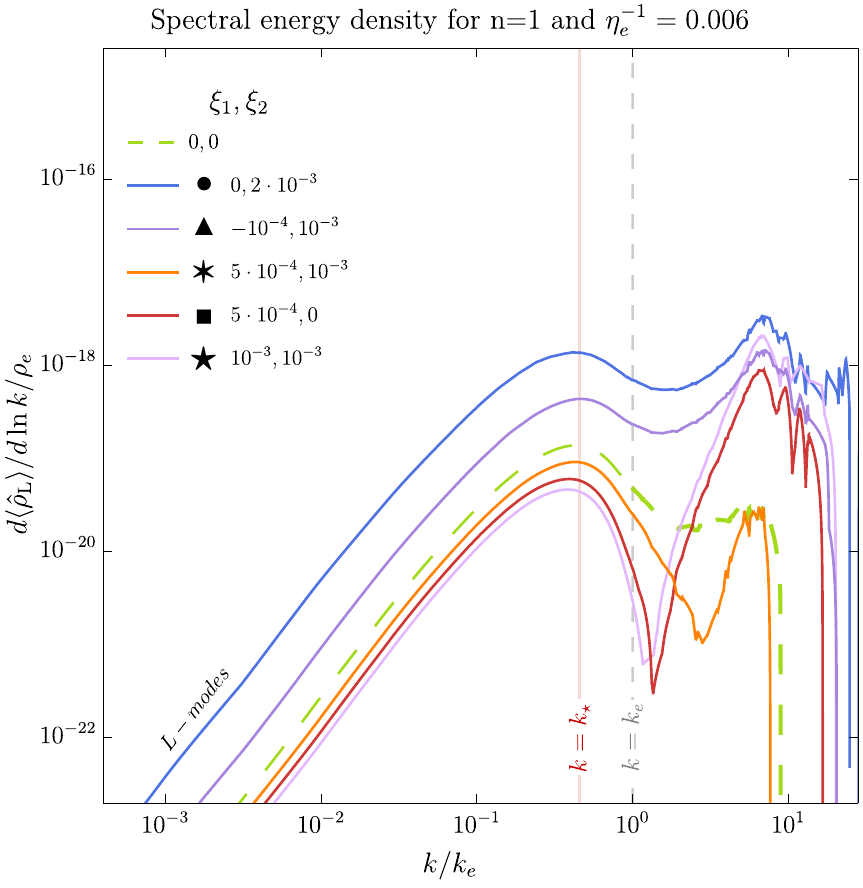}
    \caption{Left: Comparison of the unregulated (solid curves) and regularized (dashed curves) spectral energy density. Right: Normal-ordered spectral energy density for the benchmark values of the non-minimal couplings. }
\label{fig:RegUnRegCom}
\end{figure} 
In the left panel of Fig.~\ref{fig:RegUnRegCom},
we compare the unregulated (dashed curves) spectral energy density of the longitudinally polarized modes with the normal-ordered spectrum. As with the spectral number density, additional high-k peaks emerge in the regularized energy density spectrum that were not present in the unregulated spectra. This noisy pattern is induced by quantum interference effects, as discussed in Ref.\cite{Basso:2022tpd}. The regularized spectra accurately reproduce the unregulated energy densities for $k \lesssim k_\star$, with only minor deviations around $k \sim k_e$.  \\ \\
To gain a better understanding of the UV tail, in the right panel of Fig.~\ref{fig:RegUnRegCom}, we plot the spectral energy densities for five benchmark values of the non-minimal couplings and compare them with the spectrum of minimally coupled vectors (green  dashed curve). We observe that the large-k part of the spectra exhibits a multi-peak structure. For the chosen non-minimal couplings, the position of the peaks shows little dependence on $\xi_1, \xi_2$. \footnote{This observation stems from the fact, that for all considered values of $\xi_1, \xi_2$, the minimum of $c_s^2$ does not vary much. However, if $min(c_s^2)$ is considerably smaller (by at least one order of magnitude), the UV peaks shift towards larger k, see Ref.\cite{Capanelli:2024nkf}.}, while their magnitude is highly sensitive to the values of $\xi_1$, $\xi_2$. In particular, depending on $\xi_1, \xi_2$, the height of the large-k peaks can either tower or be comparable with the $k_\star$ peak. For instance, for $\xi_1 = 10^{-3} = \xi_2$ (light purple curve) and $\xi_1 = 5 \cdot 10^{-4}, \xi_2 =0$, (red curve), the production of long-wavelength modes is not enhanced compared to the minimal case, but the entire spectrum is dominated by the UV peaks, whose magnitudes are nearly two orders of magnitude larger than that of the $k_\star$ peak. On the other hand, for $\xi_1 = \xi_2/2$ (orange curve) the long-wavelength peak is slightly higher than the UV peaks. This behavior arises because the effective reduction of two non-minimal couplings to a single coupling to the Einstein tensor $G_{\mu \nu}$ hinders the resonance effects, thereby suppressing the production of short-wavelength modes. Furthermore, the spectrum is enhanced in both the super-horizon and sub-horizon regions for values of $\xi_1, \xi_2$ that amplify the $k_\star$ peak 
peak compared to the minimal case. In general, the amplification of the large-k peaks is related to the behavior of the sound speed at the onset of reheating. As first noticed in Ref.\cite{Capanelli:2024nkf}the magnitude of the UV peaks is correlated with the minimal value of $c_s^2$ for $\xi_2 \neq 0$. However, as we see, even if $X_\mu$ has no direct coupling to the Ricci tensor, implying a constant sound speed $c_s^2 =1$, the production of short-wavelength modes can still be significantly more efficient than in the minimal case, given sufficiently large values of the ratio $\xi/ \eta_e^{-1}$.
\item Adiabatic regularization \\
The best-established  approach for handling UV divergences is adiabatic regularization, developed by Fulling and Parker \cite{Fulling:1974pu, Fulling:1974zr, Parker:1974qw}. This method involves i) finding approximate solutions to the mode equations based on the 
Jeffreys-Wentzel-Kramers-Brillouin (JWKB) method, ii) expanding the obtained solutions, as well as the observables constructed from them, into an adiabatic series, iii) identifying the divergent terms of the composite quantities, and iv) subtracting the adiabatic counterterms that remove the divergent part of a given observable. Although the above procedure is well-defined, applying it to regularize the energy density integral of the non-minimally coupled spin-1 field appears to be fairly intricate, as it requires finding counterterms up to the fourth order in the adiabatic expansion. Therefore, we left this task for future work. 
\end{itemize}

\subsection{Relic abundance}
\begin{figure}[hb!]
    \centering
\includegraphics[scale=0.5]{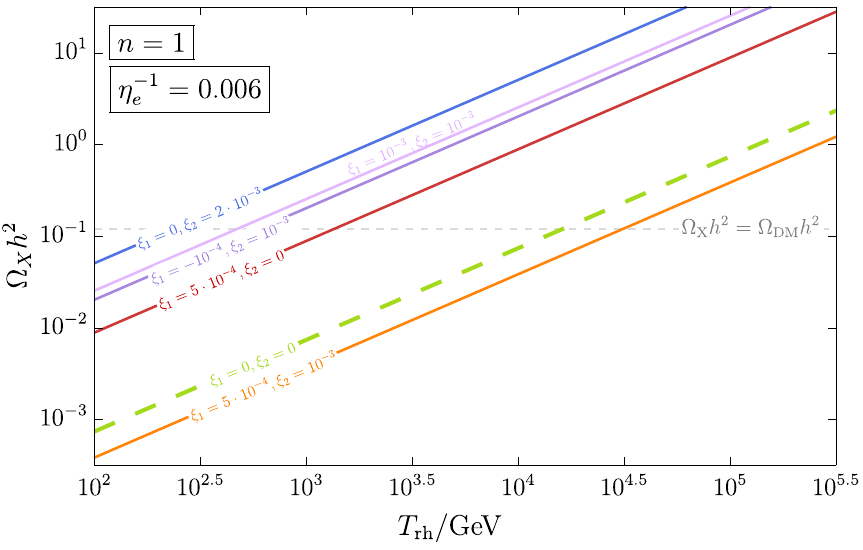}
\includegraphics[scale=0.5]{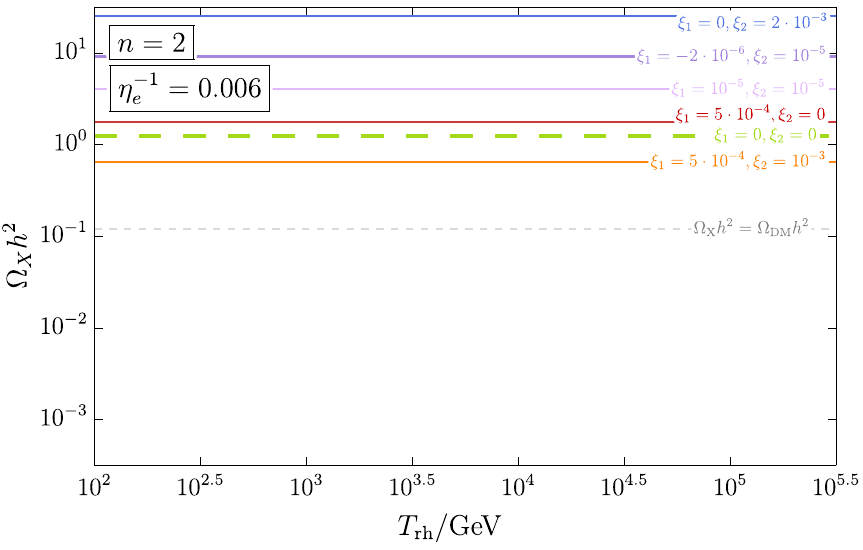}\\
\includegraphics[scale=0.5]{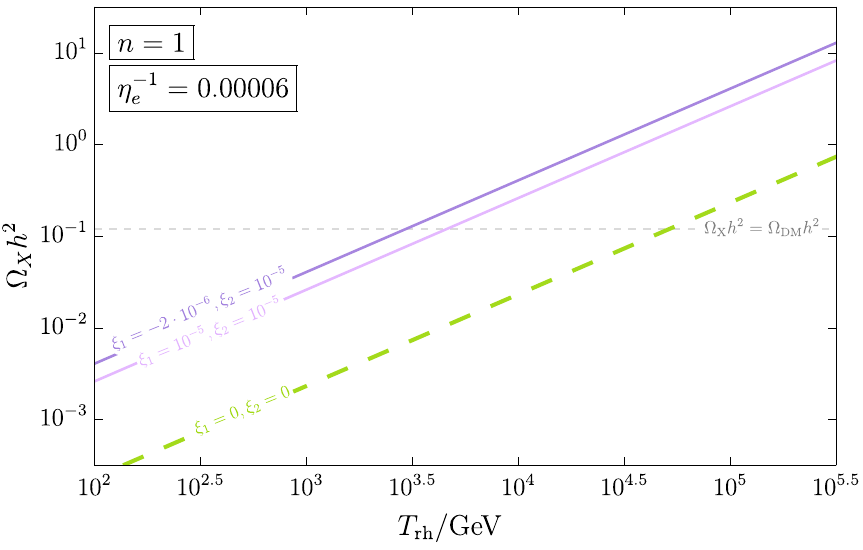}
\includegraphics[scale=0.5]{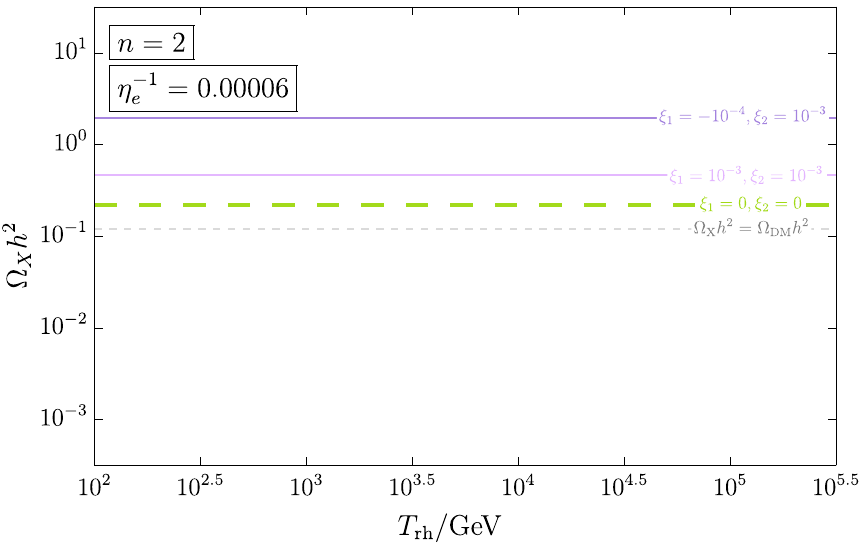}\\
    \caption{The present abundance of the $X_\mu$ field as a function of the reheating temperature $T_{\rm rh}$ for the benchmark values of the non-minimal couplings $\xi_1, \xi_2$, fixed $H_e$, and two mass $m_X = 10^{12} \,{\rm GeV}$ (upper) and $m_X = 10^{11} \,{\rm GeV}$ (lower), in the quadratic (left) and quartic (right) reheating models. }
\label{fig:OmegaTrh}
\end{figure}

\begin{figure}[htb!]
    \centering
\includegraphics[scale=0.38]{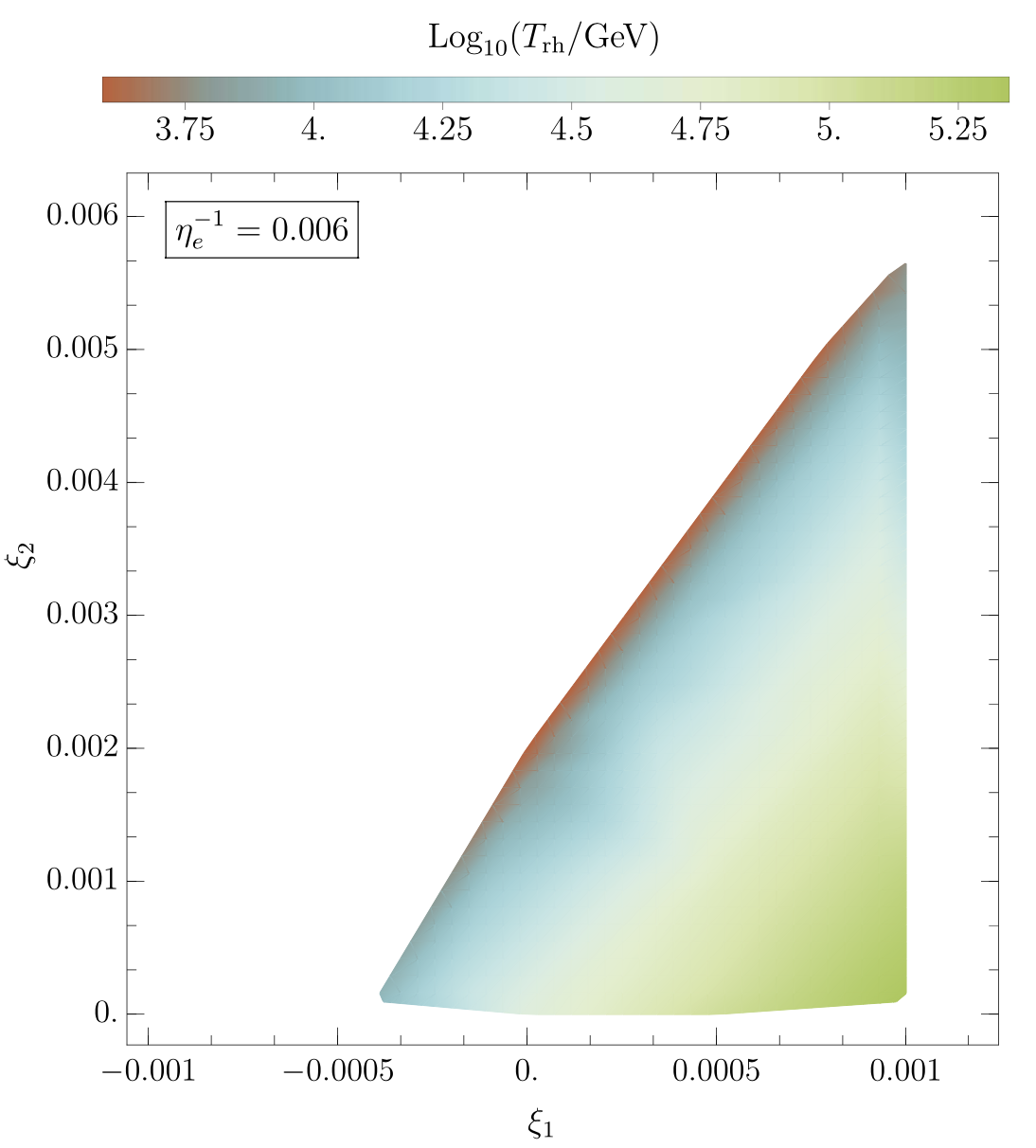}
\includegraphics[scale=0.39]{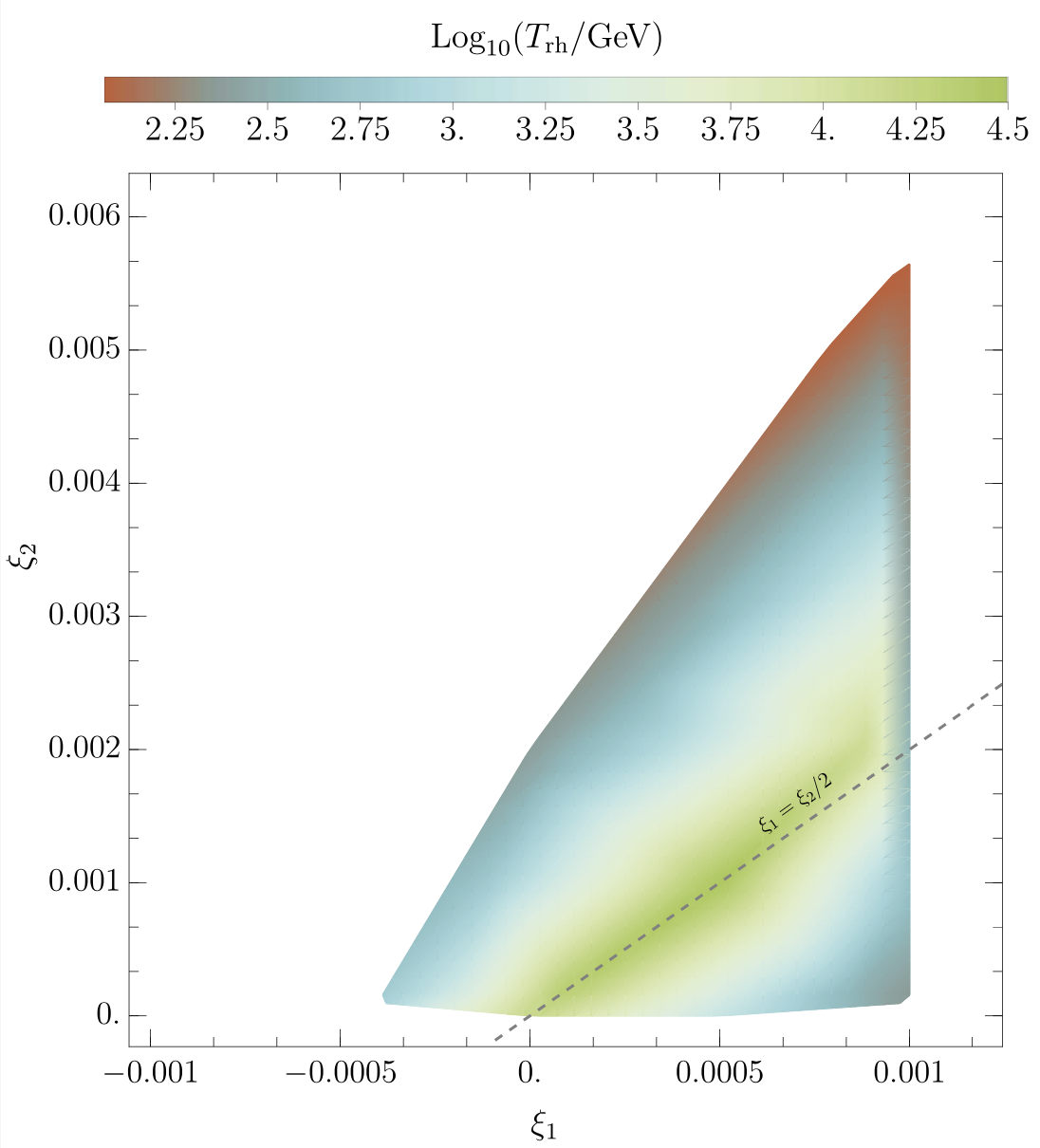}
    \caption{Left: The reheating temperature needed to saturate the observed relic abundance of dark matter, considering only contributions from super-horizon modes. Right: The same for both super-horizon and sub-horizon modes.  }
\label{fig:RelAb}
\end{figure}

Assuming that, at present, the kinetic energy of the $X_\mu$ field constitutes a negligible fraction of its total energy~\footnote{Note that this assumption implies a lower limit on the vector mass $m_X$.}, the fractional energy of that field, expressed in terms of the critical energy density $\rho_{\rm crit, 0} = 3 M_{\rm Pl}^2 H_0^2 = 1.054 \times 10^{-5} \, h^2 \; \rm{GeV \; cm^{-3}}$, can be calculated as 
\begin{align}
    \Omega_X h^2 = \frac{\rho_{X, 0}}{\rho_{\rm crit, 0}} h^2 \simeq  \frac{ m_X n_{X, 0}}{\rho_{\rm crit, 0}} h^2, 
\end{align}
where $n_{X, 0}$ denotes the present number density of $X_\mu$ particles. Leveraging the fact that at late times, i.e., for $a \gtrsim a_\star$, the integrated comoving number density $\mathcal{N}_{X} = \mathcal{N}_{\rm L} + \mathcal{N}_{\rm T } \approx \mathcal{N}_{\rm L}$ (see Eqs.\eqref{eq:NL} and \eqref{eq:NT}) of the $X_\mu$ field becomes frozen, one finds
\begin{align}
    &\mathcal{N}_{X} = \rm{const.}, &\implies &&  n_{X, 0}  = \frac{\mathcal{N}_X}{a_e^3 H_e^3} \left(\frac{a_e}{a_0} \right)^3 H_e^3, &&\text{for } a \gtrsim a_\star,
\end{align}
where we have applied normalization introduced in Ref.\cite{Kolb:2020fwh}. \\
It is convenient to recast the redshift factor as 
\begin{align}
    \left( \frac{a_e}{a_0} \right)^3 = \left(\frac{a_e}{a_{\rm rh}} \right)^3 \left(\frac{a_{\rm rh}}{a_{0}} \right)^3.  
\end{align}
Assuming that the comoving entropy density, $s a^3$, is conserved after the completion of reheating, one can express the ratio $a_{\rm rh}/a_0$ in terms of the reheating temperature
\begin{align}
    \left(\frac{a_{\rm rh}}{a_{0}} \right)^3 = s_0 \left( \frac{2 \pi^2}{45} g_{\star, rh}^{s} \right)^{-1} T_{\rm rh}^{-3} \approx (6.34 \times 10^{-29})  \cdot \left(\frac{106.75}{g_{\star, rh}^{s}} \right) \cdot \left(\frac{10^{10}\, {\rm GeV}}{T_{\rm rh}} \right)^{3} {\rm cm}^{-3} \rm{GeV}^{-3},
\end{align}
where $s_0  = 2970\; \rm{cm}^{-3}$ is the present entropy density, and $g_{\star, rh}^{s}$ denotes the effective number of relativistic degrees of freedom at the end of reheating. \\
Furthermore, from the Friedmann equation, one finds 
\begin{align}
   H_{\rm rh}^2 \simeq H_e^2 \left( \frac{a_e}{a_{\rm rh}}\right)^{3(1+\bar{w})} = \frac{\pi^2}{90} g_{\star, \rm rh} \frac{T_{\rm rh}^4}{\mpl^2},
\end{align}
which, in turn, implies
\begin{align}
    \left( \frac{a_e}{a_{\rm rh}}  \right)^3 = \left( \frac{\pi^2 g_{\star, \rm rh}}{90}  \frac{T_{\rm rh}^4}{H_e^2 \mpl^2} \right)^{\frac{1}{1+\bar{w}}} &\approx (7.71 \cdot 10^{-25})^{\frac{1}{1+ \bar{w}}}\left( \frac{g_{\star, \rm rh}}{106.75}\right)^{\frac{1}{1+\bar{w}}} \nonumber \\
    &\times \left( \frac{1.6 \cdot 10^{14} \; {\rm GeV}}{H_e}\right)^{\frac{2}{1+\bar{w}}}  \left( \frac{T_{\rm rh}}{10^{10} \; {\rm GeV}}\right)^{\frac{4}{1+\bar{w}}}.
\end{align}
Consequently, 
\begin{align}
    \Omega_{X} h^2 &\simeq 3.94 \times 10^{33} \times (7.71 \cdot 10^{-25})^{\frac{1}{1+ \bar{w}}} \times \frac{\mathcal{N}_X}{a_e^3 H_e^3} \cdot  \left( \frac{m_X}{1.6 \times 10^{14} \; {\rm GeV}} \right) \nonumber \\
    & \times \left(\frac{106.75}{g_{\star, rh}^{s}} \right) \cdot \left( \frac{g_{\star, \rm rh}}{106.75}\right)^{\frac{1}{1+\bar{w}}} \cdot \left( \frac{H_e}{1.6 \times 10^{14} \; {\rm GeV}} \right)^{\frac{1+3 \bar{w}}{1+ \bar{w}}} \cdot \left( \frac{T_{\rm rh}}{10^{10} \; {\rm GeV}}\right)^{\frac{1-3 \bar{w}}{1+\bar{w}}}.
\end{align}
Hence, for the reheating scenario with a matter and radiation-like equation-of-state one finds, 
\begin{align}
    \Omega_X  h^2 \bigg \rvert_{\bar{w}=0} &\simeq 0.12 \times \frac{\mathcal{N}_X}{a_e^3 H_e^3} \cdot  \frac{m_X}{H_e} \cdot \left( \frac{H_e}{10^{12} \; {\rm GeV}} \right)^2  \frac{T_{\rm rh}}{ 10^{4} \; {\rm GeV}}, 
\end{align}
and 
\begin{align}
    \Omega_X  h^2 \bigg \rvert_{\bar{w}=1/3} \simeq 0.12 \times \frac{\mathcal{N}_X}{ 0.12 \; a_e^3 H_e^3} \cdot   \frac{m_X}{H_e} \left( \frac{H_e}{10^{8} \; {\rm GeV}} \right)^{5/2},
\end{align}
respectively. Note that above, we have fixed $g_{\star, rh}^s = 106.75 = g_{\star, rh}$, i.e., no extra relativistic degrees of freedom beyond the SM. \\
Since the comoving number density, $\mathcal{N}_X$, depends on the mass of the $X_\mu$ field, its couplings to gravity, and the inflationary Hubble scale, $\Omega_X h^2$ is a function of six parameters $\{m_X, H_e, T_{\rm rh}, \xi_1, \xi_2, n \}$ in general. Consequently, the present abundance of the $X$ quanta pivots 
not only on the free parameters of the model but also on the evolution of the primordial universe—particularly on the transition from the inflationary to radiation-dominated phase. 
One observes that $\Omega_X h^2$ 
is sensitive to the inflationary scale, the average equation of state during reheating, and the reheating duration through $T_{\rm rh}$. However, the above findings indicate that for a radiation-like reheating scenario (n=2) the relic density of $X_\mu$ is independent on the details of reheating, see also \cite{Graham:2015rva, Kolb:2020fwh, Ahmed:2020fhc}. In Fig.~\ref{fig:OmegaTrh}, we show $\Omega_X h^2$ as a function of $T_{\rm rh}$ for benchmark values of the non-minimal couplings, two choices of $m_X$, and fixed $H_e$. Interestingly, depending on the values of $\xi_1, \xi_2$, the present density of non-minimally coupled massive spin-1 fields can be either greater or smaller than that of vectors with minimal coupling to gravity. The abundance of the former exceeds that of the latter for all $\xi_1, \xi_2$ for which the production of the long-wavelength modes is more efficient than in the minimal case. In addition, one observes the same effect for $\xi_1, \xi_2$ for which the large-k peak  dominates over the $k_\star$ peak in the number density spectrum. Comparing our results in the two considered scenarios, we find that for fixed $m_X$, $H_e$ and  low reheating temperature, the production of the $X_\mu$ quanta is more efficient in the quartic reheating model. For instance, for $\xi_1, \xi_2 =0$ and $\eta_e^{-1} = 0.006$ the present value of $\Omega_X h^2 \rvert_{\bar{w}=0}$ exceeds $\Omega_X h^2 \rvert_{\bar{w}=1/3}$ if $T_{\rm rh} \gtrsim 10^{5} \, \rm{GeV}$, and for $\xi_1, \xi_2$ for which $\Omega_X h^2 \rvert_{\bar{w}=0}(\xi_1, \xi_2) > \Omega_X h^2 \rvert_{\bar{w}=0}(0,0)$ ($\Omega_X h^2 \rvert_{\bar{w}=0}(\xi_1, \xi_2) < \Omega_X h^2 \rvert_{\bar{w}=0}(0,0)$) this happens at lower (higher) reheating temperature. We also observe that for fixed $H_e, T_{\rm rh}, \xi_1, \xi_2$, decreasing the mass leads to a reduction in the present abundance. \\ \\
 \\
Let us now consider the possibility that the $X_\mu$ field constitutes the dominant component of dark matter. Its stability, ensured by the $\mathbb{Z}_2$ invariance of the action \eqref{eq:X_action}, along with the fact that its abundance is generated solely through gravitational phenomena, makes $X_\mu$ an appealing candidate for DM. \\
It is well known (see e.q., Refs.\cite{Graham:2015rva, Kolb:2020fwh, Ahmed:2020fhc}) that the present abundance of massive vectors with minimal coupling to gravity can saturate the observed relic density of dark matter $\Omega_{\rm DM} h^2 = 0.120 \pm 0.001$ measured by the Planck Collaboration \cite{Planck:2018jri} in the vast range of parameter space. Recently, the authors of Ref.\cite{Capanelli:2024nkf} presented more general results for non-minimally coupled vectors. We aim to complement this study by examining the impact of non-zero $\xi_1, \xi_2$ on the relic abundance of
$X_\mu$ in two realistic reheating scenarios. Notably, the inclusion of non-minimal terms expands the parameter space in which $X_\mu$ can serve as a viable dark matter candidate. As shown in Fig.~\eqref{fig:OmegaTrh}, in the matter-like reheating scenario, the estimated abundance of the $X_\mu$ field aligns with $\Omega_{\rm DM} h^2$ across a range of temperatures spanning more than two orders of magnitude for the considered values of  
$\xi_1, \xi_2$. In contrast, in the quartic reheating model and for the chosen values of $m_X, H_e$, $\Omega_X h^2$ exceeds $\Omega_{\rm DM} h^2$ even in the minimal coupling case. Introducing non-zero values of $\xi_1, \xi_2$ does not substantially alter this result, as it does not significantly reduce the number density of heavy vectors. Presumably, in this model the inclusion of the non-minimal couplings might be more important for lighter species for which the minimal gravitational production is not efficient enough. However, the numerical analysis for light, non-minimally coupled vectors is computationally challenging, requiring tracking their evolution through late times (often up to dozens of e-folds after inflation). Therefore, analytical estimates for the comoving number density, including contributions from large-k modes, would be valuable; we leave this direction for future work. \\
Finally, in Fig.\eqref{fig:RelAb} we scan the instability-free region of the model, and determine the value of the reheating temperature for which the present density of the $X_\mu$ field matches DM abundance in the quadratic reheating scenario.  In the left panel, we include only the long-wavelength part of the spectrum adopting the 'standard' cut-off at $\Lambda_{\rm UV}=k_e$. The observed pattern is consistent with the scan \eqref{fig:scan}. Namely, compared to the $\xi_1, \xi_2 =0$ case, the condition $\Omega_X h^2 \rvert_{\bar{w}=0} = \Omega_{\rm DM}$ is met at higher (lower) $T_{\rm rh}$ for points lying in the bottom-right corner (along the hypotenuse), where the maximal squared amplitude is smaller (or larger) than in the minimal case. In the right panel, we also include the large-k part of the normal-ordered spectrum, 
considering larger cut-off $\Lambda_{\rm UV} = 40\, k_e$. 
Here, we observe that a lower reheating temperature is required to match the DM abundance, as including the full spectrum reduces the required $T_{\rm rh}$ by more than an order of magnitude compared to the left panel. Interestingly, a different pattern emerges: the highest $T_{rh}$ is required for points lying along the $\xi_1 = \xi_2/2$ curve, for which the $X_\mu$ field has only one non-minimal coupling to Einstein tensor.
\section{Summary} \label{sum}
We have discussed a quantum theory of massive Abelian vector bosons
in the FLRW classical gravitational background, i.e. within isotropic and homogeneous universe.
Our goal was to investigate in detail gravitational production of spin-1 fields, 
which could serve as candidates for dark matter if a stabilizing $\mathbb{Z}_2$ symmetry is imposed. We have assumed the presence of the non-minimal couplings between vectors and the Ricci scalar and the Ricci tensor. 
In order to preserve the $U(1)$ invariance of the non-minimal couplings a generalization of the Stuckelberg mechanism has been introduced. 
\\ \\ 
We have determined constraints that ensure the consistency of the model by eliminating
the ghost instability, uncontrolled ("run-away") production and super-luminal propagation of short-wavelength modes. 
It has been shown that during inflation (i.e. for $w=-1$), since $m_{\text{eff,x}}^2(a) \sim m_{\text{eff,t}}^2(a)$ as for in $k^2\to \infty$, there is no "run-away" production of short-wavelength modes. This issue may arise only during the post-inflationary epochs in models with a non-zero coupling to the Ricci tensor. Furthermore, it has been found that in the limit of massless vectors $m_X \to 0$,
there is no safe region where both effective masses, $\mefx$ and $\meft$, remain positive for an arbitrary barotropic parameter $w \in [-1,1]$. So, if the non-minimal couplings are present, then the vectors can not be arbitrarily light.  \\ \\
Canonical quantization of the Abelian gauge field in an evolving gravitational background has been employed. We have derived and solved numerically the mode equations for both longitudinal and transverse components of the vector boson. It has been shown that, when the consistency conditions are imposed, the squared frequency of the transverse polarization is always positive ($\omega_T^2(\tau,k)>0$), excluding the possibility of exponential enhancement even for non-minimal couplings. In contrast, the long-wavelength modes of the longitudinal component of the $X_\mu$ field are generally subject to exponential enhancement during certain periods of inflation and/or reheating. The tachyonic amplification is sensitive to the values of non-minimal couplings, and depending on the specific choices of $\xi_1, \xi_2$, it may intensify or attenuate the effect observed in the minimal model. Regions of exponential enhancement of the solutions have been determined numerically, illustrated, and thoroughly discussed. In addition, it has been pointed out that the inclusion of non-minimal couplings alter the dynamics of subhorizon modes that never exit the Hubble horizon. 
\\ \\ 
The spectral energy density for three physical polarizations has been calculated. 
Two peaks in the spectrum that may likely dominate the integrated energy density have been identified. The first, corresponding to lower momentum $k\star < k_e$, has already been observed for the minimally-coupled vectors. The second, appearing in the UV tail od the spectrum, could dominate the gravitational production of vectors, making its unambiguous verification crucial. Since the spectral energy density is UV divergent, a regularization method must be adopted. Here, we have compared the regularization by normal ordering~\cite{Chung:2004nh, Kolb:2023ydq} against a one with a straightforward cut-off imposed on sub-horizon modes. Both methods coincide until roughly the location of the first peak, however for higher momentum they differ substantially. In particular the second peak might be solely an artifact of regularization method adopted to cure the UV divergence of the spectral density. Hence, more comprehensive discussion involving the adiabatic regularization is needed. Therefore, a more comprehensive discussion involving adiabatic regularization is warranted. Eventually, assuming the stability of the $X_\mu$ field, we have determined its present abundance and explored the parameter space where the predicted relic density matches the observed dark matter density.\\\\
Finally, let us outline potential extensions and future directions stemming from this work. First, it would be interesting to explore the full dynamics of the vector field in the presence of the (dark) Higgs field. Additionally, obtaining analytical solutions for the mode equations of the longitudinal components could be beneficial, as it would allow for a quantitative description of the spectrum of non-minimally coupled vectors in terms of their mass, the inflationary Hubble scale, and the non-minimal couplings. Lastly, to reliably verify the existence of an additional peak in the UV tail of the spectrum, a more thorough analysis of the regularization methods is necessary, along with independent validation of the results using adiabatic regularization. 
\section*{Acknowledgements}
The authors thank Christian Capanelli, Leah Jenks, Edward W. Kolb, and Evan McDonough
for helpful discussions and correspondence. 
\\ \\
The work of B.G. is supported in part by the National Science Centre (Poland) as research projects 2020/37/B/ST2/02746 and 2023/49/B/ST2/00856. A.S. is supported by the European Union’s Horizon 2020 research and innovation programme under grant agreement No 101002846, ERC CoG CosmoChart.

\appendix
\section{Energy-momentum tensor of non-minimally coupled vector field}
\label{sec:EMT}
In this appendix, we derive the generic formula for the energy-momentum tensor of a non-minimally coupled vector field. To begin, let us recall the standard definition of the stress-energy tensor
\begin{align}
    T_{\mu \nu} := \frac{2}{\sqrt{-g}} \frac{\delta S_{\rm M}}{\delta g^{\mu \nu}},
\end{align}
with $S_{\rm M}$ being an action for a matter component. \\
To find the energy-momentum tensor of a spin-1 field $X_\mu$, whose action is given in Eq.\eqref{eq:X_action}, it is instructive to distinguish the following terms:
\begin{align}
    T_{\mu \nu}^{\rm X} = T_{\mu \nu}^{\rm{M}} + T_{\mu \nu}^{\xi_1} + T_{\mu \nu}^{\xi_2},   
\end{align}
accounting for three contributions arising from the minimal coupling, i.e., from the the first two terms in Eq.\eqref{eq:X_action}, the non-minimal coupling to the Ricci scalar (the third term in Eq.\eqref{eq:X_action}), and the non-minimal coupling to the Ricci tensor (the fourth term in Eq.\eqref{eq:X_action}). \\ Then, utilizing the following expression for the variation of the determinant of the metric tensor:
\begin{align}
    \delta(\sqrt{-g}) = - \frac{\sqrt{-g}}{2} g_{\mu \nu} \delta g^{\mu \nu},
\end{align}
one can find the well-known formula for the energy-momentum tensor of the minimally-coupled vector field
\begin{align}
    T_{\mu \nu}^{\rm M} = g_{\mu \nu} \left( \frac{1}{4} g^{\rho \sigma} g^{\alpha \beta} X_{\rho \alpha} X_{\sigma \beta} - \frac{m_X^2}{2} g^{\alpha \beta} X_\alpha X_\beta \right) - g^{\alpha \beta}X_{\mu \alpha} X_{\nu \beta} + m_X^2 X_\mu X_\nu.
\end{align}
Derivation of the remaining two contributions is less common in standard textbooks, and thus deserves special emphasis. First, we observe that the variation of the third term in Eq.\eqref{eq:X_action} is given by 
\begin{align}
    \delta \left(- \frac{\xi_1}{2} \sqrt{-g} g^{\mu \nu} R X_\mu X_\nu \right) & = - \frac{\xi_1}{2} \left[ g^{\mu \nu} R \delta(\sqrt{-g}) + \sqrt{-g} R \delta g^{\mu \nu} + \sqrt{-g} g^{\mu \nu} \delta R\right]X_\mu X_\nu \nonumber \\
    & =- \frac{\xi_1}{2} \sqrt{-g} R \left[ -\frac{1}{2} g_{\mu \nu} g^{\rho \sigma} X_\rho X_\sigma +  X_\mu X_\nu \right]  \delta g^{\mu \nu} \nonumber \\
    & \phantom{=} - \frac{\xi_1}{2} \sqrt{-g} g^{\mu \nu} \delta R X_\mu X_\nu. \label{eq:varR}
\end{align}
From the definition of the Ricci scalar, one finds
\begin{align}
    \delta R = R_{\mu \nu} \delta g^{\mu \nu} + g^{\mu \nu} \delta R_{\mu \nu}, \label{eq:VariationRicci}
\end{align}
whereas the variation of the Ricci tensor, known as the Palatini identity, reads
\begin{align}
    \delta R_{\mu \nu} = \nabla_\rho (\delta \Gamma^\rho_{\nu \mu}) - \nabla_\nu (\delta \Gamma^\rho_{\rho \mu}),  \label{eq:PalatiniI}
\end{align}
with
\begin{align}
    \delta \Gamma^\rho_{\mu \nu} = - \frac{1}{2} \left[ g_{\sigma \mu} \nabla_\nu (\delta g^{\sigma \rho}) + g_{\sigma \nu} \nabla_\mu (\delta g^{\sigma \rho}) - g_{\mu \sigma} g_{\nu \lambda} \nabla^\rho(\delta g^{\sigma \lambda}) \right]. \label{eq:GammaVar}
\end{align}
Hence,
\begin{align}
-  \frac{\xi_1}{2} \sqrt{-g} g^{\mu \nu}  X_\mu X_\nu \delta R= -  \frac{\xi_1}{2} \sqrt{-g} g^{\alpha \beta}  X_\alpha X_\beta R_{\mu \nu} \delta g^{\mu \nu} -  \frac{\xi_1}{2} \sqrt{-g} g^{\alpha \beta}  X_\alpha X_\beta g^{\mu \nu} \delta R_{\mu \nu},
\end{align}
and the variation of the second term can be explicitly written as
\begin{align}
    \mathcal{I} &\equiv -  \frac{\xi_1}{2} \sqrt{-g} g^{\alpha \beta}  X_\alpha X_\beta g^{\mu \nu} \delta R_{\mu \nu} = -  \frac{\xi_1}{2} \sqrt{-g} g^{\alpha \beta}  X_\alpha X_\beta g^{\mu \nu} \left[ \nabla_\rho (\delta \Gamma^\rho_{\nu \mu}) - \nabla_\nu (\delta \Gamma^\rho_{\rho \mu}) \right] \nonumber \\
    & =\frac{\xi_1}{4} \sqrt{-g} g^{\alpha \beta}  X_\alpha X_\beta g^{\mu \nu} \bigg\{ \nabla_\rho \left[ g_{\sigma \nu} \nabla_\mu (\delta g^{\sigma \rho}) + g_{\sigma \mu} \nabla_\nu (\delta g^{\sigma \rho}) - g_{\nu \sigma} g_{\mu \lambda} \nabla^\rho(\delta g^{\sigma \lambda}) \right] \nonumber \\
    & -\nabla_\nu \left[  g_{\sigma \rho} \nabla_\mu (\delta g^{\sigma \rho}) + g_{\sigma \mu} \nabla_\rho (\delta g^{\sigma \rho}) - g_{\rho \sigma} g_{\mu \lambda} \nabla^\rho(\delta g^{\sigma \lambda})\right] \bigg\}\non\\
    &=\frac{\xi_1}{2} \sqrt{-g} \bigg\{ \nabla_\mu \nabla_\nu (X_\sigma X^\sigma) - g_{\mu \nu} \nabla^\rho \nabla_\rho (X_\sigma X^\sigma) \bigg\}\delta g^{\mu \nu},
\end{align}
where passing from the second to third equality we have performed two integration by parts for each term. \\ 
All in all, 
\begin{align}
    \delta \left(- \frac{\xi_1}{2} \sqrt{-g} g^{\mu \nu} R X_\mu X_\nu \right) &= - \frac{\xi_1}{2} \sqrt{-g} \bigg[ (R_{\mu \nu} -\frac{1}{2} g_{\mu \nu} R) g^{\rho \sigma} X_\rho X_\sigma + R X_\mu X_\nu  \nonumber  \\
    &+ g_{\mu \nu} \nabla^\rho \nabla_\rho (X_\sigma X^\sigma) - \nabla_\mu \nabla_\nu (X_\sigma X^\sigma) \bigg]  \delta g^{\mu \nu},
\end{align}
and, thus,
\begin{align}
   T_{\mu \nu}^{\xi_1} = \xi_1 \bigg[ - R X_\mu X_\nu - G_{\mu \nu} g^{\rho \sigma} X_\rho X_\sigma - g_{\mu \nu}g^{\rho \sigma} g^{\alpha \beta} \nabla_\sigma \nabla_\rho (X_\alpha X_\beta) + g^{\rho \sigma}\nabla_\mu \nabla_\nu (X_\rho X_\sigma) \bigg].
\end{align}
Similarly, the variation of the last term in the action \eqref{eq:X_action} reads
\begin{align}
    \delta \left( \frac{\xi_2}{2} \sqrt{-g} g^{\mu \rho} g^{\nu \sigma} R_{\rho \sigma} X_\mu X_\nu\right) &= \frac{\xi_2}{2} \left[ g^{\mu \rho} g^{\nu \sigma} R_{\rho \sigma}  \delta(\sqrt{-g}) +  \sqrt{-g} g^{\nu \sigma} R_{\rho \sigma}  \delta g^{\mu \rho}\right. \nonumber \\
    &+ \left. \sqrt{-g} g^{\mu \rho} R_{\rho \sigma}  \delta g^{\nu \sigma} + \sqrt{-g} g^{\mu \rho} g^{\nu \sigma}  \delta R_{\rho \sigma}\right]X_{\mu}X_\nu \nonumber \\
    &=\frac{\xi_2}{2} \sqrt{-g}\left[ - \frac{1}{2}g^{\alpha \rho} g^{\beta \sigma}R_{\rho \sigma}X_\alpha X_\beta g_{\mu \nu} + g^{\rho \sigma}R_{\nu \sigma} X_\mu X_\rho \right. \nonumber \\
    &+ g^{\rho \sigma}R_{\mu \sigma} X_\nu X_\rho \bigg] \delta g^{\mu \nu} + \frac{\xi_2}{2} \sqrt{-g} g^{\mu \rho} g^{\nu \sigma} X_\mu X_\nu \delta R_{\rho \sigma}.
\end{align}
Inserting relation \eqref{eq:GammaVar}
into \eqref{eq:PalatiniI}, one finds that the last term above can be written as
\begin{align}
    \mathcal{J} &\equiv \frac{\xi_2}{2} \sqrt{-g} g^{\mu \rho} g^{\nu \sigma} X_\mu X_\nu \delta R_{\rho \sigma} = \frac{\xi_2}{2} \sqrt{-g} g^{\mu \rho} g^{\nu \sigma} X_\mu X_\nu \left[ \nabla_\lambda(\delta \Gamma^\lambda_{\sigma \rho}) - \nabla_\sigma (\delta \Gamma^\lambda_{\lambda \rho}) \right] \nonumber \\
    &=-\frac{\xi_2}{4} \sqrt{-g} g^{\mu \rho} g^{\nu \sigma} X_\mu X_\nu \bigg\{ \nabla_\lambda \left[g_{\alpha \sigma} \nabla_\rho( \delta g^{\alpha \lambda}) + g_{\alpha \rho} \nabla_\sigma( \delta g^{\alpha \lambda}) - g_{\sigma \alpha}g_{\rho \beta}\nabla^\lambda(\delta g^{\alpha \beta})  \right] \nonumber \\
    &-\nabla_\sigma \left[ g_{\alpha \rho} \nabla_\lambda (\delta g^{\alpha \lambda }) + g_{\alpha \lambda} \nabla_\rho (\delta g^{\alpha \lambda }) - g_{\lambda \alpha} g_{\rho \beta} \nabla^\lambda(\delta g^{\alpha \beta})\right] \bigg\} \nonumber \\
    &=-\frac{\xi_2}{4} \sqrt{-g}  \bigg\{ \nabla_\sigma \nabla_\mu(X^\sigma X_\nu) + \nabla_\sigma \nabla_\nu(X^\sigma X_\mu)  - \nabla^\sigma \nabla_\sigma (X_\mu X_\nu) - g_{\mu \nu} \nabla_\rho \nabla_\sigma(X^\rho X^\sigma) \bigg\} \delta g^{\mu \nu}.
\end{align}
The above formula can be further simplified by noting that
\begin{align}
    \bar{\mathcal{J}} &\equiv \nabla_\sigma \nabla_\mu(X^\sigma X_\nu) - \nabla_\mu \nabla_\sigma(X^\sigma X_\nu) +  \nabla_\sigma \nabla_\nu(X^\sigma X_\mu) - \nabla_\nu \nabla_\sigma(X^\sigma X_\mu) \non \\ &=
    X_\nu (\nabla_\sigma \nabla_\mu X^\sigma) - X_\nu (\nabla_\mu \nabla_\sigma X^\sigma) + X^\sigma (\nabla_\sigma \nabla_\mu X_\nu) - X^\sigma (\nabla_\mu \nabla_\sigma X_\nu) + X_\mu (\nabla_\sigma \nabla_\nu X^\sigma) \nonumber \\
    &- X_\mu (\nabla_\nu \nabla_\sigma X^\sigma) +X^\sigma(\nabla_\sigma \nabla_\nu X_\mu) - X^\sigma (\nabla_\nu \nabla_\sigma X_\mu) \non
    \\
    &= X_\nu([\nabla_\sigma, \nabla_\mu] X^\sigma) + X^\sigma([\nabla_\sigma, \nabla_\mu] X_\nu) +  X_\mu([\nabla_\sigma, \nabla_\nu] X^\sigma) + X^\sigma([\nabla_\sigma, \nabla_\nu] X_\mu). \label{eq:commDev}
\end{align}
In the absence of torsion, the commutator of covariant derivatives acting on contravariant and covariant vectors is given by
\begin{align}
    &[\nabla_\alpha, \nabla_\beta] X^\mu = R^\mu_{\nu \alpha \beta} X^\nu, && [\nabla_\alpha, \nabla_\beta] X_\mu = - R^\nu_{\mu \alpha \beta} X_\nu,
\end{align}
which, in turn, implies
\begin{align}
    \bar{\mathcal{J}} &= X_ \nu R^\sigma_{\rho \sigma \mu} X^\rho - X^\sigma R^\rho_{\nu \sigma \mu} X_\rho + X_\mu R^\sigma_{\rho \sigma \nu} X^\rho - X^\sigma R^\rho_{\mu \sigma \nu} X_\rho \non \\
    &= g^{\sigma \rho} \left(R_{\rho \mu} X_\sigma X_\nu + R_{\rho \nu} X_\sigma  X_\mu \right).
\end{align}
As a result
\begin{align}
    \mathcal{J} &= \frac{\xi_2}{4} \sqrt{-g} \bigg\{ - g^{\sigma \rho}(R_{\rho \mu} X_\sigma X_\nu + R_{\rho \nu} X_\sigma  X_\mu) +\nabla^\sigma \nabla_\sigma(X_\mu X_\nu) + g_{\mu \nu} \nabla_\rho \nabla_\sigma(X^\rho X^\sigma) \non \\
    &-\nabla_\mu \nabla_\sigma(X^\sigma X_\nu) - \nabla_\nu \nabla_\sigma(X^\sigma X_\mu) \bigg \} \delta g^{\mu \nu},
\end{align}
which, in turn, implies
\begin{align}
    \delta \left( \frac{\xi_2}{2} \sqrt{-g} g^{\mu \rho} g^{\nu \sigma} R_{\rho \sigma} X_\mu X_\nu\right) &= \frac{\xi_2}{4} \sqrt{-g}\bigg[ - g_{\mu \nu} g^{\alpha \rho} g^{\beta \sigma}R_{\rho \sigma}X_\alpha X_\beta  +  2 g^{\rho \sigma}R_{\nu \sigma} X_\mu X_\rho \non \\
    &+ 2g^{\rho \sigma}R_{\mu \sigma} X_\nu X_\rho  + \nabla^\sigma \nabla_\sigma(X_\mu X_\nu) + g_{\mu \nu} \nabla^\rho \nabla^\sigma(X_\rho X_\sigma) \non \\
    &-\nabla_\mu \nabla_\sigma(X^\sigma X_\nu) - \nabla_\nu \nabla_\sigma(X^\sigma X_\mu) \bigg] \delta g^{\mu \nu}.
\end{align}
Consequently,
\begin{align}
    T_{\mu \nu}^{\xi_2}  &=  \frac{\xi_2}{2} \bigg[ - g_{\mu \nu}g^{\alpha \rho} g^{\beta \sigma}R_{\rho \sigma}X_\alpha X_\beta  + 2 g^{\rho \sigma}R_{\nu \sigma} X_\mu X_\rho +2 g^{\rho \sigma}R_{\mu \sigma} X_\nu X_\rho  +  g^{\rho \sigma }\nabla_\rho \nabla_\sigma(X_\mu X_\nu) \non \\
    & +  g_{\mu \nu} g^{\lambda \rho}  g^{\kappa \sigma} \nabla_\lambda \nabla_\kappa(X_\rho X_\sigma)- g^{\lambda \sigma} \nabla_\mu \nabla_\sigma(X_\lambda X_\nu) - g^{\lambda \sigma}\nabla_\nu \nabla_\sigma(X_\lambda X_\mu) \bigg].
\end{align} 
\section{Energy density}
\label{sec:EnergyDensity}
The energy density of the $X_\mu$ field, $\rho_X \equiv g^{00}T_{00}^{\rm X}$, can be written as a sum of three contributions
\begin{align}
    \rho_X \equiv \rho_X^{\rm M}+\rho_X^{\xi_1} + \rho_X^{\xi_2},
\end{align}
arising from minimal coupling to gravity, non-minimal coupling to the Ricci scalar, and non-minimal coupling to the Ricci tensor, respectively.  \\
In conformal coordinates, $g_{\mu \nu} = (a^2, - a^2, -a^2, -a^2)$, $g^{\mu \nu} = (a^{-2}, - a^{-2}, -a^{-2}, -a^{-2}$), and the only non-zero components of the connection $\Gamma^\lambda_{\mu \nu}$ are
\begin{align}
    \Gamma^0_{00} = \Gamma^0_{ii} =  \Gamma^i_{i0} =  \Gamma^i_{0i} =  a H.
\end{align}
Hence, 
\begin{align}
    &R_{00} = - 3 \mathcal{H}^\prime, &&R_{ij} = \left( \mathcal{H}^\prime + 2 \mathcal{H}^2 \right)\delta_{ij},
\end{align}
with 
\begin{align}
    &\mathcal{H} \equiv \frac{a^{\prime}}{a} = H a, &&\mathcal{H}^{\prime} = - \frac{(1+3w)}{2} a^2 H^2.
\end{align}
The Ricci scalar is
\begin{align}
    R = - 6 a^{-2}\left(\mathcal{H}^\prime + \mathcal{H}^2 \right),
\end{align}
and the non-zero components of the Einstein tensor are 
\begin{align}
    &G_{00} = 3 \mathcal{H}^2,     &&G_{ij} = -\left( 2 \mathcal{H}^\prime +  \mathcal{H}^2 \right)\delta_{ij}.
\end{align}
As a result, one finds 
\begin{align}
   \rho_X^{\rm M} &=  \frac{1}{2 a^{4}} \bigg[  (X_i^\prime - \partial_i X_0)^2 + \frac{1}{2} (\partial_j X_i - \partial_i X_j)^2 + a^{2} m_X^2 X_0^2 + a^{2} m_X^2 X_i^2 \bigg], \\
    \rho_{X}^{\xi_1} 
    &= \frac{\xi_1}{a^{4}} \bigg[ - 9 w a^{2} H^2   X_0^2 + 3 H^2 a^{2}  X_i^2 + \nabla_i^2 (X_0^2) - \nabla_i^2 (X_j^2) \bigg] \label{eq:rhoxi}, \\
   \rho_{X}^{\xi_2} &= \frac{\xi_2}{2 a^4} \bigg[ \frac{9}{2}(1+3w) a^2 H^2 X_0^2 +\frac{3}{2}(-1+w) a^2 H^2 X_i^2 -  \nabla_i^2 (X_0^2)  +  \nabla_i \nabla_j (X_i X_j) \non \\
   &- [\nabla_0, \nabla_i](X_i X_0)   \bigg]. \label{eq:rhoxi2}
\end{align} 
To simplify the last two expressions, one needs to calculate the following derivatives:
\begin{align}
    \nabla_i^2(X_0^2) = 2 X_0 \nabla_i^2(X_0) + 2 (\nabla_i X_0)^2, \label{eq:ex1}
\end{align}
\begin{align}
     \nabla_i^2(X_j^2) = 2 X_j \nabla_i^2(X_j) + 2 (\nabla_i X_j)^2, \label{eq:ex2}
\end{align}
\begin{align}
     \nabla_i \nabla_j (X_i X_j) = X_j \nabla_i (\nabla_j X_i) + (\nabla_i X_i) (\nabla_j X_j) + (\nabla_i X_j) (\nabla_j X_i) + X_i \nabla_i(\nabla_j X_j), \label{eq:ex3}
\end{align}
and 
\begin{align}
    [\nabla_0, \nabla_i] (X_i X_0)  = - X_0 R^\rho_{i0i} X_\rho - X_i R^\rho_{00i} X_\rho = -   \mathcal{H}^\prime (3X_0^2 + X_i^2).
\end{align}
Employing a definition of the covariant derivative:
\begin{align}
    \nabla_\mu X_\nu = \partial_\mu X_\nu - \Gamma^\rho_{\mu \nu} X_\rho, 
\end{align}
one shows that 
\begin{align}
    \nabla_\sigma (\nabla_\mu X_\nu ) = \partial_\sigma( \nabla_\mu X_\nu) - \Gamma^\lambda_{\sigma \mu} (\nabla_\lambda X_\nu) - \Gamma^\lambda_{\sigma \nu} (\nabla_\mu X_\lambda).
\end{align}
Consequently, expressions \eqref{eq:ex1} - \eqref{eq:ex3} become
\begin{align}
    \nabla_i^2(X_0^2) &= 2 \bigg[ X_0 \partial_i^2 X_0  + (\partial_i X_0)^2 - 2 \Gamma^i_{i0} X_0 \partial_i X_i - 2 \Gamma^i_{i0} X_i \partial_i X_0 - \Gamma_{ii}^0 X_0 X_0^\prime 
    + \Gamma_{i0}^i \Gamma^0_{ii} X_0^2 \non \\
    &+  \Gamma_{i0}^i \Gamma_{i0}^i X_i^2\bigg], \\
    \nabla_i \nabla_j (X_i X_j) 
    &= X_j \partial_i \partial_j X_i + X_i \partial_i \partial_j X_j + (\partial_i X_i) (\partial_j X_j) + (\partial_i X_j) (\partial_j X_i) - 2 \Gamma^0_{ii} X_i X_i^\prime \nonumber \\
    &-4 \Gamma^0_{ii} \left(X_i \partial_i X_0+X_0 \partial_i X_i \right) + 2 \Gamma^0_{ii} \Gamma^0_{ii} X_0^2 + 2 \Gamma^0_{ii} \Gamma^i_{0i}X_i^2,
    \end{align}
Inserting the above results back into Eqs.\eqref{eq:rhoxi} and \eqref{eq:rhoxi2}, one obtains
\begin{align}
    \rho_{X}^{\xi_1}
    &= \frac{\xi_1}{a^{4}} \bigg[ 2 \bigg( X_0 \partial_i^2 X_0 -  X_j \partial_i^2 X_j +   (\partial_i X_0)^2 -  (\partial_i X_j)^2 \bigg) \non \\
    &- 3 aH (X_0 X_0^\prime + X_0^\prime X_0 - X_i X_i^\prime -  X_i^\prime X_i) - 9 w a^{2} H^2   X_0^2 - 3 H^2 a^{2}  X_i^2
    \bigg], \\
     \rho_{X}^{\xi_2}
    &=  \frac{\xi_2}{2 a^4}\bigg[ X_j \partial_i \partial_j X_i + X_i \partial_i \partial_j X_j  + (\partial_i X_j)(\partial_j X_i) +(\partial_i X_i)(\partial_j X_j)  - 2 X_0 \partial_i^2 X_0 - 2 (\partial_i X_0^2) \non \\ 
    &+ 3 aH (X_0 X_0^\prime + X_0^\prime X_0 - X_i X_i^\prime -  X_i^\prime X_i) +  9(1+w) a^2 H^2 X_0^2 +2 a^2 H^2 X_i^2  \bigg].
\end{align}
Note that for the coupling to Einstein tensor, i.e., for $ \xi_2 = 2 \xi_1$, the above  formulas simplify as
\begin{align}
    \rho_X^{\xi_2 = 2 \xi_1} &= \frac{\xi_1}{a^4} \bigg[ -2 X_j \partial_i^2 X_j - 2 (\partial_i X_j)^2 + X_j \partial_i \partial_k X_i +  X_i \partial_i \partial_k X_j + (\partial_i X_j)(\partial_j X_i) \non \\
     &+ 9 a^2 H^2 X_0^2 - a^2 H^2 X_i^2  \bigg],
\end{align}
which agrees with the result found in Ref.\cite{Ozsoy:2023gnl}.\\
 Then, using decomposition Eq.\eqref{Fourier_rep}, one finds
 \begin{align}
     \rho_X^{\rm M} &= \frac{1}{2a^4} \int \frac{d^3 k}{(2 \pi)^3} \int \frac{d^3 q}{(2 \pi)^3} e^{i(\Vec{k}+ \Vec{q})\cdot \Vec{x}} \bigg\{ \Vec{X}^\prime(k) \cdot  \Vec{X}^\prime(q) - i \left[ X_0(k) \Vec{k} \cdot \Vec{X}^\prime(q) + X_0(q) \Vec{q} \cdot \Vec{X}^\prime(k) \right] \non \\
     &+ (a^2 m_X^2 - \Vec{k} \cdot \Vec{q} )\left[ X_0(k) X_0(q) + \Vec{X}(k) \cdot \Vec{X}(q) \right] + [\Vec{k} \cdot \Vec{X}(q)] [\Vec{q} \cdot \Vec{X}(k)]  \bigg\}, \label{eq:rho1} \\
     \rho_X^{\xi_1} &=  \frac{\xi_1}{a^4} \int \frac{d^3 k}{(2 \pi)^3} \int \frac{d^3 q}{(2 \pi)^3} e^{i(\Vec{k}+ \Vec{q})\cdot \Vec{x}} \bigg\{ - 3 H^2 a^2 \left[  \Vec{X}(k) \cdot \Vec{X}(q) + 3w X_0(k) X_0(q) \right] \non \\
     &-2 \left[q^2 + \vec{k} \cdot \Vec{q} \right] \left[ X_0(k) X_0(q) - \Vec{X}(k) \cdot \Vec{X}(q)  \right] - 3 aH \left[X_0(k) X_0^\prime(q) + X_0^\prime (k)X_0(q)\right] \non \\
     &+ 3 aH \left[\vec{X}(k) \cdot \vec{X}^\prime(q) + \vec{X}^\prime(k) \cdot \vec{X}(q)\right]  \bigg\}, \label{eq:rho2}\\
     \rho_X^{\xi_2} &= \frac{\xi_2}{2 a^4} \int \frac{d^3 k}{(2 \pi)^3} \int \frac{d^3 q}{(2 \pi)^3} e^{i(\Vec{k}+ \Vec{q})\cdot \Vec{x}} \bigg\{ a^2 H^2 \bigg[ 2 \Vec{X}(k) \cdot \Vec{X}(q) + 9(1+w) X_0(k) X_0(q) \bigg] \non \\
    &+ 3 aH \left[ X_0(k) X_0^\prime(q) + X_0^\prime(k) X_0(q) \right] - 3 aH \left[ \vec{X}(k) \cdot  \vec{X}^\prime(q) + \vec{X}^\prime(k) \cdot  \vec{X}(q) \right] \non \\
    &- [ \Vec{q} \cdot \Vec{X}(k)][ \Vec{q} \cdot \Vec{X}(q)] -  [\Vec{k} \cdot \Vec{X}(k)][\Vec{k} \cdot  \Vec{X}(q)]  - [\Vec{k} \cdot \Vec{X}(q)] [\Vec{q}\cdot \Vec{X}(k)] - [\Vec{q}\cdot \Vec{X}(q)] [\Vec{k}\cdot \Vec{X}(k)] \non \\ &+ 2 \left[ k^2 +  \Vec{k}\cdot \Vec{q}\right] X_0(k) X_0(q) \bigg\},
 \end{align}
 where we have left time dependence implicit. \\\\
 The zero component of the vector field, $X_0$, does not have a kinetic term. Furthermore, it does not mix longitudinal and transverse modes, allowing it to be isolated from the other three physical components of the $X_\mu$ field. It can be shown that 
 \begin{align}
     &X_0(\tau, k) = \frac{-i \Vec{k} \cdot \Vec{X}^\prime(\tau, k)}{k^2 + a^2 m_X^2}, && X_0^\prime(\tau, k) = i \Vec{k} \cdot \Vec{X}(\tau, k) - 2 \mathcal{H}X_0(\tau, k). 
 \end{align}
Substituting the above relations into Eqs.\eqref{eq:rho1}-\eqref{eq:rho2}, one gets
\begin{align}
     \rho_X^{\rm M} &= \frac{1}{2a^4} \int \frac{d^3 k}{(2 \pi)^3} \int \frac{d^3 q}{(2 \pi)^3} e^{i(\Vec{k}+ \Vec{q})\cdot \Vec{x}} \Bigg\{ \Vec{X}^\prime(k) \cdot  \Vec{X}^\prime(q) - \frac{[\Vec{k} \cdot \Vec{X}^\prime(k) ][\Vec{k} \cdot \Vec{X}^\prime(q)]}{k^2 + a^2 m_X^2} - \frac{[\Vec{q} \cdot \Vec{X}^\prime(q)][ \Vec{q} \cdot \Vec{X}^\prime(k)]}{q^2 +a^2 m_X^2}\nonumber \\
     &+ ( a^2 m_X^2 - \Vec{k} \cdot \Vec{q}) \left[ - \frac{[\Vec{k} \cdot \Vec{X}^\prime(k)] [\Vec{q} \cdot \Vec{X}^\prime(q)] }{(k^2 + a^2 m_X^2)(q^2 + a^2 m_X^2)} + \Vec{X}(k) \cdot \Vec{X}(q) \right] + [\Vec{k} \cdot \Vec{X}(q)] [\Vec{q} \cdot \Vec{X}(k)] \Bigg\}, \\
      \rho_X^{\xi_1} &= \frac{\xi_1}{a^4} \int \frac{d^3 k}{(2 \pi)^3} \int \frac{d^3 q}{(2 \pi)^3} e^{i(\Vec{k}+ \Vec{q})\cdot \Vec{x}} \Bigg\{ - 3 H^2 a^2 \left[  \Vec{X}(k) \cdot \Vec{X}(q) - 3w \frac{[\Vec{k} \cdot \Vec{X}^\prime(k)] [\Vec{q} \cdot \Vec{X}^\prime(q)]}{(k^2 + a^2 m_X^2)(q^2 + a^2 m_X^2)} \right] \non \\
      & + 2 \left[  q^2 + \vec{k} \cdot \Vec{q} \right] \left[  \frac{[\Vec{k} \cdot \Vec{X}^\prime(k)] [\Vec{q} \cdot \Vec{X}^\prime(q)]}{(k^2 + a^2 m_X^2)(q^2 + a^2 m_X^2)} + \Vec{X}(k) \cdot \Vec{X}(q)  \right] \non \\
      &- 3 a H \Bigg[ \frac{[\vec{k} \cdot \Vec{X}^\prime(k)][\Vec{q} \cdot \Vec{X}(q)]}{k^2 + a ^2 m_X^2} +  \frac{[\vec{k} \cdot \Vec{X}(k)][\Vec{q} \cdot \Vec{X}^\prime(q)]}{q^2 + a ^2 m_X^2} + 4 a H \frac{[\Vec{k} \cdot \Vec{X}^\prime(k)] [\Vec{q} \cdot \Vec{X}^\prime(q)]}{(k^2 + a^2 m_X^2)(q^2 + a^2 m_X^2)} \non \\
      &- \Vec{X}(k) \cdot \Vec{X}^\prime(q) -  \Vec{X}(k)^\prime \cdot \Vec{X}(q) \Bigg] \Bigg\}, \\
      \rho_X^{\xi_2} &= \frac{\xi_2}{2 a^4} \int \frac{d^3 k}{(2 \pi)^3} \int \frac{d^3 q}{(2 \pi)^3} e^{i(\Vec{k}+ \Vec{q})\cdot \Vec{x}} \Bigg\{ a^2 H^2 \bigg[ 2 \Vec{X}(k) \Vec{X}(q) - 9(1+w) \frac{[\Vec{k}\cdot \Vec{X}^\prime(k)][\Vec{q}\cdot \Vec{X}^\prime(q)]}{(k^2 + a^2 m_X^2)(q^2 + a^2 m_X^2)} \bigg] \non \\
      &+3aH \Bigg[ \frac{[\Vec{k}\cdot \Vec{X}^\prime(k) ][\Vec{q}\cdot \Vec{X}(q) ]}{k^2 + a^2m_X^2} + \frac{[\Vec{k}\cdot \Vec{X}(k) ][\Vec{q}\cdot \Vec{X}^\prime(q) ]}{q^2 + a^2m_X^2}+ 4 a H \frac{[\Vec{k}\cdot \Vec{X}^\prime(k) ][\Vec{q}\cdot \Vec{X}^\prime(q) ]}{(k^2 + a^2m_X^2)(q^2 + a^2m_X^2)} \non \\
    &- \Vec{X}(k) \cdot \Vec{X}^\prime(q) - \Vec{X}^\prime(k) \cdot \Vec{X}(q) \Bigg] - [ \Vec{q} \cdot \Vec{X}(k)][ \Vec{q} \cdot \Vec{X}(q)] -  [\Vec{k} \cdot \Vec{X}(k)][\Vec{k} \cdot  \Vec{X}(q)] \non \\
    &- [\Vec{k} \cdot \Vec{X}(q)] [\Vec{q}\cdot \Vec{X}(k)] - [\Vec{q}\cdot \Vec{X}(q)] [\Vec{k}\cdot \Vec{X}(k)]-2 \left[ k^2 + \Vec{k}\cdot \Vec{q}\right] \frac{[\Vec{k} \cdot \Vec{X}^\prime(k)][\Vec{q} \cdot \Vec{X}^\prime(q)]}{(k^2 + a^2 m_X^2)(q^2 + a^2 m_X^2)} \Bigg\}.
\end{align}
Then, let us promote the classical $X_\mu$ field to the quantum operator $\hat{X}_\mu$ and compute the expectation value of the energy density operator: 
\begin{align}
    \langle \hat{\rho}_X \rangle = \langle \hat{\rho}_X^{\rm M} \rangle + \langle \hat{\rho}_X^{\xi_1 } \rangle + \langle \hat{\rho}_X^{\xi_2} \rangle. 
\end{align}
To that end, we introduce a canonical vector field variable
\begin{align}
    \hat{\vec{X}}(\tau, \Vec{x}) = \sum_{\lambda = \pm, {\rm L}} \int \frac{d^3 k}{(2 \pi)^3} \Vec{\epsilon}_\lambda(\Vec{k}) e^{i \Vec{k} \cdot \Vec{x}} \hat{\mathcal{X}}_\lambda(\tau, k), && \hat{\mathcal{X}}_\lambda(\tau, k) \equiv \hat{a}_\lambda(\Vec{k}) \mathcal{X}_{\lambda}(\tau, k) + \hat{a}^\dagger_\lambda(-\Vec{k}) \mathcal{X}^\star_{\lambda}(\tau, k),  
\end{align}
where the annihilation and creation operators satisfy the following commutation relations:
\begin{align}
    &[\hat{a}_\lambda(\Vec{k}), \hat{a}^\dagger_{\lambda^\prime}(\Vec{q})] = \delta_{\lambda \lambda^\prime} (2 \pi)^3 \delta^{(3)}(\Vec{k} - \Vec{q}), & [\hat{a}_\lambda(\Vec{k}), \hat{a}_{\lambda^\prime}(\Vec{q})] = 0, && [\hat{a}^\dagger_\lambda(\Vec{k}), \hat{a}^\dagger_{\lambda^\prime}(\Vec{q})] = 0,
\end{align}
the polarization vectors are 
\begin{align}
    &\Vec{k} \cdot \Vec{\epsilon}_\pm (k) = 0, &&\Vec{k} \cdot \Vec{\epsilon}_{\rm L} (k) = k, 
    &\Vec{k} \times \Vec{\epsilon}_\pm (k) = \mp i k \Vec{\epsilon}_\pm(k), &&\Vec{k} \times \Vec{\epsilon}_{\rm L} (k) = 0.
\end{align}
Note that the mode functions $\mathcal{X}_\lambda$ are normalized by the Wronskian condition 
\begin{align}
    W[\mathcal{X}_\lambda, \mathcal{X}_\lambda^\star] = \mathcal{X}_\lambda^\prime \mathcal{X}_\lambda^\star - \mathcal{X}_\lambda^{\star \prime} \mathcal{X}_\lambda = -i. 
\end{align}
Hence, we can identify $\mathcal{X}_\pm$ with the transverse mode function satisfying Eq.\eqref{eq:EoMT}, i.e., $\mathcal{X}_\pm = X_\pm$, whereas the mode function $\mathcal{X}_{\rm L}$ for the longitudinal polarization,  satisfying Eq.\eqref{eq:EoMT} is related to the longitudinal component of the vector field through Eq.\eqref{eq:Ltrans}. \\ 
Next, let us define the following power spectra:
\begin{align}
\langle \hat{\mathcal{X}}_{\lambda}(\tau, k) \cdot \hat{\mathcal{X}}_{\lambda^\prime}(\tau, q) \rangle &= \delta_{\lambda \lambda^\prime} (2 \pi)^3 \delta^{(3)}(\Vec{k} + \Vec{q}) \frac{2 \pi^2}{k^3} \mathcal{P}_{\mathcal{X}_\lambda}(\tau, k), \label{pow_spec_a}\\
\langle \hat{\mathcal{X}}_\lambda^\prime(\tau, k) \cdot \hat{\mathcal{X}}_{\lambda^\prime}^\prime(\tau, q) \rangle &= \delta_{\lambda \lambda^\prime}  (2 \pi)^3 \delta^{(3)}(\Vec{k} + \Vec{q}) \frac{2 \pi^2}{k^3} \mathcal{P}_{\mathcal{X}_\lambda^\prime}(\tau, k), \label{pow_spec_b}\\
\langle \hat{\mathcal{X}}_\lambda(\tau, k) \cdot \hat{\mathcal{X}}_{\lambda^\prime}^\prime(\tau, q) \rangle + \langle \hat{\mathcal{X}}_\lambda^\prime(\tau, k) \cdot \hat{\mathcal{X}}_{\lambda^\prime}(\tau, q) \rangle &= \delta_{\lambda \lambda^\prime}  (2 \pi)^3 \delta^{(3)}(\Vec{k} + \Vec{q}) \frac{2 \pi^2}{k^3} \mathcal{P}_{\mathcal{X}_\lambda\mathcal{X}_\lambda^\prime}(\tau, k).
\label{pow_spec_c}
\end{align}
The total energy density of the vector field can be expressed as the sum of the energy densities of the longitudinal and transverse components
\begin{align}
       \langle \hat{\rho}_X \rangle =    \langle \hat{\rho}_{\rm L } \rangle  +    \langle \hat{\rho}_{\pm} \rangle,
\end{align}
where 
\begin{align}
       \langle \hat{\rho}_{\rm L} \rangle &= \langle \hat{\rho}_{\rm L}^{\rm M} \rangle + \langle \hat{\rho}_{\rm L}^{\xi_1} \rangle + \langle \hat{\rho}_{\rm L}^{\xi_2} \rangle, \label{eq:edL} \\
       \langle \hat{\rho}_{\pm} \rangle &= \langle \hat{\rho}_{\pm}^{\rm M} \rangle + \langle \hat{\rho}_{\pm}^{\xi_1} \rangle + \langle \hat{\rho}_{\pm}^{\xi_2} \rangle. \label{eq:edT}
\end{align}
After some algebra, one finds
\begin{align}
    \langle \hat{\rho}_{\rm L}^{
\rm M} \rangle &= \frac{1}{2 a^4} \int \frac{d^3 k}{(2 \pi)^3} \frac{2 \pi^2}{k^3} \bigg\{\frac{a^2 m_X^2}{k^2 + a^2 m_X^2} A_{\rm L}^2 \mathcal{P}_{\mathcal{X}^\prime_{\rm L}} +   \frac{a^2 m_X^2}{k^2 + a^2 m_X^2} A_{\rm L} A_{\rm L}^\prime \mathcal{P}_{\mathcal{X}_{\rm L}\mathcal{X}_{\rm L}^\prime} \non \\
& + \left[a^2 m_X^2 A_{\rm L}^2 + \frac{a^2 m_X^2}{k^2 + a^2 m_X^2}   (A_{\rm L}^{\prime})^2 \right] \mathcal{P}_{\mathcal{X}_L} \bigg\},
\\
\langle \hat{\rho}_{\rm L}^{
\xi_1} \rangle &= \frac{\xi_1}{a^4} \int \frac{d^3 k}{(2 \pi)^3} \frac{2 \pi^2}{k^3} \Bigg\{ 3 (a H)^2  \frac{k^2}{(k^2 + a^2 m_X^2)^2} \left( 4 - 3 w \right) A_{\rm L}^2 \mathcal{P}_{\mathcal{X}_{\rm L}^\prime} \non \\
&+ \Bigg[ 3 (a H)^2 (4  - 3w) \frac{k^2}{(k^2 + a^2 m_X^2)^2} A_{\rm L}^\prime A_{\rm L}  + 3 a H  \frac{2 k^2 + a^2 m_X^2}{k^2 + a^2 m_X^2}  A_{\rm L}^2 \Bigg] \mathcal{P}_{\mathcal{X}_{\rm L} \mathcal{X}_{\rm L}^\prime} \non \\
&+ \Bigg[3 (aH)^2 \left(4-3w \right) \frac{k^2}{(k^2 + a^2 m_X^2)^2}(A_{\rm L}^\prime)^2 + 6 a H \frac{2k^2 + a^2 m_X^2}{k^2 + a^2 m_X^2} A_{\rm L} A_{\rm L}^\prime - 3 a^2 H^2 A_{\rm L}^2 \Bigg] \mathcal{P}_{\mathcal{X}_{\rm L}} \Bigg\}, \\
\langle \hat{\rho}_{\rm L}^{
\xi_2} \rangle &= \frac{\xi_2}{2 a^4} \int \frac{d^3 k}{(2 \pi)^3} \frac{2 \pi^2}{k^3} \Bigg\{ 3(aH)^2 (3w-1) \frac{k^2}{(k^2 + a^2 m_X^2)^2} A_{\rm L}^2 \mathcal{P}_{\mathcal{X}_{\rm L}^\prime} \non \\
&+ \Bigg[ 3(aH)^2 (3w -1) \frac{k^2}{(k^2 + a^2 m_X^2)^2} A_{\rm L}^\prime A_{\rm L} - 3 a H \frac{2k^2 + a^2 m_X^2}{k^2 + a^2 m_X^2} A_{\rm L}^2  \Bigg] \mathcal{P}_{\mathcal{X}_{\rm L} \mathcal{X}_{\rm L}^\prime} \non \\
&+\Bigg[ 3(aH)^2 \frac{k^2}{(k^2+ a^2 m_X^2)^2}(3w-1) (A_{\rm L}^\prime)^2 + 6 a H \frac{2k^2 + a^2 m_X^2}{(k^2 + a^2 m_X^2)^2} A_{\rm L}^\prime A_{\rm L} + 2 (a H)^2 A_{\rm L}^2  \Bigg] \mathcal{P}_{\mathcal{X}_{\rm L}} \Bigg\},
\end{align}
and 
\begin{align}
    \langle \hat{\rho}_{\pm}^{
\rm M} \rangle &= \frac{1}{2 a^4} \int \frac{d^3 k}{(2 \pi)^3} \frac{2 \pi^2}{k^3} \bigg\{ \mathcal{P}_{\mathcal{X}^\prime_{\pm}} (\tau, k) + (k^2 + a^2 m_X^2) \mathcal{P}_{\mathcal{X}_{\pm}} (\tau, k) \bigg\}, \\
\langle \hat{\rho}_{\pm}^{
\xi_1} \rangle  &=  \frac{\xi_1}{a^4} \int \frac{d^3 k}{(2 \pi)^3} \frac{2 \pi^2}{k^3} \bigg\{ -3 a^2 H^2 \mathcal{P}_{\mathcal{X}_\pm}(\tau, k) + 3 a H \mathcal{P}_{\mathcal{X}_\pm \mathcal{X}_\pm^\prime }\bigg\},\\
\langle \hat{\rho}_{\pm}^{
\xi_2} \rangle  &=  \frac{\xi_2}{a^4} \int \frac{d^3 k}{(2 \pi)^3} \frac{2 \pi^2}{k^3} \bigg\{ 2 a^2 H^2 \mathcal{P}_{\mathcal{X}_\pm}(\tau, k) - 3 a H \mathcal{P}_{\mathcal{X}_\pm \mathcal{X}_\pm^\prime }\bigg\}. 
\end{align}

\bibliography{Bibliography}
\bibliographystyle{JHEP}
\end{document}